\documentclass[12pt,a4paper]{article}
\usepackage[dvips]{epsfig}

\begin{document}
\author{\bf Yu.A. Markov$\!\,$\thanks{e-mail:markov@icc.ru}
$\,$, M.A. Markova$^*$, and A.N. Vall$\!\,$\thanks{e-mail:vall@irk.ru}}
\title{Nonlinear dynamics of soft boson\\
collective excitations in hot QCD plasma III:\\
bremsstrahlung and energy losses}
\date{\it\normalsize $^{\ast}\!$Institute of System Dynamics and Control Theory SB RAS\\
P.O. Box 1233, 664033 Irkutsk, Russia\\
$^{\dagger}\!$Irkutsk State
University, Department of Theoretical Physics,\\ 664003, Gagarin blrd,
20, Irkutsk, Russia}
\thispagestyle{empty}
\maketitle{}
\def\theequation{\arabic{section}.\arabic{equation}}
\hspace{6.8cm} {\bf Abstract} \hspace{7cm}\, 
{\small Within of the framework of semiclassical approximation 
a general formalism for deriving an
effective current generating bremsstrahlung of arbitrary number of
soft gluons (longitudinal or transverse ones) in scattering of
higher-energy parton off thermal parton in hot quark-gluon plasma
with subsequent extension to two and more scatterers, is obtained.
For the case of static color centers an expression for energy loss
induced by usual bremsstrahlung of lowest-order with allowance for
an effective temperature-induced gluon mass and finite mass of the
projectile (heavy quark), is derived. The detailed analysis of
contribution to radiation energy loss associated with existence of
effective three-gluon vertex induced by hot QCD medium, is performed.
It is shown that in general, the bremsstrahlung associated with this
vertex have no sharp direction (as in the case of usual
bremsstrahlung) and therefore here, we can expect an absence of
suppression effect due to multiple scattering. For the case of two
color static scattering centers it was shown that the problem of
calculation of bremsstrahlung induced by four-gluon hard thermal loop
(HTL) vertex correction can be reduced to the problem of the
calculation of bremsstrahlung induced by three-gluon HTL correction.
It was shown that for limiting value of soft gluon occupation number
$N_{\bf k}\sim 1/\alpha_s$ all higher processes of bremsstrahlung of
arbitrary number of soft gluons become of the same order in coupling,
and the problem of resummation of all relevant contributions to
radiation energy loss of fast parton, arises. An explicit expression
for matrix element of two soft gluon bremsstrahlung in small angles
approximation is obtained.}


\newpage

\section{Introduction}
\setcounter{equation}{0}

In the third part of our work we complete an analysis of dynamics
of boson excitations in hot QCD-medium at the soft momentum scale,
started in \cite{markov1, markov2} (to be referred to as "Paper I"
and "Paper II" through this text) in the framework of {\it hard
thermal loop} effective theory. Here, we focus our research on the
study of the bremsstrahlung of soft gluons (transverse and
longitudinal ones) by high-energy color particle (parton) induced
by collisions in the quark-gluon plasma. This color-charged parton
can be external one with respect to medium or thermal one, and
will be denoted by Greek letter $\alpha$ in the subsequent
discussion. For the sake of simplification we consider QGP
confined in an unbounded volume.

Here, we present a formalism permitting to calculate by systematic way
the probabilities of the bremsstrahlung in the
semiclassical approximation with taking account of thermal effects of
medium. These effects are appeared not only in `dressing' gluon
propagator, but also in `dressing' of vertices
of interaction by thermal corrections. At present Gyulassy-Wang model
\cite{gyulassy1,wang1} used for description of  radiation processes
in QCD medium is generally accepted one.
It presents the QGP as a system of Debye
screened color centers. The Debye screening mass $\mu_D$ plays the
role of a natural infrared cut-off that enables us to avoid infrared
divergency, which arise when we integrate over transfer momentum
${\bf q}$. Thus in this case in principal the whole effect of medium is
reduced to appearance of the only parameter $\mu_D$ (if forget at the
moment about influence of multiple scattering on bremsstrahlung). But
generally speaking, influence of medium is not restricted by this.
Fast parton can scatters not only off static color center, but also
scatters off thermal partons which form a screening `cloud' of static
center with subsequent bremsstrahlung of soft gluon. In usual plasma
such type of the bremsstrahlung was first considered by Akopian and
Tsytovich \cite{akopian,tsytovich} and named {\it transition bremsstrahlung}.
The transition bremsstrahlung is a purely
collective effect. In hot QCD plasma such type of bremsstrahlung is
produced in the leading order in coupling (in the framework of
semiclassical picture) by precession of color classical vectors of
thermal partons forming a screening cloud and
effective described by the entering of three-gluon hard thermal loop
vertex correction $\delta \Gamma_{3g}$ \cite{markov3}.
Thus the picture of `dressed' color particles
manifests itself not only in collisions and scattering, but also in
emission processes. By virtue of the fact that thermal particles of
polarization radiate, we can expect that angle distribution of such
type of the bremsstrahlung have no sharp direction as it takes place
in usual Compton bremsstrahlung, and therefore it will not to be
suppressed by multiple scattering -- Landau-Pomeranchuk-Migdal (LPM)
effect \cite{landau}. One of the aims of our work is defining
frequency range $\omega$ of emission quanta, where HTL vertex
corrections can give essential contribution to radiation energy loss.
This region is restricted by frequencies $\omega$ of order $gT$,
where $T$ is a temperature of system. In the region $\omega \gg gT$ a
contribution to energy loss induced by the effective vertex $\delta
\Gamma_{3g}$, is suppressed by coupling $g$.

Although our consideration is focused on the study of static limit of
scatterers, nevertheless obtained initial expressions for the
probabilities of bremsstrahlung enable us in principal to extend the
results of the work to more general case of moving scatterers. At
first consideration of such (more physical) situation was made by
Arnold, Moore and Yaffe \cite{arnold} starting from somewhat
different statements. A motivation to this is the fact that typical
scatterers are themselves moving at nearly the speed of light, and
they produce dynamically screened color electric and magnetic fields
formed fluctuating background field in which bremsstrahlung and the
LPM effect take place. Note that a problem of importance of
accounting of dynamical screening in the magnetic part of the
one-gluon exchange in the study of radiative energy loss of fast
parton propagating through QGP, has raised in the work \cite{wang2}.
The formalism developed by Arnold, Moore and Yaffe \cite{arnold}  on
the background of thermal field theory has valuable advantage in
comparison with different approaches (see below), since it enables us
to correctly treat the bremsstrahlung and the LPM effect at all
energies of emitted gluon $\omega \geq gT$ (as distinct from standard
requirement $\omega \gg T$). In particular it enables us to study an
phenomenon of gluon emission in a hot QCD plasma induced by
collisions of thermal partons between themselves. The estimation of
energy loss of light partons in the context of approach
\cite{arnold} has made by Jeon and Moore \cite{jeon}.

Besides our approach permits exactly to take into account so-called
double Born amplitudes \cite{zakharov1,baier1}. In the works of
Zakharov \cite{zakharov1} on the basis of path integral approach to
the induced radiation, it was shown that interaction of fast quark
and emitted gluon with each color static center will be described by
not only one- but double-gluon exchanges. The interference between the
double-gluon exchanged diagrams and the diagram without gluon
exchange is important to ensure unitarity. In our approach the
contact double Born terms arise from `non-diagonal' elements in
correlator of the product of two effective currents (Section 5).
These terms enable us to cancel exactly divergencies arising in main
`diagonal' contributions to energy loss associated with emission of
quanta lying off mass-shell.

The main application of approach to bremsstrahlung of soft gluons
developed in our work, is a problem of energy loss. The research of
the energy losses of energetic partons in QGP is of great interest
with respect to jet quenching phenomenon
including in suppression of leading hadrons from fragmentation of
hard partons due to their strong interaction with the hot and dense
QCD medium. This phenomenon was long predicted by Gyulassy {\it et
al.} \cite{gyulassy2} in the framework of QCD-based model
calculations and at present it is observed at Relativistic
Heavy Ion Collider at Brookhaven National Laboratory [15\,--\,24].

The first calculations by Gyulassy and Wang \cite{gyulassy1}, and
Wang, Gyulassy and Pl\"umer \cite{wang1} of QCD radiative parton
energy loss incorporating the LPM destructive interference effect
indicated that medium induced by non-Abelian bremsstrahlung dominates
over the elastic energy loss \cite{bjorken,braaten,thoma}. This
important conclusion of these two fundamental works stimulates
further more extensive research of mechanisms of radiative parton
energy loss in hot nuclear matter\footnote{Here, we have not
discussed the problems associated with calculations of the energy
loss of high energy partons moving through the cold nuclear matter,
which possess proper specific features
\cite{levin,baier2,kopeliovich,johnson,arleo}.}. In particular the
following important step was made in works of Baier {\it et al.}
\cite{baier3}, where it was pointed to necessity of taking into
account of the effect of radiated gluon rescattering induced by
multiple scattering in QGP. At present there are few developed
techniques for the computation of the gluon radiative spectrum and
the energy loss included a light-cone path integral approach original
proposed by Zakharov \cite{zakharov1}, and effective $2D$
Schr\"odinger (or diffusion) equation \cite{baier3}, opacity
expansion framework \cite{gyulassy3} (opacity expansion in terms of
path integral formulation \cite{wiedemann}) supplemented by exact
Reaction Operator Formalism \cite{gyulassy4} which is more suitable
for multiple parton scattering in a finite system and a twist
expansion series \cite{guo}. The detailed reviews of these techniques
one can find in \cite{baier4,gyulassy5,kovner}. Approximate ways of
estimating the observable in-medium quenching of inclusive particle
spectra include a suppression of the hard partonic cross section
\cite{baier5} or an effective attenuation of the fragmentation
function \cite{wang3,gyulassy6}. One of the further developments of
the theory of radiative energy loss is connected with
consideration of more subtle effects caused by the change in the
cut-off scale for parton splitting and running coupling constant from
the vacuum to QGP \cite{zakharov2}, the influence of finite kinematic
boundaries on the induced gluon radiation \cite{gyulassy4,zakharov3},
the influence of the color dielectric modification of the gluon
dispersion relation in QGP (Ter-Mikayelian QCD effect)
\cite{kampfer1}, allowance for mass finiteness of high energy parton
(`dead' cone effect) \cite{dokshitzer1}. Let us consider in greater
detail the reviews of the results of works concerned with research of
the two last above-mentioned effects as more close to subject of
present paper.

The Ter-Mikayelian effect (or as sometimes it calls {\it density
effect in the bremsstrahlung}) for the condensed media
\cite{ter1,ter2} or QED plasma \cite{akopian,ginzburg} is associated
with suppression of spectral density of radiation intensity of the
bremsstrahlung in frequency range $\omega \ll \omega_{\rm pl}
\gamma$, where $\omega_{\rm pl}$ is a plasma frequency and $\gamma
\equiv E/M$ is a Lorentz factor for particle with energy $E$ and mass
$M$. Thus an existence of this effect means not only the regard of
interaction of emitted quantum with medium (in particular, in gain of
medium-induced effective mass by it\footnote{Note that in the work of
Galitsky and Yakimets \cite{galitsky1} (see, also \cite{galitsky2})
it was proved the important fact that besides medium-induced mass,
the effect of absorption by medium of bremsstrahlung of
ultrarelativistic particles also can result in strong inhibition of
bremsstrahlung in a certain frequency range.}), but allowance for the
finite mass of high energy particle radiated this quantum. For hot
QCD medium this effect was first considered by K\"ampfer and Pavlenko
\cite{kampfer1} by means of the introducing of an effective gluon mass,
depending on the temperature of system. It was shown that QCD
Ter-Mikayelian effect leads to the suppression of the induced
radiation in a different phase space region, where LPM effect is
appeared. Furthermore Djordjevic and Gyulassy \cite{djordjevic1} have
considered a problem of influence dielectric properties of hot
nuclear medium on the zeroth order in opacity {\it associated}
radiation and shown that the Ter-Mikayelian effect reduces the
associated energy loss. The problem of in-medium modification of the
parton cascade without gluon exchanges between the fast parton and
thermal partons has also studied by Zakharov \cite{zakharov2}.

Now  a study of radiative energy loss of heavy partons ($c$ or $b$
quarks) propagating through hot QCD medium is of great independent
interest. At first Shuryak \cite{shuryak1} have pointed to importance of 
taking into account of the effect of the energy losses on heavy quarks in 
$A+A$ collisions in the context of their influence
on high-mass dileptons from heavy quark decays. These dileptons can
stand out as one of penetrating probes of the most dense stages of
nuclear collisions \cite{lin,kampfer2,lokhtin}. Furthermore in the
work of Mustafa, Pal and Srivastava \cite{mustafa1} semiquantity 
estimation of the radiative energy loss of heavy quarks was given,
and it was shown that this energy loss is larger than the collisional
energy loss \cite{svetitsky,mustafa2,romatschke,hees} for all energies of
practical importance (see, however \cite{mustafa3}). Finite mass
effect was taken into account in \cite{mustafa1} through kinematic
bound on the maximum transverse momentum of the emitted gluon without
modification of multiplicity distribution of the radiated gluon.

The bremsstrahlung of heavy charged particles possesses proper
specific features associated not only with different kinematic
boundaries in comparison with light charged particles. In the case of
QED interaction at first a problem of qualitative difference of the
bremsstrahlung of heavy ultrarelativistic particles (type of muon or
proton) in dense an amorphous medium was considered by Gurevich
\cite{gurevich}. It was shown that taking into account of mass of
projectile is not reduced only to replacement of electron mass by mass
of proton (for example) in formulas of the bremsstrahlung. The
main principle conclusion of the work \cite{gurevich} is consist in
the fact that from four main ranges of the bremsstrahlung with
different frequency dependence of spectral density of radiation
intensity ((I) the range of absence of medium influence
(Bethe-Heitler); (II) the range of medium polarization
(Ter-Mikayelian); (III) the range of absorption of photons
\cite{galitsky1,galitsky2}; (IV) the range of multiple scattering
(Landau-Pomeranchuk)) for heavy particles multiple scattering range
vanishes, and three first range of bremsstrahlung remain. At
qualitative level this occurrence can be explained by that in the
case of a heavy particle the formation time of photon radiation is
reduced to a light one, that in turn results in a significant
reduction of the LPM effect. This conclusion is also hold for
particles with non-Abelian type of interaction.

In the case of QCD interaction a problem of qualitative difference of
radiative heavy quark energy loss from that of the light quarks, was
considered by Dokshitzer and Kharzeev \cite{dokshitzer1}. Here, it
concerned with suppression of small-angle radiation due to so-called
`dead' cone phenomenon \cite{dokshitzer2}. The regard of the finite
mass results in appearance in the quenching factor
\cite{dokshitzer1,kharzeev} supplemented positive term specific for
heavy quarks, that leads to the  reduction of the suppression of the
heavy hadron $p_{\perp}$ distribution in comparison with one for the
light hadrons. The work of Djordjevic and Gyulassy \cite{djordjevic2}
was the next important step in the study of the problem of heavy
quark energy loss. In this paper it was be proposed an extension of
opacity expansion for massless partons to heavy quarks with mass $M$.
It was shown that a general result (taking into account the
Ter-Mikayelian plasmon effect for gluons) is obtained by a single
universal energy shift of all frequencies by $\Delta \omega =
(m_g^2+x^2M^2)/(2xE)$ in the Gyulassy-Levai-Vitev series
\cite{gyulassy3,gyulassy4}, where $m_g$ is an effective plasmon mass
and $x$ is a light-cone momentum fraction. The importance of
developed approach in the work \cite{djordjevic2} is determined by
the possibility of its subsequent application to development of the
theory of heavy quark tomography of hot nuclear matter\footnote{The
problem on modified fragmentation functions and radiative energy loss
of the heavy quarks propagating in a cold nuclear matter was
discussed by Zhang, Wang and Wang \cite{zhang,zhang1} in the framework of
the generalized factorization of multiple scattering processes.}.

In the work of Thomas, K\"ampfer and Soff \cite{thomas} a numerical
calculation of exact radiation amplitude for heavy particle as function of
radiation angle $\theta$ for the case of single scattering, has performed.
It was shown that although there is a suppression effect
due to the heavy quark mass, its simple consideration by entering
the dead cone factor is not correct in all kinematical situations.
Besides a test of the validity of the potential model
beyond the assumptions of massless scattering particles by direct
calculation of the radiation amplitude in a quark-quark collision,
when a target quark is at rest, was made. The results of the calculation
shown a departure from those based on Gyulassy-Wang model for
large angles and when the projectile is heavier then the target, i.e.
when it cannot be neglected by transfer of the energy to the target,
and here, such the necessity of improvement of potential model
arises. In the works of Armesto, Salgado and Wiedemann \cite{armesto}
and W.C. Xiang {\it et al.} \cite{xiang} similar problems were considered
within the path integral formalism.

The paper is organized as follows. In Section 2 a general formalism
of deriving an effective current generating the bremsstrahlung of
arbitrary number of soft gluons for scattering of two color partons
among themselves in hot QCD medium, is introduced. Section 3 is
concerned with deriving a formulae of radiation intensity of the
bremsstrahlung of soft gluon generated by the lowest-order process of
the induced gluon radiation. In Section 4 the expression for
radiation intensity derived in previous Section is analyzed in a
simple case of the potential model, and under condition when
HTL-correction to bare three-gluon vertex can be neglected. Section 5
presents a detailed consideration of the energy loss connected with
bremsstrahlung induced by effective three-gluon vertex $\delta
\Gamma_{3g}$. In Section 6 a problem associated with an existence of
off-diagonal contributions to radiation energy loss is considered,
and their connection with so-called double Born scattering is
discussed. In Sections 7, 8 and 9 an extension of the method of the
effective currents to high-order processes induced by soft gluon radiation:
bremsstrahlung of two gluons and soft gluon radiation in the case of
two-scattering thermal partons, is presented. In Section 7
derivation of an effective current generating bremsstrahlung of two
soft gluons is presented, and corresponding general expression of
intensity radiation in this case is written out. In Section 8 an
approximate expression in small angles approximation and in static
limit is obtained from general expression for matrix element of two
soft gluon bremsstrahlung. Section 9 is concerned with derivation of an
expression for gluon radiation intensity for the case of scattering
of a high-energy incident parton off two hard thermal partons moving
with constant velocities. The case of bremsstrahlung induced by
four-gluon HTL vertex correction is analyzed in detail. In Section
10 results of this work is briefly summarized, and some ideas of
their extension to the case of dense strongly coupled quark-gluon
plasma actually observed at present at RHIC experiments, are proposed.

\section{\bf Effective currents for induced soft gluon radiation}
\setcounter{equation}{0}

One expects the word lines of the hard color particles to obey
classical trajectories in the manner Wong \cite{wong} since their
coupling to the soft QCD-plasma modes is weak at high temperature.
Considering this circumstance, we add color classical currents of
hard point charges $\alpha$ and $\beta$\footnote{In what follows the
high-energy particle $\alpha$ will be considered as external one with
respect to medium. It either injects into medium or produces inside
whereas particle $\beta$ (or particles $\beta_1,\, \beta_2,\ldots$)
is hard thermal particle (or particles) with energy of temperature
order $T$.}
\begin{equation}
j_{Q\alpha}^{a\mu}(X)=gv^{\mu}_{\alpha}\,
Q^a_{\alpha}\!(t)\,{\delta}^{(3)}({\bf x}-
{\bf x}_{0\alpha}-{\bf v}_{\alpha}(t-t_0)),\;
j_{Q\beta}^{a\mu}(X)=gv^{\mu}_{\beta}\,
Q^a_{\beta}(t)\,{\delta}^{(3)}({\bf x}-
{\bf x}_{0\beta}-{\bf v}_{\beta}(t-t_0))
\label{eq:2q}
\end{equation}
to the right-hand side of the field equation (I.3.1). Here,
$v_{\alpha,\,\beta}^{\mu}=(1,{\bf v}_{\alpha,\,\beta})$; $Q^a_{\alpha,\,\beta}$
are color classical charges of the particles $\alpha$ and $\beta$;
${\bf x}_{0\alpha,\,\beta}$ are their initial states at time $t=t_0$. The color
charges satisfy the Wong equation (II.3.12). Its solution, for example, for
particle $\alpha$ takes the form
\[
Q^a_{\alpha}(t)=U^{ab}(t,t_0)Q_{0\alpha}^b,\quad
Q_{0\alpha}^a=Q_{\alpha}^a(t)|_{t=t_0},
\]
where
\begin{equation}
U(t,t_0) = {\rm T}\exp\{-ig\!\int_{t_0}^t\!
(v_{\alpha}\cdot A^a(\tau,{\bf v}_{\alpha}\tau))T^ad\tau\},
\quad (T^a)^{bc}=if^{abc}
\label{eq:2w}
\end{equation}
is an evolution operator taking into account the color precession
along the parton trajectory. Rewriting the field equation in momentum
space we lead to the nonlinear integral equation for gauge potential
$A_{\mu}$ instead of Eq.\,(II.3.2)\footnote{In the third part of our
work we replace a notation of 4-momentum $p$ for
Fourier-transformation by $k$. Thus our work in these
notations will be close to most of works considered QCD radiation
energy loss.}
\begin{equation}
^{\ast}{\cal D}^{-1 \, \mu \nu}(k) A^{a}_{\nu}(k) =
-J^{a\mu}_{NL}[A](k) - j_{Q\alpha}^{a\mu}[A](k) - j_{Q\beta}^{a\mu}[A](k)
\equiv -J^{(tot)a\mu}[A](k).
\label{eq:2e}
\end{equation}
Here,
\begin{equation}
J^{a\mu}_{NL}[A](k) = \sum_{s=2}^{\infty} J^{(s)a\mu}(A,\ldots,A),
\label{eq:2r}
\end{equation}
\[
J^{(s)a}_{\mu}(A,\ldots,A) =  \frac{1}{s!}\,g^{s-1}\!\int\!
\,^{\ast}\Gamma^{a a_1\ldots a_s}_{\mu\mu_1\ldots\mu_s}(k,-k_{1},\ldots,-k_{s})
A^{a_1\mu_1}(k_{1})A^{a_2\mu_2}(k_{2})\ldots A^{a_s\mu_s}(k_{s})
\]
\[
\times\,\delta (k - \sum_{i=1}^{s}k_{i})\prod_{i=1}^{s}dk_{i}
\]
is nonlinear induced color current, where the coefficient functions
$\,^{\ast}\Gamma^{a a_1\ldots a_s}_{\mu\mu_1\ldots\mu_s}$ are usual HTL
amplitudes, and
\[
j_{Q\alpha}^{a\mu}[A](k)=
\frac{g}{(2\pi)^3}\,Q_{0\alpha}^av^{\mu}_{\alpha}
\delta(v_{\alpha}\cdot k){\rm e}^{i{\bf k}\cdot{\bf x}_{0\alpha}}
\]
\begin{equation}
+\, v^{\mu}_{\alpha}\sum_{s=1}^{\infty}
\frac{g^{s+1}}{(2\pi)^3}\int\!\frac{1}
{(v_{\alpha}\cdot (k_1 +\ldots +k_s))(v_{\alpha}\cdot (k_2 + \ldots + k_s))
\ldots (v_{\alpha}\cdot k_s)}
\label{eq:2t}
\end{equation}
\[
\times (v_{\alpha}\cdot A^{a_1}(k_1))\ldots (v_{\alpha}\cdot A^{a_s}(k_s))
\delta (v_{\alpha}\cdot (k-\sum_{i=1}^{s}k_i))
{\rm e}^{i({\bf k}-\sum\limits_{i=1}^s{\bf k}_i)\cdot{\bf x}_{0\alpha}}
\prod_{i=1}^s dk_i
\,(T^{a_1}\ldots T^{a_s})^{ab}Q_{0{\alpha}}^b
\]
is Fourier transform of color current of hard particle $\alpha$
(\ref{eq:2q}), (\ref{eq:2w}) (a similar expression holds for particle
$\beta$). In derivation of two last lines in Eq.\,(\ref{eq:2t}) we
drop all terms containing initial time $t_0$. $\,^{\ast}{\cal
D}^{\,\mu\nu}(k)$ is a medium modified (retarded) gluon propagator
which in the considered problem for convenience is chosen in Coulomb
gauge $\,^{\ast}{\cal D}^{\mu\nu}(k) \equiv\,^{\ast}{\cal
D}^{\mu\nu}_C(k)$. In the rest of frame of medium we have
\begin{equation}
\,^{\ast}{\cal D}^{00}_C(k)=\biggl(\frac{k^2}{{\bf k}^2}\biggr)
\,^{\ast}\!\Delta^{l}(k);\quad \,^{\ast}{\cal D}^{0i}(k)=0,
\label{eq:2y}
\end{equation}
\[
\,^{\ast}{\cal D}^{ij}_C(k)=(\delta^{ij}-\hat{k}^i\hat{k}^j)
\,^{\ast}\!\Delta^{t}(k),\quad \hat{k}^i\equiv k^i/|{\bf k}|,
\]
where $\,^{\ast}\!\Delta^{l,t}(k)=1/(k^2-\Pi^{l,t}(k))$ are scalar longitudinal
and transverse propagators.

As in Paper I (Section 5) and Paper II (Section 3) the first step is
consideration of a solution of the nonlinear integral equation
(\ref{eq:2e}) by the approximation scheme method. Discarding the
linear and nonlinear terms in $A_{\mu}^a(k)$ on the right-hand side
of Eq.\,(\ref{eq:2e}), we obtain in the first approximation
\[
\,^{\ast}\tilde{\cal D}^{-1\,\mu \nu}(k)A_{\nu}^a(k)=
-\tilde{J}_{Q\alpha}^{(0)a\mu}(k) -
\tilde{J}_{Q\beta}^{(0)a\mu}(k),
\]
where $\tilde{J}_{Q\alpha,\beta}^{(0)a\mu}(k) =
g/(2\pi)^3\,Q_{0\alpha}^av^{\mu}_{\alpha,\beta}
\delta(v_{\alpha,\beta}\cdot k)
{\rm e}^{i{\bf k}\cdot{\bf x}_{0\alpha,\beta}}$ are initial currents of
high-energy parton ${\alpha}$ and test thermal parton ${\beta}$, respectively.
The general solution of the last equation is
\begin{equation}
A_{\mu}^a(k) = A_{\mu}^{(0)a}(k) -
\,^{\ast}{\cal D}_{C\mu\nu}(k)\tilde{J}_{Q\alpha}^{(0)a\mu}(k)-
\,^{\ast}{\cal D}_{C\mu\nu}(k)\tilde{J}_{Q\beta}^{(0)a\mu}(k),
\label{eq:2u}
\end{equation}
where $A_{\mu}^{(0)a}(k)$ is a solution of homogeneous equation (a
free field), and two last terms on the right-hand side represent a
gauge field induced by hard partons in medium.

Furthermore, we keep the term quadratic in field in nonlinear induced current
(\ref{eq:2r}) and term linear in field in `precessional' current (\ref{eq:2t}).
Substituting derived solution (\ref{eq:2u}) into the right-hand side of
Eq.\,(\ref{eq:2e}), we obtain next correction to the field (\ref{eq:2u})
\[
A_{\mu}^{(1)a}(k) =
-\!\,^{\ast}{\cal D}_{C\mu\nu}(k)J^{(2)a\nu}(A^{(0)},A^{(0)})
\]
\[
-\,^{\ast}{\cal D}_{C\mu\nu}(k)\biggl\{
(T^{a_1})^{ab}Q_{0\alpha}^b\frac{g^2}{(2\pi)^3}\!\int\!
\frac{v_{\alpha}^{\nu}}{(v_{\alpha}\cdot k_1)}\,(v_{\alpha}\cdot A^{(0)a_1}(k_1))
\delta(v_{\alpha}\cdot(k-k_1))dk_1
\]
\begin{equation}
+\,J^{(2)a\nu}(A^{(0)},-\!\,^{\ast}{\cal D}_c\tilde{J}^{(0)}_{Q\alpha})
+J^{(2)a\nu}(-\!\,^{\ast}{\cal D}_C\tilde{J}^{(0)}_{Q\alpha}),A^{(0)})
+ (\alpha\rightarrow \beta)\biggr\}
\label{eq:2i}
\end{equation}
\[
-\,^{\ast}{\cal D}_{C\mu\nu}(k)\biggl\{
(T^{a_1})^{ab}Q_{0\alpha}^b\frac{g^2}{(2\pi)^3}\!\int\!
\frac{v_{\alpha}^{\nu}}{(v_{\alpha}\cdot k_1)}\,
(v_{\alpha}^{\lambda}\,^{\ast}{\cal D}_{C\lambda\lambda^{\prime}}(k)
\tilde{J}_{Q\alpha}^{(0)a_1\lambda^{\prime}}(k_1))
\delta(v_{\alpha}\cdot(k-k_1))dk_1
\]
\[
+\,J^{(2)a\nu}(-\!\,^{\ast}{\cal D}_C\tilde{J}^{(0)}_{Q\alpha},
-\!\,^{\ast}{\cal D}_C\tilde{J}^{(0)}_{Q\alpha}) +
(\alpha\rightarrow \beta)\biggr\}
\]
\[
-\,^{\ast}{\cal D}_{C\mu\nu}(k)\biggl\{
(T^{a_1})^{ab}Q_{0\alpha}^b\frac{g^2}{(2\pi)^3}\!\int\!
\frac{v_{\alpha}^{\nu}}{(v_{\alpha}\cdot k_1)}\,
(v_{\alpha}^{\lambda}\,^{\ast}{\cal D}_{C\lambda\lambda^{\prime}}(k)
\tilde{J}_{Q\beta}^{(0)a_1\lambda^{\prime}}(k_1))
\delta(v_{\alpha}\cdot(k-k_1))dk_1
\]
\[
+\,J^{(2)a\nu}(-\!\,^{\ast}{\cal D}_C\tilde{J}^{(0)}_{Q\alpha},
-\!\,^{\ast}{\cal D}_C\tilde{J}^{(0)}_{Q\beta})
+ (\alpha\leftrightarrow\beta)\biggr\}.
\]
Here, the first term on the right-hand side is associated with a pure soft
gluon self-interaction. It was analyzed in Paper I. The second group of terms
can be rewritten in the form
\[
-\,^{\ast}{\cal D}_{C\mu\nu}(k)\Bigl\{
\tilde{J}_{Q\alpha}^{(1)a\mu}[A^{(0)}](k) + (\alpha\rightarrow \beta)\Bigr\},
\]
where effective current is
\begin{equation}
\tilde{J}_{Q\alpha}^{(1)a\mu}[A^{(0)}](k)=
(T^{a_1})^{ab}Q_{0\alpha}^b
\frac{g^2}{(2\pi)^3}
\!\int\!\!K^{\mu\mu_1}({\bf v}_{\alpha}\vert\,k,-k_1)A^{(0)\,a_1}_{\mu_1}(k_1)
\delta(v_{\alpha}\cdot(k-k_1))dk_1,
\label{eq:2o}
\end{equation}
and in turn
\begin{equation}
K^{\mu\mu_1}
\!({\bf v}_{\alpha}\vert\,k,-k_1)\equiv
\frac{v^{\mu}_{\alpha}v^{{\mu}_1}_{\alpha}}{(v_{\alpha}\cdot k_1)} +
\!\,^\ast\Gamma^{\mu\mu_1\nu}(k,-k_1,-k+k_1)
\,^\ast{\cal D}_{C\nu\nu^{\prime}}\!(k-k_1)v_{\alpha}^{\nu^{\prime}}.
\label{eq:2p}
\end{equation}
The current (\ref{eq:2o}) induces so-called nonlinear Landau damping
process, studied in detail in our previous Paper II.

\begin{figure}[hbtp]
\begin{center}
\includegraphics[width=0.95\textwidth]{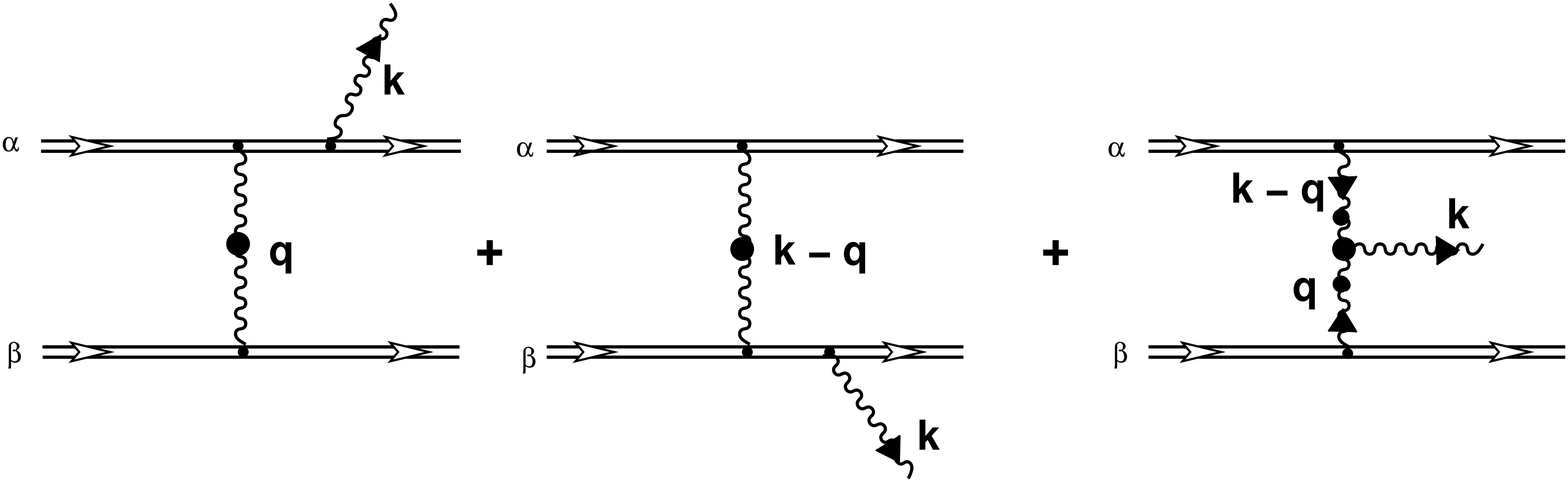}
\end{center}
\caption{\small The simplest process of soft-gluon bremsstrahlung
generating by color effective current (\ref{eq:2a}). The blob stands
for HTL resummation, and the double lines denote hard particles.}
\label{fig1}
\end{figure}

The third group of terms on the right-hand side of Eq.\,(\ref{eq:2i}) is equal
to zero by virtue of identity $f^{abc}Q_{0\alpha,\beta}^bQ_{0\alpha,\beta}^c=0$.
Now we concentrate on new nontrivial terms, forming the last fourth group.
After algebraic transformations they can be presented as
\[
-\,^{\ast}{\cal D}_{C\mu\nu}(k)
K_{\mu}^a({\bf v}_{\alpha},{\bf v}_{\beta};
{\bf x}_{0\alpha},{\bf x}_{0\beta},Q_{0\alpha},Q_{0\beta}|\,k).
\]
Here, for the case of linear dependence on $Q_{0\alpha}$ and $Q_{0\beta}$
we have
\begin{equation}
K_{\mu}^a({\bf v}_{\alpha},{\bf v}_{\beta};
{\bf x}_{0\alpha},{\bf x}_{0\beta},Q_{0\alpha},Q_{0\beta}|\,k)\cong
K_{\mu}^{abc}({\bf v}_{\alpha},{\bf v}_{\beta};
{\bf x}_{0\alpha},{\bf x}_{0\beta}|\,k)Q_{0\alpha}^bQ_{0\beta}^c,
\label{eq:2a}
\end{equation}
\[
K_{\mu}^{abc}({\bf v}_{\alpha},{\bf v}_{\beta};
{\bf x}_{0\alpha},{\bf x}_{0\beta}|\,k)=
-\,\frac{g^3}{(2\pi)^6}\,(T^a)^{bc}
K_{\mu}({\bf v}_{\alpha},{\bf v}_{\beta};
{\bf x}_{0\alpha},{\bf x}_{0\beta}|\,k),
\]
where
\begin{equation}
K_{\mu}({\bf v}_{\alpha},{\bf v}_{\beta};
{\bf x}_{0\alpha},{\bf x}_{0\beta}|\,k)=
\int\!\biggl\{
\frac{v_{\alpha\mu}}{v_{\alpha}\cdot q}\,
(v_{\alpha\nu}\!\,^{\ast}{\cal D}_C^{\nu\nu^{\prime}}(q)v_{\beta\nu^{\prime}})
-\frac{v_{\beta\mu}}{v_{\beta}\cdot (k-q)}\,
(v_{\beta\nu}\!\,^{\ast}{\cal D}_C^{\nu\nu^{\prime}}(k-q)v_{\alpha\nu^{\prime}})
\label{eq:2s}
\end{equation}
\[
+\!\,^\ast\Gamma_{\mu\nu\lambda}(k,-q,-k+q)
\!\,^{\ast}{\cal D}_C^{\nu\nu^{\prime}}(q)v_{\beta\nu^{\prime}}
\!^{\ast}{\cal D}_C^{\lambda\lambda^{\prime}}(k-q)v_{\alpha\lambda^{\prime}}
\biggr\}\,{\rm e}^{i({\bf k}-{\bf q})\cdot {\bf x}_{0\alpha}}
{\rm e}^{i{\bf q}\cdot {\bf x}_{0\beta}}
\delta(v_{\alpha}\cdot(k-q))\delta (v_{\beta}\cdot q)dq.
\]
New effective current $K_{\mu}^a({\bf v}_{\alpha},{\bf
v}_{\beta};\ldots|\,k)$ generates the simplest process of
bremsstrahlung of soft gluons. On Fig.\,\ref{fig1} diagrammatic
interpretation of three\footnote{In the semiclassical approximation
the first and the second terms on the right-hand side of Eq.\,(\ref{eq:2s}) also
contain processes, where the soft gluon is emitted prior to the
one-gluon exchange graphs which we drop on Fig.\,\ref{fig1}.}
terms on the right-hand side of Eq.\,(\ref{eq:2s}) is presented.

Thus we have shown that the effective current generating the
lowest-order process of the induced gluon radiation from fast partons
appears in the solution of the basic field equation (\ref{eq:2e})
that defines interacting soft-gluon field $A_{\mu}$ in the form of an
expansion in an initial value of color charges $Q_{0\alpha}$ and
$Q_{0\beta}$. Based on this result and on the results of Paper I
and Paper II also we can write out now more general structure of the
effective current in the form of a functional expansion in a free
field $A_{\mu}^{(0)}$ and color charges $Q_{0\alpha}$, $Q_{0\beta}$
generating the bremsstrahlung of arbitrary number of soft gluons for
scattering of two hard color partons among themselves
\begin{equation}
\tilde{J}_{Q\mu}^{({\rm br})a}[A^{(0)}](k)=
\tilde{J}_{Q\alpha\mu}^{(0)a}(k)+\tilde{J}_{Q\beta\mu}^{(0)a}(k)+
K_{\mu}^a({\bf v}_{\alpha},{\bf v}_{\beta};
{\bf x}_{0\alpha},{\bf x}_{0\beta},Q_{0\alpha},Q_{0\beta}|\,k)
\label{eq:2d}
\end{equation}
\[
+\!\sum\limits_{n=1}^{\infty}\!\int\!\!
K_{\mu\mu_1\ldots\mu_n}^{aa_1\ldots a_n}
({\bf v}_{\alpha},{\bf v}_{\beta};
{\bf x}_{0\alpha},{\bf x}_{0\beta},Q_{0\alpha},Q_{0\beta}
\vert\,k,-k_1,\ldots,-k_n)
A^{(0)a_1\mu_1}(k_1)\ldots A^{(0)a_n\mu_n}(k_n)\!\!\prod_{s=1}^{n}\!dk_s.
\]
The functions $K_{\mu\mu_1\ldots\mu_n}^{aa_1\ldots a_n},\,n=0,1,\ldots$
itself represent in general infinite series in expansion over color
charges $Q_{0\alpha}$ and $Q_{0\beta}$. Thus, for example, for $K_{\mu}^a$
we have
\begin{equation}
K_{\mu}^a({\bf v}_{\alpha},{\bf v}_{\beta};
{\bf x}_{0\alpha},{\bf x}_{0\beta},Q_{0\alpha},Q_{0\beta}|\,k) =
K_{\mu}^{abc}({\bf v}_{\alpha},{\bf v}_{\beta};
{\bf x}_{0\alpha},{\bf x}_{0\beta}|\,k)Q_{0\alpha}^bQ_{0\beta}^c,
\label{eq:2f}
\end{equation}
\[
+\frac{1}{2!}
\Bigl\{K_{\mu}^{abc_1c_2}({\bf v}_{\alpha},{\bf v}_{\beta};
{\bf x}_{0\alpha},{\bf x}_{0\beta}|\,k)Q_{0\alpha}^b
Q_{0\beta}^{c_1}Q_{0\beta}^{c_2} +
K_{\mu}^{ab_1b_2c}({\bf v}_{\alpha},{\bf v}_{\beta};
{\bf x}_{0\alpha},{\bf x}_{0\beta}|\,k)Q_{0\alpha}^{b_1}
Q_{0\alpha}^{b_2}Q_{0\beta}^{c}\Bigr\} + \ldots\,.
\]
Here, the first coefficient function
$K_{\mu}^{abc}({\bf v}_{\alpha},{\bf v}_{\beta};\ldots |\,k)$ is explicitly
defined by Eqs.\,(\ref{eq:2a}), (\ref{eq:2s}), and exact form of next-to-leading
coefficient functions
$K_{\mu}^{abc_1c_2}({\bf v}_{\alpha},{\bf v}_{\beta};\ldots|\,k)$ and
$K_{\mu}^{ab_1b_2c}({\bf v}_{\alpha},{\bf v}_{\beta};\ldots|\,k)$,
and their physical meaning will be given in Section 6.

\section{\bf Radiation intensity of soft gluon bremsstrahlung}
\setcounter{equation}{0}

In the expression for an effective current (\ref{eq:2d}) without loss
of generality one can set ${\bf x}_{0\alpha}=0$, and choose vector
${\bf x}_{0\beta}$ in the form ${\bf x}_{0\beta}=({\bf b},z_{0\beta})$,
where two-dimensional vector ${\bf b}$ is orthogonal
to relative velocity ${\bf v}_{\alpha}-{\bf v}_{\beta}$. Besides in
the subsequent discussion a longitudinal component $z_{0{\beta}}$
also does not play any role and thus it can be set equal to zero. The
energy of radiation field $E_Q^{ai}(k) = -i\omega\,^\ast{\cal
D}^{i\nu}_C(k)\tilde{J}_{Q\nu}^{({\rm br})a}[A^{(0)}] (k) + ik^i\,^\ast{\cal
D}^{0\nu}_C(k)\tilde{J}_{Q\nu}^{({\rm br})a}[A^{(0)}](k)$, generated by the
effective current, is defined by an expression
\begin{equation}
W({\bf b})=
-(2\pi)^4\!\int\!d{\bf k}d\omega
\!\int\!dQ_{0\alpha}dQ_{0\beta}\,\omega
{\rm Im}\,\langle
\tilde{J}^{({\rm br})\ast a}_{Q\mu}(k,{\bf b})
\!\,^{\ast}{\cal D}^{\mu\nu}_C(k)
\tilde{J}^{({\rm br})a}_{Q\nu}(k,{\bf b})
\rangle.
\label{eq:3q}
\end{equation}
Here,
\[
dQ_{0\alpha,\beta}=\prod_{a=1}^{d_A} dQ_{0\alpha,\beta}^a\,
\delta(Q_{0\alpha}^aQ_{0\beta}^a-C_2^{(\alpha,\beta)}),\quad
d_A = N_c^2 - 1,
\]
with the second order Casimir $C_2^{(\alpha, \beta)}$ for
$\alpha\,(\beta)$ color particles; $\langle\cdot\rangle$ is an
expectation value over the equilibrium ensemble. Eq.\,(\ref{eq:3q})
represents a minimal extension of an corresponding expression in
Abelian theory \cite{ginzburg} taking into account a color degree of
freedom of particles. The radiation energy (\ref{eq:3q}) is a
function of two-dimensional vector ${\bf b}$ being impact parameter
in this case.

Let us define an expression for radiation intensity of soft gluon
bremsstrahlung in passage of high-energy parton $\alpha$ through
medium consisting of hard thermal partons of $\beta$ sort, using
basic formula (\ref{eq:3q}). Here we follow reasoning of Ginsburg
and Tsytovich \cite{ginzburg}. For simplicity we assume that
particles $\beta$ are non-relativistic, their distribution over
momentum ${\bf p}_{\beta}$ is given by distribution function $f_{{\bf
p}_{\beta}}$. It is defined by formula
\[
n_{\beta}=\!\int\!\!f_{{\bf p}_{\beta}}\,\frac{d{\bf p}_{\beta}}{(2\pi)^3}.
\]
In this case radiation intensity is read
\begin{equation}
{\cal I} = \int\!d{\bf b}\!
\int\!f_{{\bf p}_{\beta}}\,\frac{d{\bf p}_{\beta}}{(2\pi)^3}\,
W({\bf b})\,|{\bf v}_{\alpha}-{\bf v}_{\beta}|
\equiv\left\langle\frac{dW({\bf b})}{dt}\right\rangle_{\!{\bf b}}.
\label{eq:3w}
\end{equation}

We note once again that this expression defines a change of energy
field in system for per time due to bremsstrahlung process. It equals
(with opposite sign) change of kinetic energy of $\alpha$ particle
and $\beta$ particles in medium on which nonelastic scattering
arises. In this case when both particle $\alpha$ and particles
$\beta$ radiate it is impossible in principle to
separate a contribution of energy loss of particle $\alpha$ from 
a contribution of energy loss of
particles $\beta$ in medium. In limiting case ``frozen'' $\beta$
particles only, when it can be neglected by its bremsstrahlung,
Eq.\,(\ref{eq:3w}) coincides with an expression for energy loss
${\cal I}\vert_{{\bf v}_{\beta}=0} = - dE/dt$, where $E$ is energy of
a fast parton $\alpha$.

One can define a probability of bremsstrahlung. Let us refer the
probability to unit intervals of momentum of emitted quantum 
$d{\bf k}/(2\pi)^3$
and transferred momentum $d{\bf q}/(2\pi)^3$. The probability
of bremsstrahlung ${\it w}_{{\bf p}_{\alpha},{\bf p}_{\beta}}({\bf k},{\bf q})$
is defined by a simple formula
\vspace{0.3cm}
\begin{equation}
\left\langle\frac{dW({\bf b})}{dt}\right\rangle_{\!{\bf b}}=
\int\!\omega({\bf k})
{\it w}_{{\bf p}_{\alpha},{\bf p}_{\beta}}({\bf k},{\bf q})\,
\frac{f_{{\bf p}_{\beta}}d{\bf p}_{\beta}}{(2\pi)^3}\,
\frac{d{\bf k}d{\bf q}}{(2\pi)^6}\,
\,\delta(\omega({\bf k})-{\bf k}\cdot
{\bf v}_{\alpha}-{\bf q}\cdot({\bf v}_{\beta}-{\bf v}_{\alpha})).
\label{eq:3e}
\end{equation}
\vspace{0.1cm}

Now we obtain an expression for radiation intensity generated by the
lowest-order process of the induced gluon radiation (Fig.\,\ref{fig1}).
We substitute an effective current (\ref{eq:2a}) into the right-hand side of
Eq.\,(\ref{eq:3q}). Using an expression for propagator in Coulomb gauge
(\ref{eq:2y}) and averaging rules over initial values of color charges
\[
\int\!dQ_{0\alpha}\,Q_{0\alpha}^aQ_{0\alpha}^b =
\frac{C_2^{(\alpha)}}{d_A}\,\delta^{ab},\quad
\int\!dQ_{0\beta}\,Q_{0\beta}^aQ_{0\beta}^b =
\frac{C_2^{(\beta)}}{d_A}\,\delta^{ab},
\]
from Eq.\,(\ref{eq:3q}) we find
\begin{equation}
W({\bf b})=\frac{1}{(2\pi)^2}\, \biggl(\frac{{\alpha}_s}{\pi}\biggr)^{\!3}
\!C_A\Biggl(\frac{C_2^{(\alpha)}C_2^{(\beta)}}{d_A}\Biggr)
\!\sum_{\zeta=1,2}\!\int\!d{\bf k}d\omega\,\omega\,
{\rm Im}(^{\ast}{\!\Delta}^t(k))
|\,{\rm e}^i(\hat{\bf k},\zeta)
K^i({\bf v}_{\alpha},{\bf v}_{\beta};{\bf b}|\,k)|^{\,2}
\label{eq:3r}
\end{equation}
\[
\hspace{0.3cm}
-\,\frac{1}{(2\pi)^2}\,\biggl(\frac{{\alpha}_s}{\pi}\biggr)^{\!3}
\!C_A\Biggl(\frac{C_2^{(\alpha)}C_2^{(\beta)}}{d_A}\Biggr)
\!\sum_{\zeta=1,2}\!\int\!d{\bf k}d\omega\,\omega\,
{\rm Im}(^{\ast}{\!\Delta}^l(k))
|\,K^0({\bf v}_{\alpha},{\bf v}_{\beta};{\bf b}|\,k)|^{\,2}.
\]
Here, $\alpha_s\equiv g^2/4\pi$. The first term on the right-hand side is
connected with bremsstrahlung of transverse gluon, and a second one is
associated with bremsstrahlung of longitudinal gluon. For transverse mode we
introduce polarization vectors ${\rm e}^i(\hat{\bf k},\zeta),\,\zeta=1,2$
possessing properties
\[
{\bf k}\cdot{\bf e}(\hat{\bf k},\zeta)=0,\quad
{\bf e}^{\ast}(\hat{\bf k},\zeta)\cdot
{\bf e}(\hat{\bf k},\zeta^{\prime})=
\delta_{\zeta{\zeta}^{\prime}}.
\]
Three-dimensional transverse projector in (\ref{eq:2y}) is
associated with polarization vectors by relation
\begin{equation}
P^{ij}(\hat{\bf k})=(\delta^{ij}-\hat{k}^i\hat{k}^j)=
\sum_{\zeta = 1,2}{\rm e}^{\ast i}(\hat{\bf k},\zeta)\,
{\rm e}^j(\hat{\bf k},\zeta).
\label{eq:3t}
\end{equation}

Let us consider in more detail an expressions $|{\rm e}^i K^i|^{\,2}$ and
$|K^0|^{\,2}$ on the right-hand side of Eq.\,(\ref{eq:3r}). Due to
Eq.\,(\ref{eq:2s}) the first expression can be presented in the form
\[
|{\rm e}^i(\hat{\bf k},\zeta)
K^i({\bf v}_{\alpha},{\bf v}_{\beta};{\bf b}|\,k)|^{\,2}\!=\!
\left|\int\!{\rm e}^{i{\bf q}\cdot{\bf b}}
{\rm e}^i(\hat{\bf k},\zeta)
{\cal K}^i({\bf v}_{\alpha},{\bf v}_{\beta}|\,k,q)
\delta(\omega\!-\!{\bf v}_{\beta}\!\cdot\!{\bf q}\!-\!{\bf v}_{\alpha}
\!\cdot\!
({\bf k}\!-\!{\bf q})) d{\bf q}\right|^{\,2}_{q_0={\bf v}_{\beta}\cdot{\bf q}}
\]
\begin{equation}
=\int\!{\rm e}^{i({\bf q}-{\bf q}^{\prime})\cdot{\bf b}}
\Bigl[({\rm e}^i(\hat{\bf k},\zeta)
{\cal K}^i({\bf v}_{\alpha},{\bf v}_{\beta}|\,k,q))
({\rm e}^{j}(\hat{\bf k},\zeta)
{\cal K}^{j}({\bf v}_{\alpha},{\bf v}_{\beta}|\,k,q))^{\ast}
\Bigr]_{q_0={\bf v}_{\beta}\cdot{\bf q}}
\label{eq:3y}
\end{equation}
\[
\times\,
\delta(\omega-{\bf v}_{\beta}\cdot{\bf q}-{\bf v}_{\alpha}\cdot
({\bf k}-{\bf q}))\,
\delta(({\bf v}_{\alpha}-{\bf v}_{\beta})\cdot({\bf q}-{\bf q}^{\prime}))
d{\bf q}d{\bf q}^{\prime},
\]
where
\[
{\cal K}^{\mu}({\bf v}_{\alpha},{\bf v}_{\beta}|\,k,q)=
\frac{v_{\alpha}^{\mu}}{v_{\alpha}\cdot q}\,
(v_{\alpha\nu}\!\,^{\ast}{\cal D}_C^{\nu\nu^{\prime}}(q)v_{\beta\nu^{\prime}})
-\frac{v_{\beta}^{\mu}}{v_{\beta}\cdot (k-q)}\,
(v_{\beta\nu}\!\,^{\ast}{\cal D}_C^{\nu\nu^{\prime}}(k-q)v_{\alpha\nu^{\prime}})
\]
\begin{equation}
+\,^\ast\Gamma^{\mu\nu\lambda}(k,-q,-k+q)
\,\,^{\ast}{\cal D}_{C\nu\nu^{\prime}}(q)v_{\beta}^{\nu^{\prime}}
\,^{\ast}{\cal D}_{C\lambda\lambda^{\prime}}(k-q)v_{\alpha}^{\lambda^{\prime}}.
\label{eq:3u}
\end{equation}
In the last line of Eq.\,(\ref{eq:3y}) delta-function $\delta(({\bf
v}_{\alpha}-{\bf v}_{\beta})\cdot({\bf q}-{\bf q}^{\prime}))=
\delta(q_{\|}-q^{\prime}_{\|})/|{\bf v}_{\alpha}-{\bf v}_{\beta}|$
enables the integration over longitudinal component of
vector ${\bf q}^{\prime}=({\bf q}_{\perp}^{\prime},q_{\|}^{\prime})$
to be performed. For deriving
radiation intensity we multiply (\ref{eq:3r}) by $f_{{\bf
p}_{\beta}}d{\bf p}_{\beta}/(2\pi)^3 |{\bf v}_{\alpha}-{\bf
v}_{\beta}|d{\bf b}$ and integrate over ${\bf p}_{\beta}$ and ${\bf
b}$. We perform integration over impact parameter ${\bf b}$ with the
help of expression
\begin{equation}
\int\!{\rm e}^{i({\bf q}-{\bf q}^{\prime})\cdot{\bf b}}d{\bf b}
= (2\pi)^2\delta^{(2)}(({\bf q}-{\bf q}^{\prime})_{\perp}).
\label{eq:3i}
\end{equation}
The Eq.\,(\ref{eq:3i}) enables us to performe complete integration over
$d{\bf q}^{\prime}$ in (\ref{eq:3y}) and thus to define initial for
subsequent analysis expression of radiation intensity for the bremsstrahlung
process dipected on Fig.\,\ref{fig1}
\begin{equation}
\left\langle\frac{dW({\bf b})}{dt}\right\rangle_{\!{\bf b}}=
-\biggl(\frac{{\alpha}_s}{\pi}\biggr)^{\!3}
\!C_A\Biggl(\frac{C_2^{(\alpha)}C_2^{(\beta)}}{d_A}\Biggr)
\Biggl(\int\!{\bf p}_{\beta}^2f_{|{\bf p}_{\beta}|}
\,\frac{d|{\bf p}_{\beta}|}{2\pi^2}\Biggr)
\label{eq:3o}
\end{equation}
\[
\times\,
\Biggl[\int\!\frac{d\Omega_{{\bf v}_{\beta}}}{4\pi}
\!\sum_{\zeta=1,2}\!\int\!d{\bf k}d\omega\,\omega\,
{\rm Im}(^{\ast}{\!\Delta}^t(k))\int\!d{\bf q}\,
|{\rm e}^i(\hat{\bf k},\zeta)
{\cal K}^i({\bf v}_{\alpha},{\bf v}_{\beta}|\,k,q)|^{\,2}
\]
\[
\times\,
\delta(\omega-{\bf v}_{\beta}\cdot{\bf q}-{\bf v}_{\alpha}\cdot
({\bf k}-{\bf q}))+
\Bigl(\,^{\ast}{\!\Delta}^t(k)\rightarrow \,^{\ast}{\!\Delta}^l(k),\;
{\rm e}^i{\cal K}^i\rightarrow \sqrt{\frac{k^2}{{\bf k}^2}}\,{\cal K}^0\Bigr)
\Biggr]_{q^0=({\bf v}_{\beta}\cdot{\bf q})}.
\]
If there are several sorts of thermal partons, then on the right-hand side of
Eq.\,(\ref{eq:3o}) it should be summarize over $\beta$. In notation
(\ref{eq:3o}) we assume isotropy of a medium, i.e. $f_{{\bf p}_{\beta}}=
f_{|{\bf p}_{\beta}|}$.

Now we consider a special case of Eq.\,(\ref{eq:3o}). Let us define the
radiation intensity caused by bremsstrahlung of real quantum of oscillations,
i.e. oscillations lying on mass-shell. For this purpose, for a weak-absorption
medium, when ${\rm Im}(^{\ast}{\!\Delta}^{\!-1\,t,\,l}(k))\rightarrow 0$,
we can approximate imaginary part of scalar propagators in the following way
\begin{equation}
{\rm Im}(^{\ast}{\!\Delta}^{t,\,l}(k))\simeq
-\pi\,{\rm sign}(\omega)\,
\frac{{\rm Z}_{t,\,l}({\bf k})}{2\omega_{\bf k}^{t,\,l}}\,
[\,\delta(\omega-\omega_{\bf k}^{t,\,l}) +
\delta(\omega+\omega_{\bf k}^{t,\,l})],
\label{eq:3p}
\end{equation}
where ${\rm Z}_{t,\,l}({\bf k})$ are the residues of appropriate
scalar propagators $^{\ast}{\!\Delta}^{\!-1\,t,\,l}(k)$ at the poles
and $\omega_{\bf k}^{t,\,l}\equiv{\omega}^{t,\,l}({\bf k})$ are the
dispersion relations for transverse and longitudinal modes. The term
with $\delta(\omega+\omega_{\bf k}^{t,\,l})$ on the right-hand side
of Eq.\,(\ref{eq:3p}) takes into consideration not emission, but
absorption of oscillation. Substituting (\ref{eq:3p}) into
(\ref{eq:3o}) and omitting the last contribution after integration
over $d\omega$, we find instead of (\ref{eq:3o})
\begin{equation}
\left\langle\frac{dW({\bf b})}{dt}\right\rangle_{\!{\bf b}}=
\frac{{\alpha}_s^3}{\pi^2}\!\sum_{\beta = q,\,\bar{q},\,{\rm g}}
\!C_A\Biggl(\frac{C_2^{(\alpha)}C_2^{(\beta)}}{d_A}\Biggr)
\Biggl(\int\!{\bf p}_{\beta}^2f_{|{\bf p}_{\beta}|}
\,\frac{d|{\bf p}_{\beta}|}{2\pi^2}\Biggr)
\label{eq:3a}
\end{equation}
\[
\times
\Biggl[\int\!\frac{d\Omega_{{\bf v}_{\beta}}}{4\pi}
\!\sum_{\zeta=1,2}\!\int\!d{\bf k}\,\omega_{\bf k}^t
\Biggl(\frac{{\rm Z}_{t}({\bf k})}{2\omega_{\bf k}^{t}}\Biggr)
\!\int\!d{\bf q}\,
|\,{\rm e}^i(\hat{\bf k},\zeta)
{\cal K}^i({\bf v}_{\alpha},{\bf v}_{\beta}|\,k,q)|^{\,2}_{\,{\rm on-shell}}
\]
\[
\times\,
\delta(\omega_{\bf k}^t-{\bf v}_{\beta}\cdot{\bf q}-{\bf v}_{\alpha}\cdot
({\bf k}-{\bf q}))+
\Bigl(t\rightarrow l,\;
{\rm e}^i{\cal K}^i\rightarrow \sqrt{\frac{k^2}{{\bf k}^2}}\,{\cal K}^0\Bigr)
\Biggr]_{q^0=({\bf v}_{\beta}\cdot{\bf q})}.
\]
Comparing the last expression with (\ref{eq:3e}) we derive an
explicit form for the probabilities of soft-gluon bremsstrahlung
\begin{equation}
{\it w}_{{\bf p}_{\alpha},{\bf p}_{\beta}}^t({\bf k},{\bf q})=
4(2\pi)^4\alpha_s^3
C_A\Biggl(\frac{C_2^{(\alpha)}C_2^{(\beta)}}{d_A}\Biggr)
\Biggl(\frac{{\rm Z}_{t}({\bf k})}{2\omega_{\bf k}^{t}}\Biggr)\!
\sum_{\zeta=1,2}\!
|\,{\rm e}^i(\hat{\bf k},\zeta)
{\cal K}^i({\bf v}_{\alpha},{\bf v}_{\beta}|\,k,q)
|^{\,2}_{\,\omega=\omega_{\bf k}^t,\,q_0={\bf v}_{\beta}\cdot {\bf q}},
\label{eq:3s}
\end{equation}
\[
{\it w}_{{\bf p}_{\alpha},{\bf p}_{\beta}}^l({\bf k},{\bf q})=
4(2\pi)^4\alpha_s^3
C_A\Biggl(\frac{C_2^{(\alpha)}C_2^{(\beta)}}{d_A}\Biggr)
\Biggl(\frac{{\rm Z}_{l}({\bf k})}{2\omega_{\bf k}^{l}}\Biggr)\!
\Biggl(\frac{k^2}{{\bf k}^2}\Biggr)
|{\cal K}^0({\bf v}_{\alpha},{\bf v}_{\beta}|\,k,q)
|^{\,2}_{\,\omega=\omega_{\bf k}^l,\,q_0={\bf v}_{\beta}\cdot {\bf q}}.
\]
The expressions (\ref{eq:3a}), (\ref{eq:3s}) were obtained with the
assumption that QGP represents a system of non-relativistic thermal
partons. In this case only the use of the impact parameter and
averaging over it, is valid. It enables us to diagonalize the product
of amplitude ${\rm e}^i{\cal K}^i$ and complex conjugate amplitude 
(Eq.\,(\ref{eq:3y})) in the ${\bf q},\,{\bf q}^{\prime}$ variables, 
and to write the
probability in the simple form (\ref{eq:3s}). In the work 
\cite{ginzburg} a possible way of extension of this approach to the
case of arbitrary moving thermal particles, was suggested. It
consists in construction of subsequent Lorentz transformations for
particles group with momenta in interval $({\bf p}_{\beta},{\bf
p}_{\beta} + d{\bf p}_{\beta})$ at rest, where Eq.\,(\ref{eq:3a}) is
true. However practical use of this idea is very difficult. In work
by Akopian and Tsytovich \cite{akopian} for the case when both
particle $\alpha$ and particles $\beta$ are ultra-relativistic, the
probability of bremsstrahlung (\ref{eq:3s})  for ordinary plasma was
derived by direct computations with use of dynamical equation without
notion of impact parameter. It gives some ground to suppose that
expressions (\ref{eq:3a}), (\ref{eq:3s}) hold for ultra-relativistic
high-temperature plasma also.

\section{\bf Approximation of static color center}
\setcounter{equation}{0}

Let us analyze an expression for gluon radiation intensity
(\ref{eq:3a}) in the case when target particle is modeled by static
screened potential\footnote{Remind that in this case (\ref{eq:3a})
correct to a sign, coincides with the expression for energy loss of
high-energy parton $\alpha$.} $({\bf {\bf v}}_{\beta}=0)$, and
HTL-correction $\delta\Gamma_{3g}$ to bare three-gluon vertex can be
neglected. For simplicity we restrict our consideration to the case
of transverse gluon radiation. At first we consider integral over
momentum transfer ${\bf q}$ on the right-hand side of
Eq.\,(\ref{eq:3a}) presented as
\begin{equation}
\int\! d{\bf q}_{\perp}dq_{\|}
\left|\, {\rm e}^i(\hat{\bf k},\zeta)
{\cal K}^i({\bf v}_{\alpha},0|\,k,q)
\right|^{\,2}_{\,\omega=\omega^t_{\bf k},\;q_0=0}
\delta(\omega^t_{\bf k}-{\bf v}_{\alpha}\cdot {\bf k} +
v_{\alpha}q_{\|}),
\label{eq:4q}
\end{equation}
where ${\bf q}_{\perp}$ and $q_{\|}$ are transverse and longitudinal
components of momentum transfer with respect to velocity ${\bf v}_{\alpha}$,
correspondingly, and $v_{\alpha}\equiv |{\bf v}_{\alpha}|.$ Based on
an explicit expression for ${\cal K}^i({\bf v}_{\alpha},{\bf v}_{\beta}|\,k,q)$
(Eq.\,(\ref{eq:3u})) and limiting static expression for propagator
\begin{equation}
\,^{\ast}{\cal D}^{00}_C(0,{\bf q})
= \frac{1}{{\bf q}^2 + \mu_D^2},
\label{eq:4w}
\end{equation}
where $\mu_D^2$ is Debay screening mass, and performing integration
over $dq_{\|}$, we find instead of (\ref{eq:4q})
\begin{equation}
\frac{1}{v_{\alpha}}\int\!d{\bf q}_{\perp}
\frac{1}{\left({\bf q}_{\perp}^2 + \mu_D^2 +
\displaystyle\frac{1}{v_{\alpha}^2}
(\omega_{\bf k}^t - {\bf v}_{\alpha}\cdot {\bf k})^2\right)^2}
\label{eq:4e}
\end{equation}
\[
\times
\left|\, \frac{{\bf e}\cdot {\bf v}_{\alpha}}
{\omega_{\bf k}^t - {\bf v}_{\alpha}\cdot {\bf k}} +
{\rm e}^i(\hat{\bf k},\zeta) \Gamma^{i0\mu}(k,-q,-k+q)
\!\,^{\ast}{\cal D}_{C\mu\mu^{\prime}}(k-q)v_{\alpha}^{\mu^{\prime}}
\right|^{\,2}_{\, q_0=0,\; q_{\|}= -\frac{1}{v_{\alpha}}
(\omega_{\bf k}^t - {\bf v}_{\alpha}\cdot {\bf k})}.
\]
Here, for brevity we use a notation ${\bf e}\equiv{\bf e}(\hat{\bf
k},\zeta)$. We note that in our approach in deciding on Coulomb
gauge, a contribution of thermal parton $\beta$ (the second term in 
function ${\cal K}^i$, Eq.\,(\ref{eq:3u})) to bremsstrahlung of 
transverse soft gluon is exactly equal to zero. However here, there is
a possibility of radiation of longitudinal plasmon from target
$\beta$. Let us consider a difference $\omega_{\bf k}^t - {\bf
v}_{\alpha}\cdot {\bf k}$. We approximate a spectrum of transverse
oscillations $\omega_{\bf k}^t$ in the limit of $|{\bf k}_{\perp}|\ll
k_{\|}$ by standard expression
\[
\omega_{\bf k}^t \simeq \sqrt{{\bf k}^2 + m_g^2}
\simeq k_{\|} + \frac{{\bf k}^2_{\perp} + m_g^2}{2k_{\|}}.
\]
Here, $m_g^2$ is an effective gluon mass square,
depending on temperature of the system, $m_g^2=3\omega_{pl}^2/2$,
where $\omega_{pl}^2=g^2T^2(N_f + 2N_c)/18$ is plasma frequency square.

Besides, for ultrarelativistic particle $\alpha$ with energy $E$ and finite
mass $M$, we have
\begin{equation}
v_{\alpha} \simeq 1 - \frac{M^2}{2E^2}.
\label{eq:4r}
\end{equation}
Based on above-mentioned we can write
\begin{equation}
\omega_{\bf k}^t - {\bf v}_{\alpha}\cdot {\bf k}
\simeq\frac{{\bf k}_{\perp}^2 + m_g^2 + x^2M^2}{2\omega},\;
x\equiv\frac{\omega}{E}.
\label{eq:4t}
\end{equation}
Furthermore from a condition of transversity we obtain $e_{\|}\simeq
-({\bf k}_{\perp}\cdot{\bf {\bf e}}_{\perp})/\omega$ whence it follows
\begin{equation}
{\bf e}\cdot {\bf v}_{\alpha}\simeq
-v_{\alpha}\,\frac{{\bf k}_{\perp}\cdot {\bf e}_{\perp}}{\omega}.
\label{eq:4y}
\end{equation}
Taking into account (\ref{eq:4t}) and (\ref{eq:4y}) we can write a first term
in expression under module squared in (\ref{eq:4e}) as
\begin{equation}
\frac{{\bf e}\cdot {\bf v}_{\alpha}}
{\omega_{\bf k}^t - {\bf v}_{\alpha}\cdot {\bf k}} \simeq
-2v_{\alpha}\,\frac{{\bf k}_{\perp}\cdot {\bf e}_{\perp}}
{{\bf k}_{\perp}^2 + m_g^2 + x^2M^2},
\label{eq:4u}
\end{equation}
and Coulomb factor reads
\begin{equation}
\frac{1}{\left({\bf q}_{\perp}^2 + \mu_D^2 +
\displaystyle\frac{1}{v_{\alpha}^2} (\omega_{\bf k}^t - {\bf
v}_{\alpha}\cdot {\bf k})^2\right)^2} \simeq \frac{1}{({\bf
q}^2_{\perp} + \mu_D^2 + l_f^{-2})^2}.
\label{eq:4i}
\end{equation}
Here, $l_f=v_{\alpha}\tau_f$, where $\tau_f\equiv2\omega/({\bf
k}_{\perp}^2 + m_g^2 + x^2M^2)$ is finite formation time with regard
to temperature induced gluon mass and finite mass of projectile
$\alpha$.

Let us consider now a contribution to matrix element associated with
bare three-gluon vertex. Taking into account structure of propagator
in the Coulomb gauge (Eq.\,(\ref{eq:2y})) and approximation
(\ref{eq:4y}), we can present this contribution as
\begin{equation}
-2\,\frac{\{\omega^2-({\bf k}-{\bf q})^2\}}{({\bf k}-{\bf q})^2}\,
({\bf e}\cdot {\bf q})\,^{\ast}\!\Delta^{l}(k-q)
+ 2\omega
\left\{\,-v_{\alpha}\,\frac{{\bf k}_{\perp}\cdot {\bf e}_{\perp}}{\omega} +
\omega\,\frac{{\bf e}\cdot {\bf q}}{({\bf k}-{\bf q})^2}
\right\}\!\,^{\ast}\!\Delta^{t}(k-q).
\label{eq:4o}
\end{equation}

Furthermore by virtue of (\ref{eq:4t}) we have
\[
{\bf e}\cdot {\bf q}=
{\bf e}_{\perp}\cdot {\bf q}_{\perp} +
e_{\|}q_{\|} \simeq
{\bf e}_{\perp}\cdot \left(\,{\bf q}_{\perp}+
\frac{{\bf k}_{\perp}}{\omega}\frac{1}{l_f}\right)
\simeq {\bf e}_{\perp}\cdot {\bf q}_{\perp}
\]
under condition $|{\bf k}_{\perp}|/{\omega} l_f\ll |{\bf q}_{\perp}|$.
Besides we have $({\bf k}-{\bf q})^2=({\bf k}-{\bf q})^2_{\perp}
+\omega^2/v_{\alpha}^2 \simeq \omega^2/v_{\alpha}^2$, when
$\omega^2/v_{\alpha}^2\!\gg\!({\bf k}-{\bf q})^2_{\perp}$. By this mean expression
(\ref{eq:4o}) can be replaced by approximate one
\[
2\,\frac{x^2M^2 + ({\bf k}-{\bf q})^2_{\perp}}{\omega^2}\,
({\bf e}_{\perp}\cdot {\bf q}_{\perp})\,^{\ast}\!\Delta^{l}(k-q)
+2v_{\alpha} ({\bf e}_{\perp}\cdot (v_{\alpha}{\bf q} - {\bf k})_{\perp})
\,^{\ast}\!\Delta^{t}(k-q).
\]
Here, a first term connected with a contribution from longitudinal
virtual oscillation is suppressed with respect to a second term by
small factor and it can be omitted.

Let us consider transverse gluon propagator
$\,^{\ast}\!\Delta^{t}(\omega,{\bf k}-{\bf q})$. Its explicit expression
in hard thermal loop approximation is

\[
\,^{\ast}\!\Delta^{t}(\omega,{\bf k}-{\bf q})=
\left[\,\omega^2 - ({\bf k}-{\bf q})^2 - \frac{1}{2}\,\mu_D^2
\Biggl\{\frac{\omega^2}{({\bf k}-{\bf q})^2}
-\frac{\omega^2 - ({\bf k}-{\bf q})^2}{({\bf k}-{\bf q})^2}
\,F\Biggl(\frac{\omega}{\vert {\bf k}-{\bf q}\vert}\Biggr)\!\Biggr\}
\right]^{-1},
\]
where
\[
F(z)\equiv\frac{z}{2}\,\bigg[\ln\bigg\vert\frac{1 + z}{1 - z}
\bigg \vert - i \pi
\theta ( 1 - \vert z \vert)\bigg].
\]
By making use approximations
\begin{equation}
({\bf k}-{\bf q})^2\simeq {\bf k}^2 + ({\bf k}-{\bf q})^2_{\perp} +
m_g^2 + x^2M^2,\quad \frac{\omega}{\vert {\bf k}-{\bf q}\vert} \simeq
1-\,\frac{[({\bf k}-{\bf q})^2_{\perp} + x^2M^2]}{2{\bf k}^2},
\label{eq:4oo}
\end{equation}
we derive for transverse propagator
\[
\,^{\ast}\!\Delta^{t}(\omega,{\bf k}-{\bf q})\!\simeq\!
-\!\left[({\bf k}\!-\!{\bf q})^2_{\perp}\!+\!x^2M^2\!
+\!\frac{1}{2}\mu_D^2\Biggl\{1\!+\!
\frac{[({\bf k}\!-\!{\bf q})^2_{\perp}\!+\!x^2M^2]}{2{\bf k}^2}
\Biggl(\!\ln\frac{4{\bf k}^2}{[({\bf k}\!-\!{\bf q})^2_{\perp}\!+\!x^2M^2]}
\!-\!i\pi\!\Biggr)\!\!\Biggr\}\right]^{-1}\!\!.
\]
The factor before a logarithm on the right-hand side of the last
expression is a small owing to ${\bf k}^2\gg ({\bf k}-{\bf
q})^2_{\perp} + x^2M^2$. On the other hand, the logarithm can be 
large by the same condition. Therefore this term can be discarded under
more severe constraint for the magnitude of momentum $|{\bf k}|$:
\[
\epsilon\ln\epsilon\ll 1,\quad
\epsilon\equiv[({\bf k}-{\bf q})^2_{\perp} + x^2M^2]/{\bf k}^2.
\]
Under the last condition with allowance for $\mu_D^2/2=m_g^2$ we
derive finally expression for transverse gluon propagator
\begin{equation}
\,^{\ast}\!\Delta^{t}(\omega,{\bf k}-{\bf q})\simeq
-\frac{1}{({\bf k}-{\bf q})^2_{\perp} + m_g^2 + x^2M^2}.
\label{eq:4ooo}
\end{equation}

By this means a contribution to matrix element connected with bare
three-gluon vertex in this high-frequency limit can be presented as
\begin{equation}
{\rm e}^i\Gamma^{i0\mu}(k,-q,-k+q)
\!\,^{\ast}{\cal D}_{C\mu\mu^{\prime}}(k-q)v_{\alpha}^{\mu^{\prime}}
\simeq -2v_{\alpha}\,
\frac{{\bf e}_{\perp}\cdot (v_{\alpha}{\bf q} - {\bf k})_{\perp}}
{({\bf k}-{\bf q})^2_{\perp} + m_g^2 + x^2M^2}.
\label{eq:4p}
\end{equation}
In this approximation the residue $Z_t({\bf k})$ can be replaced by unit.

With regard to (\ref{eq:4u}), (\ref{eq:4i}) and (\ref{eq:4p}) the expression
for energy loss of parton $\alpha$, associated with bremsstrahlung of soft
transverse gluon, takes following form within the potential model
\begin{equation}
\left(-\frac{dE}{dt}\right)^{\!t} =
2\pi\biggl(\,\frac{\alpha_s}{\pi}\biggr)^{\!3}v_{\alpha}E\, C_{\! A}\!
\sum\limits_{\beta=q\!,\,\bar{q}\!,\,{\rm g}}\left(
\frac{C_2^{(\alpha)}C_2^{(\beta)}}{d_A}\right) n_{\beta}
\sum\limits_{\zeta = 1,2}\int\!d{\bf k}_{\perp}\!\int\!dx
\label{eq:4a}
\end{equation}
\[
\times\!
\int\!d{\bf q}_{\perp}\,\frac{1}{({\bf q}^2_{\perp} + \mu_D^2 + l_f^{-2})^2}\,
\left|\,{\bf e}_{\perp}({\bf k},\zeta)\cdot
\Biggl(\frac{{\bf k}_{\perp}}{{\bf k}_{\perp}^2 + m_g^2 + x^2M^2} +
\frac{(v_{\alpha}{\bf q} - {\bf k})_{\perp}}
{({\bf k}-{\bf q})^2_{\perp} + m_g^2 + x^2M^2}
\Biggr)\right|^{\,2}.
\]
The expression (\ref{eq:4a}) takes into account the QCD Ter-Mikaelian
effect associated with existence of an effective gluon mass
$m_g$, depending on the temperature $T$. As was mentioned in
Introduction for hot QCD plasma this effect was first considered by
K\"ampfer and Pavlenko \cite{kampfer1}. As distinct from this work,
in our paper the effective gluon mass appears not only in the first
term of matrix element, but also in the second one corresponding to
the radiation from gluon line via the triple vertex. Besides this
expression takes into account mass finiteness of the projectile (mass
effect) in according with expression obtained by Djordjevic and
Gyulassy \cite{djordjevic2} for energy loss of heavy quarks.

Let us consider more closely the expression (\ref{eq:4a}). First of
all we define conditions wherein a term $l_f^{-2}$ in Coulomb factor
$1/({\bf q}^2_{\perp} + \mu_D^2 + l_f^{-2})^2$ can be neglected. The
requirement $\mu_D^2\gg l_f^{-2}$ results in
\begin{equation}
\frac{{\bf k}_{\perp}^2}{2\omega}\ll \mu_D,\quad
\frac{1}{4}\,\mu_D\ll\omega\ll 2\gamma^2\mu_D,
\label{eq:4s}
\end{equation}
where we enter Lorentz-factor $\gamma=E/M$. Performing summation over polarization
state of radiated transverse gluon, we rewrite the Eq.\,(\ref{eq:4a}) as
integral over frequency $\omega$
\[
\left(-\frac{dE}{dt}\right)^{\!t} =
2\pi\biggl(\,\frac{\alpha_s}{\pi}\biggr)^{\!3}v_{\alpha}C_{\! A}\!
\sum\limits_{\beta=q,\,\bar{q},\,{\rm g}}\left(
\frac{C_2^{(\alpha)}C_2^{(\beta)}}{d_A}\right)\!n_{\beta}
\!\int\!d\omega\,\Bigl\{{\cal T}_1(\omega) + {\cal T}_2(\omega) +
{\cal T}_3(\omega)\Bigr\},
\]
where
\[
{\cal T}_1(\omega) = \int\!d{\bf k}_{\perp}\!\int\!d{\bf q}_{\perp}\,
\frac{1}{({\bf q}^2_{\perp} + \mu_D^2 + l_f^{-2})^2}\,
\frac{{\bf k}^2_{\perp}}{({\bf k}^2_{\perp} + \tilde{m}_g^2)^2},
\hspace{4.3cm}
\]
\begin{equation}
{\cal T}_2(\omega) = \int\!d{\bf k}_{\perp}\!\int\!d{\bf q}_{\perp}\,
\frac{-2}{({\bf q}^2_{\perp} + \mu_D^2 + l_f^{-2})^2}\,
\frac{{\bf k}_{\perp}\cdot({\bf k}-{\bf q})_{\perp}}
{({\bf k}^2_{\perp} + \tilde{m}_g^2)
(({\bf k}-{\bf q})_{\perp}^2 + \tilde{m}_g^2)},
\hspace{0.5cm}
\label{eq:4d}
\end{equation}
\[
{\cal T}_3(\omega) = \!\int\!d{\bf k}_{\perp}\!\int\!d{\bf q}_{\perp}\,
\frac{1}{({\bf q}^2_{\perp} + \mu_D^2 + l_f^{-2})^2}\,
\frac{({\bf k}-{\bf q})_{\perp}^2}
{(({\bf k}-{\bf q})_{\perp}^2 + \tilde{m}_g^2)^2}.
\hspace{3.3cm}
\]
Here, for brevity we denote $\tilde{m}_g^2\equiv m_g^2 +
\omega^2\gamma^{-2}$. Now we consider a function ${\cal
T}_1(\omega)$. In conditions of (\ref{eq:4s}) a simple calculation of
two-dimensional integrals $\int\!d{\bf k}_{\perp}\!\int\!d{\bf
q}_{\perp}$ leads to
\begin{equation}
{\cal T}_1(\omega) \simeq \frac{\pi^2}{\mu_D^2}
\left\{\ln\Biggl[\frac{{\bf k}^2_{\perp\,{\rm max}} + \tilde{m}_g^2}
{\tilde{m}_g^2}\Biggr] - 1\right\}.
\label{eq:4f}
\end{equation}
We introduce cut-off ${\bf k}^2_{\perp\,{\rm max}}$ on upper limit
for integration over $d{\bf k}^2_{\perp}$. As will be shown below the
divergent term in a sum of three expressions (\ref{eq:4d}), is
exactly cancelled. Here we draw attention to the following
circumstance. If we keep the term $l_f^{-2}$ in Coulomb factor, then
in this case integrating leads to finite expression by convergence of
integral over $d|{\bf k}_{\perp}|$. For conditions (\ref{eq:4s}) we will have
instead of (\ref{eq:4f})
\[
{\cal T}_1(\omega) \simeq \frac{\pi^2}{\mu_D^2}
\left\{\ln\Biggl[\frac{2\gamma^2\omega\mu_D}
{\omega^2 + \gamma^2 m_g^2}\Biggr] - 1 +\,
\frac{\omega^2 + \gamma^2 m_g^2}{2\gamma^2\omega\mu_D}\,
\Biggl(\frac{\pi}{2} - \arctan\Biggl[
\frac{\omega^2 + \gamma^2 m_g^2}{2\gamma^2\omega\mu_D}\Biggr]\Biggr)
\!\right\}.
\]
In particular in a frequencies region $\mu_D/4\ll\omega\ll\gamma m_g$ it
follows\footnote{Note that precisely in this frequencies region Ter-Mikaelian
effect in usual plasma \cite{akopian} is manifested. It is associated with
appearance of suppression factor $\omega^2/(\omega^2+\gamma^2m_g^2)\simeq
(\omega/\gamma m_g)^2\ll 1$ before logarithm in spectral radiation density.
In the case of QCD plasma such suppression is not observed.}
\[
{\cal T}_1(\omega) \simeq \frac{\pi^2}{\mu_D^2}
\left\{\ln\Biggl(\frac{4\omega}{\mu_D}\Biggr) -1 +\,
\Biggl(\frac{\mu_D}{4\omega}\Biggr)^{\!2} + \,\ldots\right\}.
\]
In this region a dependence on Lorentz-factor vanishes.

Finally, in asymptotic region $\omega\gg 2\gamma^2\mu_D$ we have
\[
{\cal T}_1(\omega) \simeq \frac{\pi^2}{\mu_D^2}\,
\frac{1}{6}\Biggl(\frac{2\gamma^2\mu_D}{\omega}\Biggr)^{\!2}.
\]
Here, we note a strong suppression of spectral density.

Such allowance for finiteness (inverse) formation length enables us
not only to perform more accurate integration providing its
finiteness, but gives also a possibility to analyze a behavior of
spectral density outside of framework of restriction (\ref{eq:4s}).
However here, a problem connected with necessity of the keeping of terms of
higher order smallness over ${\bf k}^2_{\perp},\,{\bf
q}^2_{\perp}\,etc$ arises in an approximation scheme, that results in
Eq.\,(\ref{eq:4a}). Besides for allowance of finiteness of $l_f^{-2}$
for deriving of the remaining expressions of ${\cal T}_2(\omega)$ and
${\cal T}_3(\omega)$, calculation complexities arise. In conditions
of requirements (\ref{eq:4s}) before integrating, these formulas can
be introduced in the form
\[
{\cal T}_2(\omega) \simeq -2\frac{\pi^2}{\mu_D^2}
\ln\Biggl[\frac{{\bf k}^2_{\perp\,{\rm max}} + \tilde{m}_g^2}
{\tilde{m}_g^2}\Biggr] - 2\frac{\pi^2}{\mu_D^2}
\left\{\Biggl(\frac{\mu_D^2 - 3\tilde{m}_g^2}{\mu_D^2 - 4\tilde{m}_g^2}\Biggr)
\ln \frac{\tilde{m}_g^2}{|\tilde{{m}_g^2}-\mu_D^2|}\, +
\right.
\]
\begin{equation}
\left.
\frac{2\tilde{m}_g^3}{\mu_D(\mu_D^2 - 4\tilde{m}_g^2)}
\ln\left|\frac{\tilde{m}_g + \mu_D}{\tilde{m}_g-\mu_D}\right|\right\} -
4\pi^2\tilde{m}_g\!\int\limits_{0}^{\infty}\!
\frac{{\bf k}^2_{\perp}d{\bf k}^2_{\perp}}
{{\bf k}^2_{\perp} + \tilde{m}_g^2}\,
\Phi({\bf k}^2_{\perp},\tilde{m}_g^2,\mu_D^2),
\label{eq:4g}
\end{equation}
\[
{\cal T}_3(\omega) = \frac{\pi^2}{\mu_D^2}
\ln\frac{\sqrt{{\bf k}^4_{\perp\,{\rm max}}+b{\bf k}^2_{\perp\,{\rm max}}+c}}
{|\tilde{m}_g^2-\mu_D^2|} +
\frac{\pi^2}{\mu_D^2}\left\{
-\frac{\tilde{m}_g^2}{\tilde{{m}_g^2}-\mu_D^2}+
{\cal K}(\omega,\tilde{{m}_g^2},\mu_D^2) +
\tilde{m}_g^2\frac{\partial{\cal K}(\omega,\tilde{{m}_g^2},\mu_D^2)}
{\partial \tilde{m}_g^2}\right\},
\]
where
\[
\Phi({\bf k}^2_{\perp},\tilde{m}_g^2,\mu_D^2)=
\frac{1}{D^{3/2}}\left\{\ln\,\Biggl|1
- \frac{{\bf k}^2_{\perp} - \tilde{m}_g^2 + \mu_D^2}{D^{1/2}}\Biggr| -
\ln\Biggl|\frac{D^{1/2} + {\bf k}^2_{\perp} + \tilde{m}_g^2}{\mu_D^2}-
\frac{{\bf k}^2_{\perp} - \tilde{m}_g^2 + \mu_D^2}{D^{1/2}}\Biggr|
\right\},
\]
\[
D={\bf k}^4_{\perp}+b{\bf k}^2_{\perp}+c,\quad
b=2(\tilde{m}_g^2 + \mu_D^2),\quad c=(\tilde{m}_g^2 - \mu_D^2)^2,
\quad \tilde{m}_g^2= m_g^2 + \omega^2\gamma^{-2},
\]
\[
{\cal K}(\omega,\tilde{{m}_g^2},\mu_D^2)=
-\frac{\mu_D}{\tilde{m}_g}
\ln\left|\frac{\tilde{m}_g + \mu_D}{\tilde{m}_g-\mu_D}\right|
-\mu_D^2\!\int\limits_0^{\infty}\!({\bf k}^2_{\perp} -
\tilde{m}_g^2 + \mu_D^2)
\Phi({\bf k}^2_{\perp},\tilde{m}_g^2,\mu_D^2)d{\bf k}^2_{\perp}.
\]
As we see from Eqs.\,(\ref{eq:4f}) and (\ref{eq:4g}) in limit
${\bf k}^2_{\perp\,{\rm max}}\rightarrow\infty$ cut-off disappears in a
sum ${\cal T}_1(\omega)+{\cal T}_2(\omega)+{\cal T}_3(\omega)$ as expected.
In region of $\omega\ll\gamma m_g$ from explicit analytic expressions there is
no any suppression by factor $\omega/\gamma m_g$.

\section{\bf Bremsstrahlung induced by effective vertex correction
$\delta\Gamma_{3g}$}
\setcounter{equation}{0}

Let us consider now a contribution to radiation energy loss related
to existence of medium induced HTL-correction $\delta \Gamma_{3g}$ to
bare three gluon vertex. According to Eq.\,(\ref{eq:3u}) amplitude of
HTL-induced bremsstrahlung of transverse soft gluon takes a form
\begin{equation}
{\rm e}^i(\hat{\bf k},\zeta)
\delta\Gamma^{i\nu\lambda}(k,-q,-k+q)
\,\,^{\ast}{\cal D}_{C\nu\nu^{\prime}}(q)v_{\beta}^{\nu^{\prime}}
\,^{\ast}{\cal D}_{C\lambda\lambda^{\prime}}(k-q)v_{\alpha}^{\lambda^{\prime}}
|_{\,\omega=\omega_{\bf k}^t,\,q_0={\bf v}_{\beta}\cdot {\bf q}}\,.
\label{eq:5q}
\end{equation}
In what follows we restrict our consideration to approximation of
static color target, i.e. we set ${\bf v}_{\beta}=0$. As it will be shown
below the amplitude (\ref{eq:5q}) gives a main contribution to energy
loss. However it is necessary to note that generally speaking,
approximation of static color target is in contradiction with
conception of {\it hard thermal loops}. As known HTL amplitude is
calculated within high-temperature QCD plasma, when it is assumed
that all thermal partons $\beta$ constituting plasma, are
ultra-relativistic (and massless). Nevertheless in this Section we
consider the medium-induced bremsstrahlung in the framework of the
potential model, to compare with usual bremsstrahlung within the same
model, keeping in mind above-mentioned remark. Now we turn to
analysis of amplitude ({\ref{eq:5q}). In the expression (\ref{eq:5q})
in the propagator $^{\ast}{\cal D}_{C\lambda\lambda^{\prime}}(k - q)$
we keep only transverse part, that gives also leading contribution in
medium-induced bremsstrahlung (see below). Then (\ref{eq:5q}) in
statistic limit reads
\begin{equation}
\frac{1}{{\bf q}^2+\mu_D^2}
\,^{\ast}{\!\Delta}^t(k-q)
{\rm e}^i(\hat{\bf k},\zeta)\!\left[\,
\delta\Gamma^{i0j}(k,-q,-k+q)\Biggl(\delta^{jj^{\,{\prime}}}-
\frac{({\bf k}-{\bf q})^j({\bf k}-{\bf q})^{j^{\,{\prime}}}}
{({\bf k}-{\bf q})^2}
\Biggr) v_{\alpha}^{j^{\,{\prime}}}\right]_{q_0=0}.
\label{eq:5w}
\end{equation}
The explicit expression for HTL-correction for $q_0=0$ is
\[
\delta\Gamma^{i0j}(k,-q,-k+q)|_{\,q_0=0}=
\mu_D^2\,{\omega}\!\int\!\frac{d{\Omega}_{\bf v}}{4\pi}\,
\frac{v^iv^j}{(v\cdot k)(v\cdot (k-q) + i\epsilon)},\;\,\epsilon\rightarrow +0.
\]
Following Frenkel and Taylor \cite{frenkel} we present integral on the right-hand
side as an expansion in basis
\begin{equation}
\int\!\frac{d{\Omega}_{\bf v}}{4\pi}\,
\frac{v^iv^j}{(v\cdot k)(v\cdot (k-q) + i\epsilon)} =
Xn^in^j + Yl^il^j + Y^{\prime}l^{\prime i}l^{\prime j} +
Z(l^il^{\prime j} + l^{\prime i}l^j),
\label{eq:5e}
\end{equation}
where ${\bf n}=[{\bf q},{\bf k}]=[{\bf k},{\bf k}-{\bf q}],\,
{\bf l}=[{\bf n},{\bf k}]$ and ${\bf l}^{\prime}=[{\bf n},{\bf k}-{\bf q}]$.
The coefficient functions in this expansion in static limit $q_0=0$
according to Eqs.\,(3.32), (3.33) in \cite{frenkel}, have a form
\[
X=\frac{1}{\delta^2}\,
\Bigl[\,-\Delta\,M(k,k-q) + \omega({\bf k}\cdot{\bf q})L(k) -
\omega(({\bf k}-{\bf q})\cdot {\bf q}) L(k-q)\Bigr],
\hspace{1,3cm}
\]
\begin{equation}
Y=\frac{1}{\delta^2}\,
\left[\,-\omega^2 M(k,k-q) -
\omega\Biggl(1+\frac{{\bf k}\cdot({\bf k}-{\bf q})}
{{\bf k}^2}\Biggr)L(k) + \frac{{\bf k}\cdot({\bf k}-{\bf q})}
{{\bf k}^2}\right],
\hspace{0.23
cm}
\label{eq:5r}
\end{equation}
\[
Y^{\prime}=\frac{1}{\delta^2}
\left[\,-\omega^2 M(k,k-q) -
\omega\Biggl(1+\frac{{\bf k}\cdot({\bf k}-{\bf q})}
{({\bf k}-{\bf q})^2}\Biggr)L(k-q)
+ \frac{{\bf k}\cdot({\bf k}-{\bf q})}{({\bf k}-{\bf q})^2}
\right],
\hspace{0.2cm}
\]
\[
Z=\frac{1}{\delta^2}\,
\Bigl[\,-\omega^2\,M(k,k-q)+\omega L(k)+\omega L(k-q)-1\Bigr].
\hspace{3.8cm}
\]
Here,
\[
\delta={\bf n}^2,\quad\Delta = k^2q^2-(k\cdot q)^2=
-{\bf q}^2\omega^2+\delta^2,
\]
\begin{equation}
L(k)=\frac{1}{2|{\bf k}|}\,
\ln\!\left(\frac{\omega+|{\bf k}|}{\omega-|{\bf k}|}
\right) - \,\frac{i\pi}{2|{\bf k}|}\,\theta\Bigl(1-\frac{\omega}{|{\bf k}|}
\Bigr),
\label{eq:5t}
\end{equation}
\[
M(k,k-q)=\,\frac{1}{2\sqrt{-\Delta}}\,
\ln\!\left(\frac{k\cdot(k-q)+\sqrt{-\Delta}}{k\cdot(k-q)-\sqrt{-\Delta}}
\right)-
\frac{i\pi}{2\sqrt{-\Delta}}\,\theta(-\Delta).
\]

Even in static limit of amplitude (\ref{eq:5w}) analysis of energy
loss induced by medium-induced bremsstrahlung in closed analytical
form is impossible by virtue of awkwardness of the expressions
(\ref{eq:5e}), (\ref{eq:5r}). Therefore we make some simplifying
assumption for subsequent analysis. Note that coefficient functions
(\ref{eq:5r}) contain  a factor $1/\delta^2$ as a common one. Let us
assume that main contribution to HTL-induced bremsstrahlung follows
from kinematic region of momentum variables ${\bf k}$ and ${\bf q}$,
where $\delta$ vanishes, i.e. when vectors ${\bf k}$ and ${\bf q}$
are collinearity. We approximate all functions entering into amplitude
(\ref{eq:5w}) in kinematical region $\delta = \delta({\bf k}, {\bf
q}) = 0$.

In the amplitude (\ref{eq:5w}) in three-dimensional transverse
projector, there is also contracting HTL-correction with 
$({\bf k}-{\bf q})^j({\bf k}-{\bf q})^{\,{j^{\,{\prime}}}}/({\bf k}
-{\bf q})^2$. By virtue of Ward identity for HTL amplitudes
\cite{frenkel,pisarski} here, we have
\[
\delta\Gamma^{i0j}(k,-q,-k+q)({\bf k}-{\bf q})^j=-\omega\,
\delta\Gamma^{i00}(k,-q,-k+q) + \delta\Pi^{i0}(q) - \delta\Pi^{i0}(k),
\]
where $\delta\Pi^{\mu\nu}$ is polarization tensor. The vertex
$\delta\Gamma^{i00}(k,-q,-k+q)$ on the right-hand side contains in
coefficient functions the factor $1/\delta$ only in the first power
(Eq.\,(3.30) in \cite{frenkel}) and such is ``less singularity'',
then vertex $\delta\Gamma^{i0j}(k,-q,-k+q)$. Therefore this
contribution can be omitted. For the same reason in propagator
$^{\ast}{\cal D}_{C\lambda\lambda^{\prime}}(k-q)$ we neglected by the
term containing longitudinal scalar propagator
$^{\ast}{\!\Delta}^l(k-q)$ and in matrix element we dropped
medium-induced bremsstrahlung of longitudinal oscillations quantum
(plasmon). All these contributions are proportional either zero or
first power of $1/\delta$.

We introduce the coordinate system in which axis $0Z$ is aligned with the
velocity ${\bf v}_{\alpha}$ (Fig.\,\ref{fig2}). It is convenient to enter a new
variable ${\bf q}^{\prime}\equiv{\bf k}-{\bf q}=({\bf q}_{\perp}^{\prime},
q_{\|}^{\prime})$.
\begin{figure}[hbtp]
\begin{center}
\includegraphics*[height=8cm]{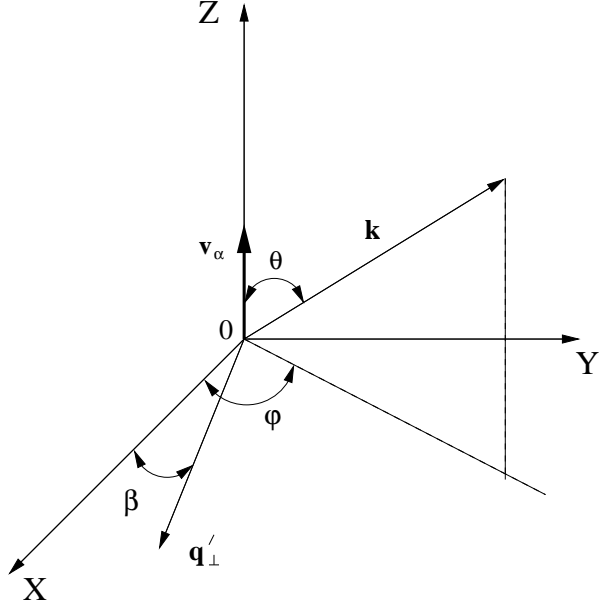}
\end{center}
\caption{\small The coordinate system for analysis of
HTL-induced bremsstrahlung.}
\label{fig2}
\end{figure}
The longitudinal component $q_{\|}^{\prime}$ by virtue of delta-function
in integrand (\ref{eq:3a}) (for ${\bf v}_{\beta}=0$) equals $\omega/v_{\alpha}$.
In future discussion we will write $\omega$ instead of
$\omega^t_{\bf k}$. In chosen coordinate system we have
${\bf k}=(|{\bf k}|,\theta,\varphi),\,{\bf q}_{\perp}^{\prime}=
(|{\bf q}_{\perp}^{\prime}|,\pi/2,\beta)$, and
integration measures are
\[
d{\bf k}={\bf k}^2d|{\bf k}|\sin\theta d\theta d\varphi,\quad
d{\bf q}_{\perp}^{\prime}= |{\bf q}_{\perp}^{\prime}|
d|{\bf q}_{\perp}^{\prime}|d \beta.
\]
It is not difficult to see that in this coordinate system one have
\[
{\bf k}\cdot{\bf q}^{\prime} =
\frac{\omega|{\bf k}|}{v_{\alpha}}\,\cos\theta +
|{\bf k}||{\bf q}_{\perp}^{\prime}|\sin\theta\cos(\varphi-\beta).
\]

Subsequently computation azimuth angles $\varphi$ and $\beta$ will
be enter as a difference $\varphi-\beta$. Therefore we remove
integration over $\beta$: $\int d\beta=2\pi$, replacing
$\varphi-\beta\rightarrow\varphi$. The kinematic region of variables
${\bf k}$ and ${\bf q}$ that is of our interest, is defined by
equation
\begin{equation}
\delta = {\bf k}^2{\bf q}^{\,\prime 2}-({\bf k}\cdot{\bf q}^{\prime})^2=0,
\label{eq:5y}
\end{equation}
and using a scalar product defined above, we have
\begin{equation}
\delta={\bf k}^2\Biggl[\,\Biggl(\frac{\omega^2}{v_{\alpha}^2}
+{\bf q}_{\perp}^{\prime\,2}\Biggr) -
\Biggl(\frac{\omega}{v_{\alpha}}\,\cos\theta +
|{\bf q}_{\perp}^{\prime}|\sin\theta\cos\varphi\Biggr)^{\!2}\,\Biggr]=0.
\label{eq:5u}
\end{equation}
We consider (\ref{eq:5u}) as equation with respect to variable $\cos\varphi$.
Eq.\,(\ref{eq:5u}) defines two solutions
\begin{equation}
\cos\varphi_{\pm}=-\,\frac{1}{\chi\sin\theta}\,
\Bigl(\cos\theta\mp\sqrt{1+{\chi}^2}\,\Bigr),
\label{eq:5i}
\end{equation}
where we have introduced a new variable
$\chi=v_{\alpha}|{\bf q}_{\perp}^{\prime}|/\omega$. The condition
$\cos^2\varphi_{\pm}\leq 1$ results in restriction
\[
\Bigl(1 \mp \cos\theta\sqrt{1+{\chi}^2}\,\Bigr)^{\,2}\leq 0.
\]
It is evident that the restriction is valid only for two relations between
polar angle $\theta$ and variable $\chi$
\begin{equation}
\cos\theta_{+}=\,\frac{1}{\sqrt{1+{\chi}^2}},\quad
\cos\theta_{-}=-\,\frac{1}{\sqrt{1+{\chi}^2}}.
\label{eq:5o}
\end{equation}

Substituting these relations into (\ref{eq:5i}), we obtain that
condition $\delta=0$ is true only in ``end'' points:
$\cos^2\varphi_{\pm}=1$. For definiteness we choose
$\cos\varphi_{+}=1,$ i.e. $\varphi_{+}=0$, and $\cos\varphi_{-}=-1,$
i.e. $\varphi_{-}=\pi$. It follows that
$\sin\theta_{\pm}=\chi/\sqrt{1+{\chi}^2}$, and therefore by virtue of
(\ref{eq:5o}) we have
\[
\tan\theta_{\pm}=\pm\chi.
\]
The graphs of functions $\theta_{\pm}=\theta_{\pm}(\chi)$ are
given on Fig.\,\ref{fig3}. Such the equation (\ref{eq:5y}) defines
two non-overlapping
\begin{figure}[hbtp]
\begin{center}
\includegraphics*[height=8cm]{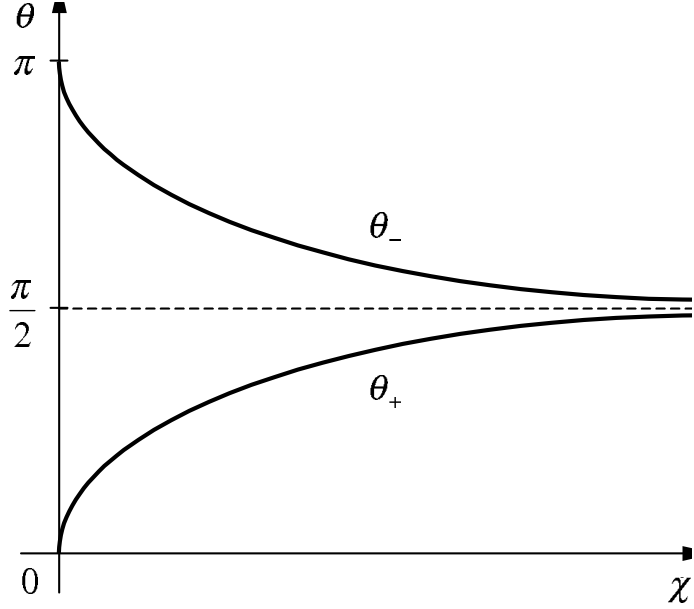}
\end{center}
\caption{\small The dependence of angles $\theta_{\pm}$ on $\chi$.}
\label{fig3}
\end{figure}
cones of medium-induced bremsstrahlung: (1) along a particle velocity
${\bf v}_{\alpha}$ $(0\leq\theta_{+}\leq\pi/2)$ and (2) in opposite direction
$(\pi/2\leq\theta_{-}\leq\pi)$. It is clear that such separation exists only
within assumption on leading contribution to medium-induced bremsstrahlung
discussed above.

Now we define module squared of amplitude (\ref{eq:5w}) in region
(\ref{eq:5y}). Taking into account an expansion (\ref{eq:5e}) and
summing over polarization state of radiated quantum with relation
\[
\sum_{\zeta=1,\,2}|\,{\bf e}({\hat{\bf k}},\zeta)\cdot {\bf a}|^{\,2}=
[\hat{\bf k},{\bf a}]^2
\]
valid for arbitrary vector ${\bf a}$, after simple algebraic transformations
we have
\[
\frac{1}{({\bf q}^2+\mu_D^2)^2}\,
|\,^{\ast}{\!\Delta}^t(k-q)|^{\,2}
\sum_{\zeta=1,2}
|\,{\rm e}^i(\hat{\bf k},\zeta)
\delta\Gamma^{i0j}(k,-q,-k+q)v_{\alpha}^{j}|^{\,2},
\]
where
\[
\sum_{\zeta=1,2}|\,{\rm e}^i\delta\Gamma^{i0j}v_{\alpha}^{j}|^{\,2}=
\delta\Bigl[\,
({\bf v}_{\alpha}\cdot{\bf n})^2 |X|^2 +
({\bf v}_{\alpha}\cdot{\bf l})^2
\Bigl| |{\bf k}|Y + (\hat{\bf k}\cdot{\bf q}^{\prime})Z\Bigr|^{\,2}
+({\bf v}_{\alpha}\cdot{\bf l}^{\prime})^2
\Bigl| |{\bf k}|Z + (\hat{\bf k}\cdot{\bf q}^{\prime})Y^{\prime}\Bigr|^{\,2}
\]
\begin{equation}
+\,
2\,({\bf v}_{\alpha}\cdot{\bf l})({\bf v}_{\alpha}\cdot{\bf l}^{\prime})\,
{\rm Re}\Bigl(\!\{|{\bf k}|Y + (\hat{\bf k}\cdot{\bf q}^{\prime})Z\}
\{|{\bf k}|Z + (\hat{\bf k}\cdot{\bf q}^{\prime})Y^{\prime}\}^{\ast}\Bigr)\Bigr].
\label{eq:5p}
\end{equation}
In chosen coordinate system scalar productions of velocity ${\bf
v}_{\alpha}$ with vectors ${\bf n}$, ${\bf l}$ and ${\bf l}^{\prime}$
entering into the right-hand side of Eq.\,(\ref{eq:5p}), are
represented in the form
\begin{equation}
{\bf v}_{\alpha}\cdot{\bf n} = \omega|{\bf k}|\chi
\sin\theta\sin\varphi,\;\;
{\bf v}_{\alpha}\cdot{\bf l} =
\omega{\bf k}^2\sin\theta\cos\theta\,
[\,\tan\theta - \chi\cos\varphi],
\label{eq:5a}
\end{equation}
\[
{\bf v}_{\alpha}\cdot{\bf l}^{\prime} =
-\omega|{\bf k}||{\bf q}_{\perp}^{\prime}|\cos\theta\,
[\,\chi - \tan\theta\cos\varphi].
\]
They vanish for $\cos\varphi_{\pm}=\pm1,\;\tan\theta_{\pm}=\pm\chi$. By using
explicit expressions for coefficient functions (\ref{eq:5r}) one can derive
limiting values of its quadratic combinations which
enter into the right-hand side of Eq.\,(\ref{eq:5p}). Rather cumbersome
calculations lead to the following result
\begin{equation}
\lim\limits_{\delta\rightarrow 0}|X|^{\,2}\cong
\frac{1}{\delta^4}\,\omega^2\biggl(|{\bf k}|\mp\frac{\omega}{v_{\alpha}}
\cosh\xi\biggr)^{\!2}\,\biggl[{\cal M}^2_{\pm}(\omega,|{\bf k}|,\xi) +
\pi^2\left(
\begin{array}{c}
0 \\ 1
\end{array}
\right)\biggr],
\label{eq:5s}
\end{equation}
\[
\lim\limits_{\delta\rightarrow 0\,}
\Bigl| |{\bf k}|Y + (\hat{\bf k}\cdot{\bf q}^{\prime})Z\Bigr|^{\,2}
\cong
\frac{1}{\delta^4}\,\frac{\omega^2}{\biggl(|{\bf k}|\mp
\displaystyle\frac{\omega}{v_{\alpha}}
\cosh\xi\biggr)^{2}}\,
\biggl[\,{\bf k}^2{\cal M}^2_{\pm}(\omega,|{\bf k}|,\xi) +
\pi^2\!\left(
\begin{array}{c}
\displaystyle\frac{\omega^2}{v_{\alpha^2}}\,\cosh^2\xi \\ {\bf k}^2
\end{array}
\right)\biggr],
\]
\[
\lim\limits_{\delta\rightarrow 0}\,
\Bigl| |{\bf k}|Z + (\hat{\bf k}\cdot{\bf q}^{\prime})Y^{\prime}\Bigr|^{\,2}
\!\cong\!
\frac{1}{\delta^4}\,\frac{\omega^2}{\biggl(|{\bf k}|\mp
\displaystyle\frac{\omega}{v_{\alpha}}
\cosh\xi\biggr)^{2}}\,
\biggl[\,\frac{\omega^2}{v_{\alpha}^2}\,\cosh^2\xi\,
{\cal M}^2_{\pm}(\omega,|{\bf k}|,\xi) +
\pi^2\!\left(\!
\begin{array}{c}
{\bf k}^2 \\ \displaystyle\frac{\omega^2}{v_{\alpha^2}}\,\cosh^2\xi
\end{array}
\!\!\right)\biggr],
\]
\[
\lim\limits_{\delta\rightarrow 0}\,
\Bigl(\{|{\bf k}|Y + (\hat{\bf k}\cdot{\bf q}^{\prime})Z\}
\{|{\bf k}|Z + (\hat{\bf k}\cdot{\bf q}^{\prime})Y^{\prime}\}^{\ast}\Bigr)
\]
\[
\cong\pm\,
\frac{1}{\delta^4}\,\frac{\omega^2}{\biggl(|{\bf k}|\mp
\displaystyle\frac{\omega}{v_{\alpha}}\cosh\xi\biggr)^{2}}\,
\biggl(\frac{\omega|{\bf k}|}{v_{\alpha}}\,\cosh\xi\biggr)
\biggl[\,{\cal M}^2_{\pm}(\omega,|{\bf k}|,\xi) - \pi^2\biggr],
\]
where
\[
{\cal M}_{\pm}(\omega,|{\bf k}|,\xi)=
\ln\!\left\{\!\Biggl(\frac{\omega+|{\bf k}|}{\omega-|{\bf k}|}\Biggr)
\Biggl(\frac{v_{\alpha}\mp\cosh\xi}{v_{\alpha}\pm\cosh\xi}
\Biggr)\!\!\right\},\quad
\xi\equiv\ln\Bigl(\chi + \sqrt{1+\chi^2}\,\Bigr),\quad\chi=\sinh\xi.
\]
For convenience of the subsequent reference we write out once more the
expression for energy loss of particle $\alpha$ connected with medium-induced
bremsstrahlung of soft transverse gluon
\begin{equation}
\left(-\frac{dE}{dt}\right)^{\!t}=
\frac{{\alpha}_s^3}{\pi}\,\frac{2}{v_{\alpha}}\,C_A\!
\sum_{\beta=q,\,\bar{q},\,{\rm g}}
\Biggl(\frac{C_2^{(\alpha)}C_2^{(\beta)}}{d_A}\Biggr)n_{\beta}\!
\int\limits_{0}^{\infty}\!{\bf k}^2d|{\bf k}|\,
(\omega_{\bf k}^t)^3\Biggl(\frac{Z_t({\bf k})}{2\omega_{\bf k}^t}\Biggr)
\int\limits_{0}^{\pi}\sin\theta
d\theta\!\int\limits_{0}^{2\pi}d\varphi
\label{eq:5d}
\end{equation}
\[
\times\!\int\!|{\bf q}_{\perp}^{\prime}|
\,d|{\bf q}_{\perp}^{\prime}|
\,\frac{{\mu}_D^4}{({\bf q}^2+\mu_D^2)^2}\,
\sum\limits_{\zeta=1,2}\!|\,{\rm e}^i(\hat{\bf k},\zeta)
\delta\Gamma^{i0j}(k,-q,-k+q)v_{\alpha}^{j}|^{\,2}\,
|\,^{\ast}{\!\Delta}^t(\omega_{\bf k}^t,{\bf q}_{\perp}^{\prime},
q_{\|}^{\prime})|^{\,2}_{q_{\|}^{\,\prime}=\omega_{\bf k}^t/v_{\alpha}}.
\]
Here, an expression $\sum_{\zeta=1,2}|\,{\rm
e}^i\delta\Gamma^{i0j}v_{\alpha}^{j}|^{\,2}$ in limit
$\delta\rightarrow 0$, is defined by equations
(\ref{eq:5p})\,--\,(\ref{eq:5s}). Coulomb factor in integrand
(\ref{eq:5d}) have limiting value
\[
\lim\limits_{\delta\rightarrow 0}
\,\frac{1}{({\bf q}^2+\mu_D^2)^2}=
\frac{1}{\Biggl[\biggl(|{\bf k}|\mp
\displaystyle\frac{\omega}{v_{\alpha}}\cosh\xi\biggr)^{\,2}
+\mu_D^2\Biggr]^{\,2}}\,.
\]
From the equations (\ref{eq:5p})\,--\,(\ref{eq:5s}) we see that there
are four types of integrals over azimuth angle $\varphi$, having
singular integrands in limit $\delta\rightarrow 0$:
\begin{equation}
\int\limits_{0}^{2\pi}\frac{\sin^2\varphi\,d\varphi}{\delta^3}\,,\quad
\int\limits_{0}^{2\pi}\,\frac{[\tan\theta - \chi\cos\varphi]
[\chi -\tan\theta\cos\varphi]\,d\varphi}
{\delta^3}\,,
\label{eq:5f}
\end{equation}
\[
\int\limits_{0}^{2\pi}\,\frac{[\tan\theta - \chi\cos\varphi]^2\,d\varphi}
{\delta^3}\,,\quad
\int\limits_{0}^{2\pi}\,\frac{[\chi -\tan\theta\cos\varphi]^2\,d\varphi}
{\delta^3}\,.
\hspace{0.9cm}
\]

Let us consider for example the first integral. By using an explicit expression
for function $\delta$ (the left-hand side of Eq.\,(\ref{eq:5u})) the factor
$1/\delta^3$ can be presented as
\[
\frac{1}{\delta^3}=
\Biggl(\frac{v_{\alpha}}{\omega|{\bf k}|}\Biggr)^{\!6}
\frac{1}{(2\cosh\xi)^3}\,\left\{
\frac{1}{(a_{+} + b_{+}\cos\varphi)^3}
+3\,\frac{1}{(a_{+} + b_{+}\cos\varphi)^2(a_{-} + b_{-}\cos\varphi)}
\right.
\]
\[
\left.
+\,3\,\frac{1}{(a_{+} + b_{+}\cos\varphi)(a_{-} + b_{-}\cos\varphi)^2}
+\frac{1}{(a_{-} + b_{-}\cos\varphi)^3}\right\},
\]
where
$a_{\pm}=\cosh\xi\mp\,\cos\theta,\;b_{\pm}=\mp\sinh\xi\sin\theta$.
The main contribution is connected with terms $1/(a_{\pm} +
b_{\pm}\cos\varphi)^3$. Using table integrals \cite{gradshteyn}, we
derive
\[
\int\limits_{0}^{2\pi}\frac{\sin^2\!\varphi\,d\varphi}
{(a_{\pm} + b_{\pm}\cos\varphi)^3}=
\frac{\pi}{\Bigl[(\cosh\xi\mp\cos\theta)^2-
\sinh^2\xi\sin^2\theta\Bigr]^{\,3/2}}=
\frac{\pi}{\Bigl[\cosh\xi\cos\theta \mp 1\Bigr]^{\,3}}\,.
\]
The integral diverges for values: $\cos\theta_{\pm}=\pm\,1/\cosh\xi$.

Furthermore we consider for concreteness the energy losses associated
with bremsstrah\-lung to hemisphere along velocity ${\bf v}_{\alpha}$
($0\leq\theta\leq\pi/2$), i.e. in previous expressions we choose
upper sign and besides for the sake of simplification we restrict our
consideration to contribution connected with function $X$ only. By
virtue of (\ref{eq:5p}) this contribution enters by independent
fashion in the squared amplitude, i.e. it doesn't interfere with
other contributions: $Y,\,Y^{\prime}$ and $Z$. In this case taking
into account above-mentioned, we have instead of (\ref{eq:5p})
\[
\frac{v_{\alpha}^6}{\omega^2|{\bf k}|^4}\,
\sinh^2\!\xi\sin^2\!\theta\,\biggl(|{\bf k}|-
\frac{\omega}{v_{\alpha}}\cosh\xi\biggr)^2
{\cal M}_{+}^2(\omega,|{\bf k}|,\xi)\,
\frac{1}{(2\cosh\xi\cos\theta)^3}\,
\frac{\pi}{\left[\cosh\xi-\displaystyle\frac{1}{\cos\theta}\right]^{\,3}}\,.
\]

By singular factor $1/[\cosh\xi-1/\!\cos\theta]^3$ for rough
estimation of energy loss we set $\cosh\xi=1/\!\cos\theta$ in
integrand of Eq.\,(\ref{eq:5d}) (or in terms of $|{\bf
q}^{\prime}_{\perp}|:\; |{\bf
q}^{\prime}_{\perp}|=(\omega/v_{\alpha})\tan\theta$). The
integration measure over $|{\bf q}^{\prime}_{\perp}|$ will be
presented as
\[
\int\!|\,{\bf q}_{\perp}^{\prime}|\,d|{\bf q}_{\perp}^{\prime}|
=\left(\frac{\omega}{v_{\alpha}}\right)^{\!2}\frac{1}{\cos\theta}\!
\int\!\sinh\xi\,d\xi.
\]
We perform the integration of singular factor by the principal-value
prescription
\[
{\rm P}\!\int\limits_0^{\infty}\!\frac{\sinh\xi\,d\xi}
{\left(\cosh\xi-1/\!\cos\theta\right)^{\,3}}
=\lim\limits_{\varepsilon\rightarrow 0}
\left(\!\int\limits_0^{1/\!\cos\theta - \varepsilon}\!\!\frac{dz}
{\left(z-1/\cos\theta\right)^{\,3}}\,+
\int\limits_{1/\!\cos\theta + \varepsilon}^{\infty}\!\frac{dz}
{\left(z-1/\!\cos\theta\right)^{\,3}}\!\right)
= \frac{1}{2}\,\cos^2\!\theta.
\]
In the context of this approximation the expression for energy loss
(\ref{eq:5d}) associated with contribution of function $X$ takes a form
\begin{equation}
\left(-\frac{dE}{dt}\right)^{\!t}_{\! X}=
\frac{1}{4}\,{\alpha}_s^3v_{\alpha}^5\,C_A\!\!
\sum_{\beta=q\!,\,\bar{q}\!,\,{\rm g}}
\Biggl(\frac{C_2^{(\alpha)}C_2^{(\beta)}}{d_A}\Biggr)n_{\beta}\!
\int\limits_{0}^{\infty}\!\frac{d|{\bf k}|}{{\bf k}^2}\,
(\omega_{\bf k}^t)^3\Biggl(\frac{Z_t({\bf k})}{2\omega_{\bf k}^t}\Biggr)
\label{eq:5g}
\end{equation}
\[
\times\!\int\limits_{0}^{\pi/\!2}\!d\theta\cos\theta\sin^5\!\theta
\,\frac{\mu_D^4\,(\omega_{\bf k}^t-|{\bf k}|v_{\alpha}\cos\theta)^2}
{[(\omega_{\bf k}^t-|{\bf k}|v_{\alpha}\cos\theta)^2 +
v_{\alpha}^2\cos^2\!\theta\mu_D^2]^{2}}\,
{\cal M}_{+}^2(\omega_{\bf k}^t,|{\bf k}|,\theta)\,
|\,^{\ast}{\!\Delta}^t(\omega_{\bf k}^t,|{\bf k}|,\theta)|^{\,2},
\]
where now
\[
{\cal M}_{+}(\omega_{\bf k}^t,|{\bf k}|,\theta)=
\ln\!\left\{\!\Biggl(\frac{\omega_{\bf k}^t +|{\bf k}|}{\omega_{\bf
k}^t-|{\bf k}|}\Biggr)
\Biggl(\frac{1-v_{\alpha}\cos\theta}{1+v_{\alpha}\cos\theta}
\Biggr)\!\!\right\},
\]
and a transverse propagator is
\[
\,^{\ast}{\!\Delta}^{\!-1\,t}(\omega_{\bf k}^{t},|{\bf k}|,\theta)=
-(\omega_{\bf k}^{t})^2\,\frac{1-v_{\alpha}^2\cos^2\!\theta}
{v_{\alpha}^2\cos^2\!\theta} - m_g^2\,v_{\alpha}^2\cos^2\!\theta
\left\{1 +\frac{1-v_{\alpha}^2\cos^2\!\theta}
{v_{\alpha}^2\cos^2\!\theta}\,F(v_{\alpha}\cos\theta)\right\}.
\]
Here, the function $F(z)$ was defined in Section 4.

Let us estimate the expression (\ref{eq:5g}) in the long-wave limit
$|{\bf k}|\rightarrow 0,\;\omega_{\bf k}^{t}\sim\omega_{pl}$ and a
small-angle approximation $\theta\leq\gamma^{-1}$, where
$\gamma=E/M$. Taking into account
$1-v_{\alpha}\cos\theta\simeq(\gamma^{-2}+\theta^2)/2$ and
integrating over angle $\theta$ in range $(0,\gamma^{-1})$, we derive
from (\ref{eq:5g})
\begin{equation}
\left.\left(-\frac{dE}{dt}\right)^{\!t}_{\! X}
\right|_{\,\theta\leq\gamma^{-1}}\simeq
\frac{1}{3\cdot 2^5}\,{\alpha}_s^3\,C_A\!\!
\sum_{\beta=q\!,\,\bar{q}\!,\,{\rm g}}
\Biggl(\frac{C_2^{(\alpha)}C_2^{(\beta)}}{d_A}\Biggr)n_{\beta}\!
\left(\int\limits_{0}^{\omega_{pl}}
\!\frac{d|{\bf k}|}{{\bf k}^2}\right)
\label{eq:5h}
\end{equation}
\[
\times(\gamma^{-2})^3\!
\left\{\!\Bigl[\,\ln(\gamma^{-2}/4) -  C_E\Bigr]
\Bigr[\,2\ln 2 -5/6\Bigr] + \frac{1}{2}\,\ln^2(\gamma^{-2}/4) +
3\sum\limits_{k=1}^{\infty}\frac{(-1)^k\psi(k)}{(3+k)k}\!\right\},
\]
where $C_E$ is an Euler's constant and
$\psi(z)=(\ln\Gamma(z))^{\prime}$, $\Gamma(z)$ is a
gamma-function. The integral over $d|{\bf k}|$ infrared diverges
in this approximation. It should be cut-off in lower limit on
ultrasoft scale $g^2T$. It follows
\[
\int\limits_{g^2T}^{\omega_{pl}}\!\frac{d|{\bf k}|}{{\bf k}^2}
\sim\frac{1}{g^2T}.
\]
With allowance for the last expression and the fact that for
ultrarelativistic plasma densities $n_\beta\sim T^3$, we derive
from Eq.\,(\ref{eq:5h}) that energy loss connected with
bremsstrahlung of soft transverse gluons $(\omega_{\bf k}^{t}\sim
gT)$ in small-angle approximation in value order is
\begin{equation}
\left.\left(-\frac{dE}{dt}\right)^{\!t}_{\! X}
\right|_{\,\theta\leq\gamma^{-1}}\!\sim\,
{\alpha}_s^2\, T^2 (\gamma^{-2})^3\ln^2\!\left(\frac{\gamma^{-2}}{4}\right).
\label{eq:5j}
\end{equation}
Although here, we have the same order in coupling constant as for
usual bremsstrahlung, however derived expression is strongly
suppressed by Lorentz-factor $\gamma$. Such in a small-angle
approximation the energy loss connected with existence of an
effective vertex $\delta\Gamma_{3g}$ induced by medium, gives negligible
small contribution in comparison with usual energy loss mechanism for
high-energy parton. The most of energy loss, as we expected, is
defined by radiation of soft gluons for large angles $\theta\gg
\gamma^{-1}$.

Now we consider an opposite case of short-wave limit, $|{\bf
k}|\gg\omega_{pl}$, when $\omega_{\bf k}^{t}\simeq|{\bf
k}|+m_g^2/2|{\bf k}|$. In approach of small angles and
$v_{\alpha}\simeq 1-\gamma^{-2}/2$ the following approximations are
obeyed
\[
\omega_{\bf k}^t-|{\bf k}|v_{\alpha}\cos\theta\simeq\frac{1}{2}\,
|{\bf k}|\left(\theta^2+\gamma^{-2} + \frac{m_g^2}{{\bf k}^2}\right),\quad
{\cal M}_{+}(\omega_{\bf k}^t,|{\bf k}|,\theta)\simeq
\ln\left(\frac{{\bf k}^2}{m_g^2}(\theta^2+\gamma^{-2})\right),
\]
\[
\,^{\ast}{\!\Delta}^{\!-1\,t}(\omega_{\bf k}^{t},|{\bf k}|,\theta)\simeq
-\Bigl(\,{\bf k}^2(\theta^2+\gamma^{-2}) + m_g^2\Bigr).
\]
The Coulomb factor in this case is
\begin{equation}
\frac{1}{[(\omega_{\bf k}^t-|{\bf k}|v_{\alpha}\cos\theta)^2 +
v_{\alpha}^2\cos^2\!\theta\mu_D^2]^{2}}\simeq\frac{1}{\mu_D^4}.
\label{eq:5k}
\end{equation}

After integrating over angle $\theta$ in range
$(0,\gamma^{-1})$ we will have from Eq.\,(\ref{eq:5g})
\begin{equation}
\left.\left(-\frac{dE}{dt}\right)^{\!t}_{\! X}
\right|_{\,\theta\leq\gamma^{-1}}\simeq
\frac{1}{3\cdot 2^7}\,{\alpha}_s^3\,C_A\!\!
\sum_{\beta=q\!,\,\bar{q}\!,\,{\rm g}}
\Biggl(\frac{C_2^{(\alpha)}C_2^{(\beta)}}{d_A}\Biggr)n_{\beta}\!
\int\limits_{\omega_{pl}}^{\infty}
\!\frac{d|{\bf k}|}{{\bf k}^2}
\label{eq:5l}
\end{equation}
\[
\times(\gamma^{-2})^3\!
\left\{\!\Biggl(\ln\biggl[\frac{{\bf k}\gamma^{-2}}{m_g^2}\biggr]
 -  C_E\Biggr)
\Bigr(2\ln 2 -5/6\Bigr) + \frac{1}{2}\,
\ln^2\biggl[\frac{{\bf k}^2\gamma^{-2}}{m_g^2}\biggr] +
3\sum\limits_{k=1}^{\infty}\frac{(-1)^k\psi(k)}{(3+k)k}\!\right\},
\]
The remaining integrals over $d|{\bf k}|$ can be estimated by formula
\[
\int\limits_{\omega_{pl}}^{\infty}
\!\frac{d|{\bf k}|}{{\bf k}^2}
\ln^n\left[\frac{{\bf k}^2\gamma^{-2}}{m_g^2}\right]\sim
\frac{1}{gT}\,\ln^n\!\gamma^{-2},\quad n=0,1,2.
\]
Such in the higher-frequency approximation $\omega\gg gT$ we have
besides similar suppression by Lorentz-factor as in expression
(\ref{eq:5j}), an suppression by one power of $g$. This is a
consequence of the fact that if one of lines incoming (or
outgoing) in HTL vertex $\delta\Gamma_{3g}$ is hard, in our case
this is external leg of radiation gluon, then this effective
vertex will be usual perturbative correction to bare three-gluon
vertex. This conclusion is certainly correct also for energy loss
connected with large angle bremsstrahlung.

\section{\bf Off-diagonal contribution to radiation energy loss.
Connection with double Born scattering}
\setcounter{equation}{0}

The previous Sections were concerned with analysis of radiation
intensity of soft-gluon brems\-st\-rahlung generated the lowest-order
process induced by effective current (\ref{eq:2a}). It is associated
with `diagonal' contribution
$\langle\tilde{J}^{\ast(1)a}_{Q\mu}(k,{\bf b})\!\,^{\ast}{\cal
D}^{\mu\nu}_C(k) \tilde{J}^{(1)a}_{Q\nu}(k,{\bf b})\rangle$ to
radiation field energy $W({\bf b})$ (Eq.\,(\ref{eq:3q})), where
\[
\tilde{J}^{(1)a}_{Q\mu}(k,{\bf b})\equiv
K_{\mu}^{abc}({\bf v}_{\alpha},{\bf v}_{\beta};
{\bf b}|\,k)Q_{0\alpha}^bQ_{0\beta}^c,
\]
and function $K_{\mu}^{abc}$ is given by Eqs.\,(\ref{eq:2a}), (\ref{eq:2s}).
In this Section
we briefly analyze a role of simplest `off-diagonal' terms\footnote{The
`diagonal' and `off-diagonal' terms are defined with respect to those of 
product of two effective currents expansion of (\ref{eq:2d}) type.}
\begin{equation}
\langle\tilde{J}^{\ast(0)a}_{Q\mu}(k,{\bf b})\!\,^{\ast}{\cal D}^{\mu\nu}_C(k)
\tilde{J}^{(2)a}_{Q\nu}(k,{\bf b})\rangle +
\langle\tilde{J}^{\ast(2)a}_{Q\mu}(k,{\bf b})\!\,^{\ast}{\cal D}^{\mu\nu}_C(k)
\tilde{J}^{(0)a}_{Q\nu}(k,{\bf b})\rangle\,,
\label{eq:6q}
\end{equation}
where
\begin{equation}
\tilde{J}^{(0)a}_{Q\mu}(k,{\bf b})=
\frac{g}{(2\pi)^3}\,Q_{0\alpha}^av_{\alpha\mu}
\delta(v_{\alpha}\cdot k)+
\frac{g}{(2\pi)^3}\,Q_{0\beta}^av_{\beta\mu}
\delta(v_{\beta}\cdot k)
{\rm e}^{i{\bf k}\cdot{\bf b}}
\label{eq:6w}
\end{equation}
is initial ``bare'' color effective current, and
$\tilde{J}^{(2)a}_{Q\mu}(k,{\bf b})$ is effective current of the next
higher-order in coupling constant by comparison with 
$\tilde{J}^{(1)a}_{Q\mu}(k,{\bf b})$ in the considered problem of 
bremsstrahlung.

Let us clear up first of all what effective currents of the second order
(and the processes associated with them) can give nontrivial
off-diagonal contribution to energy of field radiation (\ref{eq:3q}).
We consider, for example, a term in the expansion
(\ref{eq:2d}) linear in a free field $A^{(0)}$
\begin{equation}
\tilde{J}^{(2)a}_{\mu}[Q_{0\alpha},Q_{0\beta},A^{(0)}](k)=
\int\!\! K_{\mu\mu_1}^{aa_1}
({\bf v}_{\alpha},{\bf v}_{\beta};
{\bf x}_{0\alpha},{\bf x}_{0\beta},Q_{0\alpha},Q_{0\beta}
\vert\,k,-k_1)A^{(0)a_1\mu_1}(k_1)dk_1.
\label{eq:6e}
\end{equation}
As will be shown in next Section, this effective current generates
bremsstrahlung of two gluons in scattering of particle $\alpha$ off
thermal parton $\beta$. In the leading order in coupling in the expansion
of coefficient function $K_{\mu\mu_1}^{a a_1}$ in initial values of
color charges we keep term linear in $Q_{0\alpha}$ and $Q_{0\beta}$
only, (see next Section, Eq.\,(\ref{eq:7w})). Substituting (\ref{eq:6e}) 
into (\ref{eq:6q})
and then into (\ref{eq:3q}), taking into account (\ref{eq:7w}), after
averaging over initial values of color charges, we obtain that by
virtue of equality $\int dQ_{0\alpha,\,\beta}Q_{0\alpha,\,\beta}=0$
the off-diagonal contribution to the radiation field vanishes.

Furthermore  one can consider an effective current generating
bremsstrahlung process of soft gluon in scattering  of projectile
parton $\alpha$ off two thermal partons $\beta_1$ and $\beta_2$. This
process will be considered in Section 9. The effective current here,
have (in leading order) a color structure given in (\ref{eq:9e}). If
one substitutes this effective current  instead of
$\tilde{J}_{Q\mu}^{(2)}$ with initial one (\ref{eq:6w}) (the last one
should be supplemented by initial current of second thermal parton)
into (\ref{eq:6q}) and then into (\ref{eq:3q}), at that time after 
averaging over
color charges we obtain that this off-diagonal contribution will be
also vanishing.

The only non-trivial `off-diagonal' contribution to radiation field
energy arises from the terms of higher-order in color charges
$Q_{0\alpha}$ and $Q_{0\beta}$ in the expansion of color effective
current $K_{\mu}^a({\bf v}_{\alpha},{\bf v}_{\beta},...,
Q_{0\alpha},Q_{0\beta}\vert\,k)$. In this case it is  second and
third terms on the right-hand side of Eq.\,(\ref{eq:2f}). These terms
define ``classical'' soft one-loop corrections to bremsstrahlung
depicted in Fig.\,\ref{fig1}. We already faced with corrections of
this type in research of the scattering processes of Compton type
studied in our previous Paper II. Let us substitute the corrections
(non-linear in color charges) and
initial current (\ref{eq:6w}) into (\ref{eq:6q}) and then into
(\ref{eq:3q}). Performing average over color charges we lead to the
expression for off-diagonal contribution to energy of radiation field
\[
\Bigl(W({\bf b})\bigr)_{\rm off-diag.}^t=
2\pi g\,\Biggl(\frac{C_2^{(\alpha)}C_2^{(\beta)}}{d_A}\Biggr)
\!\sum_{\zeta=1,2}\!\int\!d{\bf k}d\omega\,
({\bf e}(\hat{\bf k},\zeta)\cdot {\bf v}_{\alpha})\,
\omega\,{\rm Im}(^{\ast}{\!\Delta}^t(k))
\]
\begin{equation}
\times\,
{\rm Re}\Bigl[\,
K^{aaccj}({\bf v}_{\alpha},{\bf v}_{\beta};{\bf b}|\,k)
{\rm e}^{j}(\hat{\bf k},\zeta)
\Bigr]\delta(v_{\alpha}\cdot k)
\label{eq:6r}
\end{equation}
\[
+\,
2\pi g\,\Biggl(\frac{C_2^{(\alpha)}C_2^{(\beta)}}{d_A}\Biggr)
\!\sum_{\zeta=1,2}\!\int\!d{\bf k}d\omega\,
({\bf e}(\hat{\bf k},\zeta)\cdot {\bf v}_{\beta})\,
\omega\,{\rm Im}(^{\ast}{\!\Delta}^t(k))
\]
\[
\times\,
{\rm Re}\Bigl[\,{\rm e}^{-i{\bf k}\cdot{\bf b}}
K^{abbaj}({\bf v}_{\alpha},{\bf v}_{\beta};{\bf b}|\,k)
{\rm e}^{j}(\hat{\bf k},\zeta)
\Bigr]\delta(v_{\beta}\cdot k).
\]
Here, we keep a contribution of transverse part of propagator
$^{\ast}{\cal D}_C^{\mu\mu^{\prime}}(k)$ only and take into account
expansion (\ref{eq:3t}). If the function $K^{abbaj}({\bf
v}_{\alpha},{\bf v}_{\beta},{\bf b}\vert\, k){\rm e}^j({\bf k},\xi)$
is not singular for ${\bf v}_{\beta} = 0$, then within potential
model the second term on the right-hand side of Eq.\,(\ref{eq:6r})
vanishes. Leaving the calculation details of higher coefficient
functions $K_{\mu}^{abc_1c_2}$ and $K_{\mu}^{ab_1b_2c}$, which are
similar ones considered in Paper I and Paper II, we give at once
an explicit form of the first of them
\[
K^{abc_1c_2}_{\mu}({\bf v}_{\alpha},{\bf v}_{\beta},{\bf b}\vert\,k)
=\frac{g^5}{(2\pi)^9}\,\{T^c,T^{c_1}\}^{ab}\!
\int\!\biggl\{
-\frac{v_{\beta\mu}v_{\beta\nu}}{v_{\beta}\cdot(k - q_1)}
\,^{\ast}{\cal D}_C^{\nu\nu^{\prime}}\!(k-q_1)
{\cal K}_{\nu^{\prime}}({\bf v}_{\alpha},{\bf v}_{\beta}|\,k-q_1,q_2)
\]
\[
+\,\frac{v_{\alpha\mu}}{(v_{\alpha}\cdot(q_1+q_2))(v_{\alpha}\cdot q_2)}
\,(v_{\alpha\nu}\!\,^{\ast}{\cal D}_C^{\nu\nu^{\prime}}\!(q_1)
v_{\beta\nu^{\prime}})
\,(v_{\alpha\lambda}\!
\,^{\ast}{\cal D}_C^{\lambda\lambda^{\prime}}\!(q_2)
v_{\beta\lambda^{\prime}})
\]
\begin{equation}
-\,\frac{v_{\beta\mu}}{(v_{\beta}\cdot(k-q_2))(v_{\beta}\cdot(k-q_1-q_2))}
\,(v_{\beta\nu}\!\,^{\ast}{\cal D}_C^{\nu\nu^{\prime}}\!(q_1)
v_{\beta\nu^{\prime}})
\,(v_{\beta\lambda}\!
\,^{\ast}{\cal D}_C^{\lambda\lambda^{\prime}}\!(k-q_1-q_2)
v_{\alpha\lambda^{\prime}})
\label{eq:6t}
\end{equation}
\[
+\,^{\ast}\Gamma_{\mu\nu\lambda}(k,-q_1,-k+q_1)
\,^{\ast}{\cal D}_C^{\nu\nu^{\prime}}\!(q_1)v_{\beta\nu^{\prime}}
\,^{\ast}{\cal D}_C^{\lambda\lambda^{\prime}}\!(k-q_1)
{\cal K}_{\lambda^{\prime}}({\bf v}_{\alpha},{\bf v}_{\beta}|\,k-q_1,q_2)
\]
\[
-\,\frac{1}{2}\,
\,^{\ast}\Gamma_{\mu\nu\lambda\sigma}(k,-q_2,-k+q_1+q_2,-q_1)
\,^{\ast}{\cal D}_C^{\nu\nu^{\prime}}\!(q_2)v_{\beta\nu^{\prime}}
\,^{\ast}{\cal D}_C^{\lambda\lambda^{\prime}}\!(k-q_1-q_2)
v_{\alpha\lambda^{\prime}}
\,^{\ast}{\cal D}_C^{\sigma\sigma^{\prime}}\!(q_1)v_{\beta\sigma^{\prime}}
\biggr\}
\]
\[
\times\,{\rm e}^{-i({\bf q}_1+{\bf q}_2)\cdot{\bf b}}
\delta(v_{\beta}\cdot q_1)\delta(v_{\beta}\cdot q_2)
\delta(v_{\alpha}\cdot (k-q_1-q_2))dq_1dq_2,
\]
where function ${\cal K}^{\mu}({\bf v}_{\alpha},{\bf
v}_{\beta}|\,k,q)$ is defined by Eq.\,(\ref{eq:3u}). Note that
integrand is not symmetric with respect to replacement
$q_1\rightleftharpoons q_2$. The diagrammatic interpretation for the
first term on the right-hand side of Eq.\,(\ref{eq:6t}) is depicted
in Fig.\,\ref{fig4}.
\begin{figure}[hbtp]
\begin{center}
\includegraphics[width=0.95\textwidth]{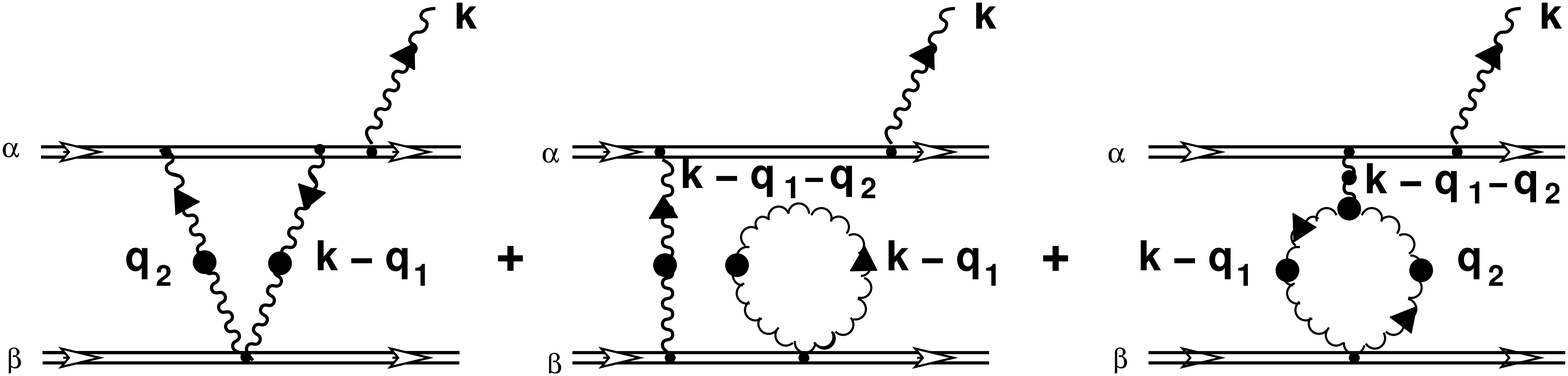}
\end{center}
\caption{\small Some soft one-loop corrections to bremsstrahlung
depicted in Fig.\,\ref{fig1}.}
\label{fig4}
\end{figure}

We consider further approximation of static color centers, i.e. we set
${\bf v}_{\beta}=0$. We substitute an expression (\ref{eq:6r}) with partial
coefficient function (\ref{eq:6t}) into formula for radiation intensity
(\ref{eq:3w}). Delta-functions in integrands (\ref{eq:6r}) and (\ref{eq:6t})
in static limit results in integration measure of the form
\begin{equation}
\delta(v_{\alpha}\cdot k)[\delta(q_{10})\delta(q_{20})dq_{10}dq_{20}]
\delta({\bf v}_{\alpha}\cdot({\bf q}_1+{\bf q}_2))d{\bf q}_1d{\bf q}_2\,
d{\bf k}d\omega.
\label{eq:6y}
\end{equation}
Integration over impact parameter ${\bf b}$ leads to additional delta-function
in integrand
\[
\int\!\!d{\bf b}\,{\rm e}^{-i({\bf q}_1+{\bf q}_2)\cdot{\bf b}}
=(2\pi)^2\delta^{(2)}(({\bf q}_1+{\bf q}_2)_{\perp}).
\]
This expression together with (\ref{eq:6y}) enables us to integrate over
$d{\bf q}_2$.
Performing replacement of variable ${\bf q}_1\rightarrow {\bf q}$, after some
algebraic transformations and regrouping the terms, we result in final
expression for off-diagonal contribution to radiation energy loss of high-energy
color particle $\alpha$ within static approximation
\[
\left(\!-\frac{dE}{dx}\right)_{\!{\rm off-diag.}}^{\!t} = \Lambda_1 + \Lambda_2.
\]
Here, on the right-hand side the function $\Lambda_1$ is
\begin{equation}
\Lambda_1 =
\biggl(\frac{{\alpha}_s}{\pi}\biggr)^{\!3}
\!C_A\sum\limits_{\beta=q\!,\,\bar{q}\!,\,{\rm g}}\, \Biggl(\frac{C_2^{(\alpha)}
C_2^{(\beta)}}{d_A}\Biggr)n_{\beta}
\!\sum_{\zeta=1,2}\!\int\!d{\bf k}d\omega\,\omega\,
{\rm Im}(^{\ast}{\!\Delta}^t(k))\!
\int\!d{\bf q}\,\frac{1}{({\bf q}^2+\mu^2)^2}
\label{eq:6u}
\end{equation}
\[
\times\,
\Biggl[\,\frac{({\bf e}\cdot{\bf v}_{\alpha})^2}
{({\bf v}_{\alpha}\cdot{\bf q})^2} - 2\,
\frac{({\bf e}\cdot{\bf v}_{\alpha})}
{({\bf v}_{\alpha}\cdot{\bf q})}\,
{\rm Re}\,\Bigl\{
{\rm e}^i(\hat{\bf k},\zeta)\,^{\ast}\Gamma^{i0\mu}(k,-q,-k+q)
\!\,^{\ast}{\cal D}_{C\mu\mu^{\prime}}(k-q)v_{\alpha}^{\mu^{\prime}}
\Bigr\}\!\Biggr]
\delta(v_{\alpha}\cdot k),
\]
and the function $\Lambda_2$ has a form
\begin{equation}
\Lambda_2 =-
\biggl(\frac{{\alpha}_s}{\pi}\biggr)^{\!3}
\!C_A\!\sum\limits_{\beta=q,\,\bar{q},\,{\rm g}}\, 
\!\Biggl(\frac{C_2^{(\alpha)}
C_2^{(\beta)}}{d_A}\Biggr)n_{\beta}
\!\!\sum_{\zeta,\,\zeta^{\prime}=1,2}\!\int\!d{\bf k}d\omega\,\omega\,
{\rm Im}(^{\ast}{\!\Delta}^t(k))
\Bigl\{\!({\bf e}(\hat{\bf k},\zeta)\cdot {\bf v}_{\alpha})
({\bf e}(\hat{\bf k},{\zeta}^{\prime})\cdot {\bf v}_{\alpha})\!\Bigr\}
\label{eq:6i}
\end{equation}
\[
\times
\int\!d{\bf q}\,\frac{1}{({\bf q}^2+\mu^2)^2}\,
{\rm Re}\,\Bigl\{\,^{\ast}{\!\Delta}^t(k)
\Bigl({\rm e}^i(\hat{\bf k},\zeta)
\,^{\ast}\tilde{\Gamma}^{i0j0}(k,q,-k,-q)
{\rm e}^j(\hat{\bf k},\zeta^{\prime})\Bigr)\!\Bigr\}
\delta(v_{\alpha}\cdot k).
\]
In the last expression $^{\ast}\tilde{\Gamma}^{i0j0}$ is an effective
four-gluon vertex entered in Paper I (Eq.(I.5.6)). The diagrammatic
interpretation of terms in $\Lambda_1$ and $\Lambda_2$ that easily
follows from initial diagrams of Fig.\,\ref{fig4} type, is presented
in Fig.\,\ref{fig5}.
\begin{figure}[hbtp]
\begin{center}
\includegraphics[width=0.95\textwidth]{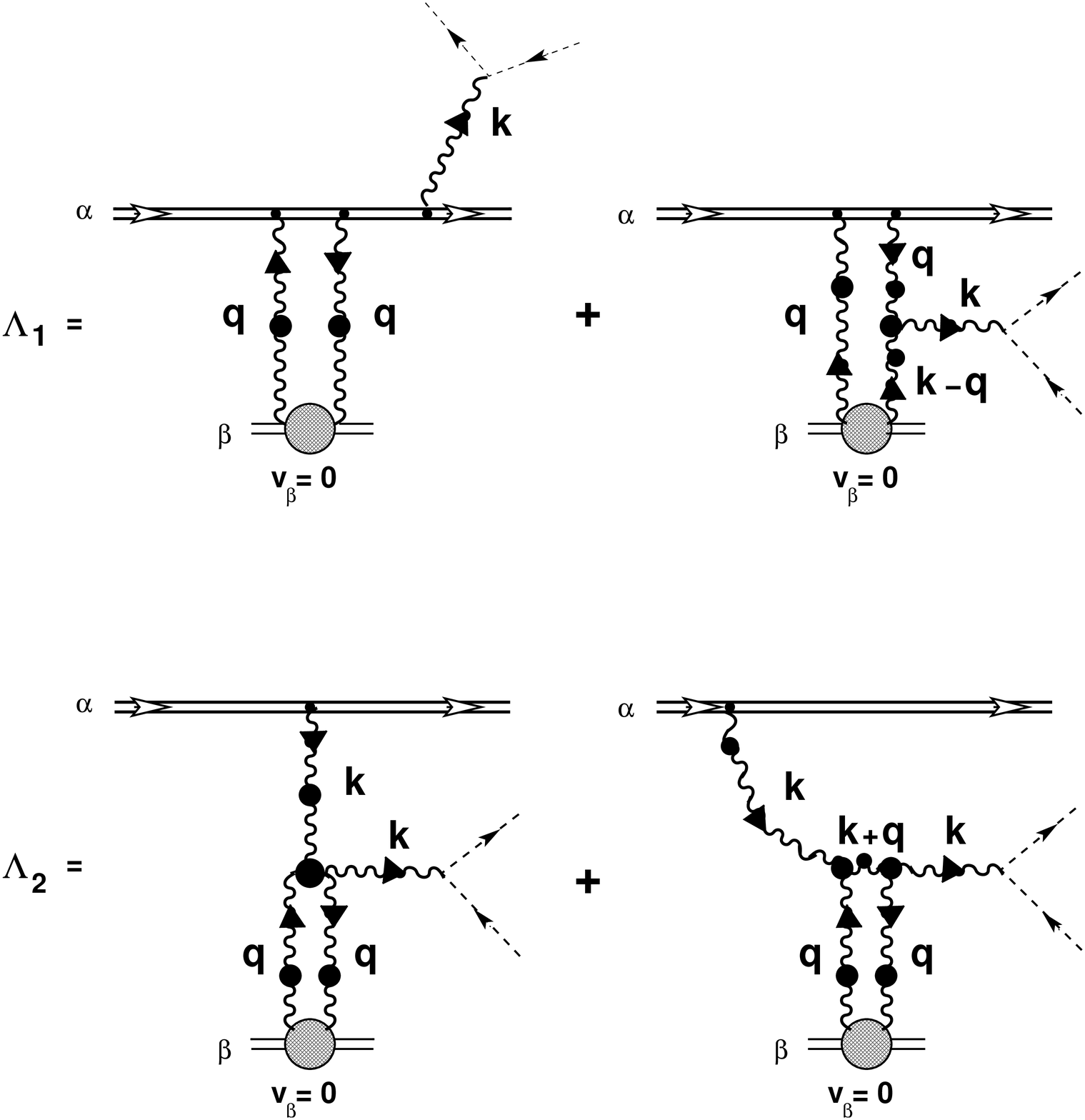}
\end{center}
\caption{\small The diagrammatic interpretation of terms defining
`off-diagonal' contribution to radiation energy loss. The dotted
lines denote thermal partons absorbing virtual bremsstrahlung
gluons. Here, ${\bf q}$ is three-dimensional vector.}
\label{fig5}
\end{figure}
Remind that two diagrams corresponding to radiation of soft gluon prior
to one-gluon exchange and after it, are determined by the
first term on the right-hand side of Eq.\,(\ref{eq:6u}).
We depicted the second diagram in Fig.\,\ref{fig5} only. 
Note also that functions $\Lambda_1$
and $\Lambda_2$ are not vanishing for plasma excitations lying
off mass-shell due to delta-function in integrands.

From the form of graphs in Fig.\,\ref{fig5} it is evident that they represent
the {\it contact double Born graphs}, first considered by Zakharov
\cite{zakharov1}
to ensure unitarity as was already mentioned in Introduction. However in our
case propagators and vertices are effective (i.e. we take
into account hard thermal loops effects) and besides additional
contribution (the first diagram for $\Lambda_2$ in Fig.\,\ref{fig5})
connected with existence of four-gluon HTL amplitude
$^{\ast}\Gamma_{4g} = \Gamma_{4g} + \delta\Gamma_{4g}$, appears.

Usually the contributions to radiation energy loss containing bare
four-gluon vertex $\Gamma_{4g}$, are omitted for kinematical reasons
\cite{gyulassy1,wang1} (the absence of momentum dependence). However
HTL-correction $\delta \Gamma_{4g}$ have highly nontrivial momentum
dependence and therefore in advance, it is not evident that this term can give
negligible small contribution, at least to region of soft transfer and
soft gluon bremsstrahlung.

Let us analyze the role of contribution $\Lambda_1$ to the
theory under consideration of brems\-s\-trahlung. For this purpose we compare it
with basic diagonal contribution (\ref{eq:3o}) (more exactly with the
first term on the right-hand side of (\ref{eq:3o})). Setting
${\bf v}_{\beta} = 0$ we rewrite this diagonal contribution once more,
considered module squared $\vert{\rm e}^i{\cal K}^i\vert^{\,2}$
\[
\left(\!-\frac{dE}{dx}\right)_{\!{\rm diag.}}^{\!t} =
-\biggl(\frac{{\alpha}_s}{\pi}\biggr)^{\!3}
\!C_A\sum\limits_{\beta=q,\,\bar{q},\,{\rm g}}\, 
\Biggl(\frac{C_2^{(\alpha)}
C_2^{(\beta)}}{d_A}\Biggr)n_{\beta}
\!\sum_{\zeta=1,2}\!\int\!d{\bf k}d\omega\,\omega\,
{\rm Im}(^{\ast}{\!\Delta}^t(k))\!
\int\!d{\bf q}\,\frac{1}{({\bf q}^2+\mu^2)^2}
\]
\begin{equation}
\times\,
\Biggl[\,\frac{({\bf e}\cdot{\bf v}_{\alpha})^2}
{({\bf v}_{\alpha}\cdot{\bf q})^2} - 2\,
\frac{({\bf e}\cdot{\bf v}_{\alpha})}
{({\bf v}_{\alpha}\cdot{\bf q})}\,
{\rm Re}\,\Bigl\{
{\rm e}^i(\hat{\bf k},\zeta)\,^{\ast}\Gamma^{i0\mu}(k,-q,-k+q)
\!\,^{\ast}{\cal D}_{C\mu\mu^{\prime}}(k-q)v_{\alpha}^{\mu^{\prime}}
\Bigr\}
\label{eq:6o}
\end{equation}
\[
+\,\Bigl|\,{\rm e}^i(\hat{\bf k},\zeta)\,^{\ast}\Gamma^{i0\mu}(k,-q,-k+q)
\!\,^{\ast}{\cal D}_{C\mu\mu^{\prime}}(k-q)v_{\alpha}^{\mu^{\prime}}
\Bigr|^{\,2}\Biggr]
\delta(v_{\alpha}\cdot k + {\bf v}_{\alpha}\cdot{\bf q}).
\]
If one defines the last expression on mass-shell of plasma oscillations using
formula (\ref{eq:3p}), then factors
\[
\frac{1}{({\bf v}_{\alpha}\cdot{\bf q})^2}\equiv
\frac{1}{(v_{\alpha}\cdot k)^2},\quad
\frac{1}{({\bf v}_{\alpha}\cdot{\bf q})}\equiv
-\,\frac{1}{(v_{\alpha}\cdot k)}
\]
occurring in the first two terms in square brackets of integrand
will not to be singular due to the fact that linear Landau damping
process is absent in QGP. However for off mass-shell excitations of
medium, when frequency and momentum of plasma excitations approach to
``Cherenkov cone''
\begin{equation}
(v_{\alpha}\cdot k) \rightarrow 0,
\label{eq:6p}
\end{equation}
these factors become singular, that results in divergence of the
integral on the right-hand side of Eq.\,(\ref{eq:6o}). From
comparison of the expressions (\ref{eq:6o}) and (\ref{eq:6u}) we see
that these singularities  are exactly compensated by appropriate
terms on the right-hand side of Eq.\,(\ref{eq:6u}). Thus, the only
complete sum of all contributions (`diagonal' and `off-diagonal',
on-shell and off-shell) to radiation energy loss of energetic parton
$\alpha$ is a finite value.

The meaning of the second contribution $\Lambda_2$ (Eq.\,(\ref{eq:6i}))
is less evident. We can only say that following reasoning in
Section 9 of Paper II, this contribution to energy loss takes into
account such subtle collective effect as change of chromodielectric
properties of hot QCD medium induced by soft gluon self-interaction.
This effect partially can be taken into account by replacement of the
HTL-resummed scalar propagator $^{\ast}{\!\Delta}^t(k)$ by effective one
$^{\ast}\tilde{\Delta}^t(k)$ taking into account nonlinear effects
of soft gluon self-interaction in lowest order (elastic rescattering
of two soft transverse gluons $gg \leftrightarrow gg$)
\[
^{\ast}\tilde{\Delta}^{\!-\!1\,t}(k)\simeq \,^{\ast}{\!\Delta}^{\!-\!1\,t}(k) -
\Pi^{(1)\,t}(k).
\]
Here, $\Pi^{(1)\,t}(k)$ represents the correction to transverse part of
gluon self-energy in HTL-approximation, taking into account a change of
dielectric properties of medium by the action of the processes of
nonlinear interaction (transverse) plasma excitations among
themselves. Its explicit form is similar to (II.9.5) for longitudinal
part of correction.

\section{Bremsstrahlung of two soft gluons}
\setcounter{equation}{0}

Hereafter we are concerned with higher-order processes
produced soft gluon radiation. Here, we consider scattering process of
color particle $\alpha$ off hard thermal parton  $\beta$, when
brems\-st\-rahlung of two gluons, arises. Here, we are mainly interested in
question for what typical amplitudes the soft gluon field (or, in
other words, the density of soft gluon radiation) in medium, the
process of brems\-st\-rahlung of higher order will give a contribution to
gluon radiation intensity of the same order as the process of
bremsstrahlung discussed in previous Sections. As was mentioned in
Section 6, this process is defined by the effective current
\begin{equation}
\tilde{J}^{(2)a}_{\mu}[Q_{0\alpha},Q_{0\beta},A^{(0)}](k)=
\int\!\! K_{\mu\mu_1}^{aa_1}
({\bf v}_{\alpha},{\bf v}_{\beta};{\bf b},Q_{0\alpha},Q_{0\beta}
\vert\,k,-k_1)A^{(0)a_1\mu_1}(k_1)dk_1.
\label{eq:7q}
\end{equation}
We take the coefficient function in integrand in the leading order in
coupling constant, i.e.
\begin{equation}
K_{\mu\mu_1}^{aa_1}
({\bf v}_{\alpha},{\bf v}_{\beta};{\bf b},Q_{0\alpha},Q_{0\beta}
\vert\,k,-k_1)\cong
K_{\mu\mu_1}^{aa_1bc}
({\bf v}_{\alpha},{\bf v}_{\beta};{\bf b}\vert\,k,-k_1)
Q_{0\alpha}^bQ_{0\beta}^c.
\label{eq:7w}
\end{equation}
For the deriving of function $K_{\mu\mu_1}^{aa_1bc} ({\bf v}_{\alpha},{\bf
v}_{\beta};{\bf b}\vert\,k,-k_1)$ we use simple and effective
procedure suggested in Paper I and Paper II and extended for this
case. The calculations shown that this function has a
following color structure
\[
K_{\mu\mu_1}^{aa_1bc}
({\bf v}_{\alpha},{\bf v}_{\beta};{\bf b}\vert\,k,\!-k_1)\!=\!
(T^aT^{a_1})^{bc}\!
K_{\mu\mu_1}({\bf v}_{\alpha},{\bf v}_{\beta};{\bf b}\vert\,k,\!-k_1)+
(T^{a_1}T^{a})^{bc}\!
K_{\mu_1\mu}({\bf v}_{\alpha},{\bf v}_{\beta};{\bf b}\vert\,-k_1,k).
\]
The right-hand side of this expression is automatically symmetric with respect
to permutation of external soft gluon legs: $k\rightleftharpoons -k_1,\;
\mu\rightleftharpoons \mu_1,\; a\rightleftharpoons a_1$. The requirement of
symmetry over permutation of hard lines $(\alpha\rightleftharpoons \beta,\;
c\rightleftharpoons b)$ leads to additional condition for partial
coefficient function $K_{\mu\mu_1}$:
\[
K_{\mu\mu_1}({\bf v}_{\alpha},{\bf v}_{\beta};{\bf b}\vert\,k,-k_1)=
K_{\mu_1\mu}({\bf v}_{\beta},{\bf v}_{\alpha};-{\bf b}\vert\,-k_1,k).
\]
The explicit form of this function, and also graphic interpretation of
different terms is given in Appendix A.

Let us substitute an expression for an effective current (\ref{eq:7q}),
(\ref{eq:7w}) into formula (\ref{eq:3q}) defining total energy of radiation
field in single scattering. Performing average over color charges, we
obtain
\begin{equation}
W_{2{\rm g}}({\bf b})=\biggl(\frac{{\alpha}_s}{\pi}\biggr)^{\!4}
\Biggl(\frac{C_2^{(\alpha)}C_2^{(\beta)}}{d_A}\Biggr)
\!\int\!d{\bf k}d\omega\,\omega\,
{\rm Im}\Bigl(K_{\mu\mu_1}^{\ast aa_1bc}
({\bf v}_{\alpha},{\bf v}_{\beta};{\bf b}\vert\,k,\!-k_1)
\!\,^{\ast}{\cal D}^{\mu\mu^{\prime}}_C(k)
\label{eq:7e}
\end{equation}
\[
\times
K_{\mu^{\prime}\mu_1^{\prime}}^{aa_1^{\prime}b^{\prime}c^{\prime}}
({\bf v}_{\alpha},{\bf v}_{\beta};{\bf b}\vert\,k,\!-k_1)\Bigr)
\langle A^{\ast(0)\,a_1\mu_1}(k_1)
A^{(0)\,a_1^{\prime}\mu_1^{\prime}}(k_1^{\prime})\rangle
\,dk_1dk_1^{\prime}.
\]
For the conditions of stationary and homogeneity of hot QCD plasma we
have for correlation function of the soft boson excitations
\[
\langle A^{\ast(0)\,a_1\mu_1}(k_1)
A^{(0)\,a_1^{\prime}\mu_1^{\prime}}(k_1^{\prime})\rangle
=\delta^{a_1a_1^{\prime}}I^{\mu_1\mu_1^{\prime}}(k_1)\delta(k_1-k_1^{\prime}).
\]
One defines the spectral density $I^{\mu\nu}(k)$ in the form of an expansion
\[
I^{\mu\nu}(k)= P^{\mu\nu}(k)I_{k}^t + \tilde{Q}^{\mu\nu}I_{k}^l,\
\]
where in turn the scalar functions $I_k^t$ and $I_k^l$ are taken in the form
of the {\it quasiparticle approximation}
\begin{equation}
I_{k}^{t,\,l} = I_{\bf k}^{t,\,l} \delta(\omega - \omega_{\bf k}^{t,\,l})
+ I_{-{\bf k}}^{t,\,l}\delta(\omega + \omega_{\bf k}^{t,\,l})
\label{eq:7r}
\end{equation}
\[
=-\,\frac{1}{(2\pi)^3}
\left(\frac{{\rm Z}_{t,\,l}({\bf k})}{2\omega_{\bf k}^{t,\,l}}\right)\!
\Bigl\{N_{\bf k}^{t,\,l} \delta(\omega - \omega_{\bf k}^{t,\,l})
+ N_{-{\bf k}}^{t,\,l}\delta(\omega + \omega_{\bf p}^{t,\,l})\!\Bigr\}.
\]
In the last line we pass from functions $I_{\bf k}^{t,\,l}$ to
the number densities of soft plasma oscillations $N_{\bf k}^{t,\,l}$.
In Eq.\,(\ref{eq:7r}) the term containing delta-function
$\delta(\omega + \omega_{\bf k}^{t,\,l})$ only will define
contribution to energy of radiation field (\ref{eq:7e}) connected
with bremsstrahlung of two gluons. The
term containing $\delta(\omega-\omega_{\bf k}^{t,\,l})$ gives
contribution associated with process extending nonlinear Landau
damping process (Section 3, Paper II) to the case of two hard color
partons. Therefore hereafter we drop the last contribution and focus
our attention on pure bremsstrahlung. Note that to take into account
effects of weakly inhomogeneous and weakly non-stationary state of
QGP induced by the effects of the nonlinear interaction of waves and
particles, it is sufficient in the first approximation to assume that
functions $N_{\bf k}^{t,\,l}$ are slowly dependent on $x = (t,{\bf
x})$. In this case the functions obey proper kinetic equations, one
of which will be given below.

For simplicity we restrict our consideration to bremsstrahlung of real
transverse gluons. With allowance for above-mentioned in (\ref{eq:7e})
we make following substitution
\[
I^{\mu_1\mu_1^{\prime}}(k_1)\Rightarrow
\frac{1}{(2\pi)^3}
\left(\frac{{\rm Z}_{t}({\bf k}_1)}{2\omega_{{\bf k}_1}^{t}}\right)\!
N_{-{\bf k}_1}^{t}\delta(\omega_1 + \omega_{{\bf k}_1}^{t})
\!\!\sum_{\zeta_1 = 1,2}{\rm e}^{\ast i_1}(\hat{\bf k}_1,\zeta_1)\,
{\rm e}^{i_1^{\prime}}(\hat{\bf k}_1,\zeta_1),
\]
\[
\!\,^{\ast}{\cal D}^{\mu\mu^{\prime}}_C(k)\Rightarrow
\,^{\ast}{\!\Delta}^t(k)
\!\sum_{\zeta = 1,2}{\rm e}^{\ast i}(\hat{\bf k},\zeta)\,
{\rm e}^{i^{\prime}}(\hat{\bf k},\zeta),
\hspace{1.5cm}
\]
\[
{\rm Im}(^{\ast}{\!\Delta}^{\!-1\,t}(k))\Rightarrow
-\pi\,{\rm sign}(\omega)
\left(\frac{{\rm Z}_{t}({\bf k})}{2\omega_{\bf k}^{t}}\right)
\!\delta(\omega-\omega_{\bf k}^{t}).
\]
Here, we have taken into account representation (\ref{eq:3t}). Performing trivial
integration over $dk_1^{\prime}$, $d\omega$ and $d\omega_1$ and replacing
variable ${\bf k}_1\rightarrow -{\bf k}_1$
$(\omega_{{\bf k}_1}^{t}\rightarrow -\omega_{{\bf k}_1}^{t})$ we derive
instead of (\ref{eq:7e})
\[
(W_{2{\rm g}}({\bf b}))^t
=\frac{1}{2(2\pi)^2}\,
\biggl(\frac{{\alpha}_s}{\pi}\biggr)^{\!4}
\Biggl(\frac{C_2^{(\alpha)}C_2^{(\beta)}}{d_A}\Biggr)
\!\sum_{\zeta,\,\zeta_1=1,2}\!\int\!d{\bf k}d{\bf k}_1\,
\omega_{\bf k}^{t}N_{{\bf k}_1}^{t}
\left(\frac{{\rm Z}_{t}({\bf k})}{2\omega_{\bf k}^{t}}\right)\!
\left(\frac{{\rm Z}_{t}({\bf k}_1)}{2\omega_{{\bf k}_1}^{t}}\right)
\]
\begin{equation}
\times
|\,K^{aa_1bc\,ii_1}({\bf v}_{\alpha},{\bf v}_{\beta};{\bf b}\vert\,k,k_1)
{\rm e}^{i}(\hat{\bf k},\zeta)\,
{\rm e}^{i_1}(\hat{\bf k}_1,\zeta_1)|^{\,2}_{\,{\rm on-shell}}.
\label{eq:7t}
\end{equation}
Let us define an expression for gluon radiation intensity. For this
purpose we write partial coefficient function $K_{\mu\mu_1}$ in the
form of integral representation
\[
K_{\mu\mu_1}({\bf v}_{\alpha},\!{\bf v}_{\beta};\!{\bf b}\vert\,k,k_1)
\!=\!\!\int\!\!{\rm e}^{i{\bf q}\cdot{\bf b}}
[{\cal K}_{\mu\mu_1}({\bf v}_{\alpha},\!{\bf v}_{\beta}|\,k,k_1;q)
]_{q_0={\bf v}_{\beta}\cdot{\bf q}}
\delta(\omega^t_{\bf k}+\omega^t_{{\bf k}_1}\!
-{\bf v}_{\beta}\cdot{\bf q}-{\bf v}_{\alpha}\cdot
({\bf k}+{\bf k}_1-{\bf q})\!)d{\bf q}.
\]
The function ${\cal K}_{\mu\mu_1}$ is defined by integrand in
Eq.\,(A.1). Subsequent reasoning are completely similar to those in
Section 3. Performing integration over impact parameter with 
(\ref{eq:3i}) and allowing for the following equalities
\begin{equation}
(T^aT^{a_1})^{\ast bc}(T^aT^{a_1})^{bc}=C_A^2d_A,\quad
(T^aT^{a_1})^{\ast bc}(T^{a_1}T^a)^{bc}=\frac{1}{2}\,C_A^2d_A,
\label{eq:7y}
\end{equation}
we obtain expressions for intensity radiation in this case
\[
\left\langle\frac{dW_{2{\rm g}}({\bf b})}{dt}\right\rangle_{\!{\bf b}}=
\frac{1}{2}\,
\biggl(\frac{{\alpha}_s}{\pi}\biggr)^{\!4}\,C_A^{\,2}\!
\sum_{\beta=q\!,\,\bar{q}\!,\,{\rm g}}
\Biggl(\frac{C_2^{(\alpha)}C_2^{(\beta)}}{d_A}\Biggr)
\!\sum_{\zeta,\,\zeta_1=1,2}\!\int\!d{\bf k}d{\bf k}_1\,
\omega_{\bf k}^{t}N_{{\bf k}_1}^{t}
\left(\frac{{\rm Z}_{t}({\bf k})}{2\omega_{\bf k}^{t}}\right)\!
\left(\frac{{\rm Z}_{t}({\bf k}_1)}{2\omega_{{\bf k}_1}^{t}}\right)
\]
\begin{equation}
\times
\Biggl(\int\!{\bf p}_{\beta}^2f_{|{\bf p}_{\beta}|}
\,\frac{d|{\bf p}_{\beta}|}{2\pi^2}\Biggr)
\int\!\frac{d\Omega_{{\bf v}_{\beta}}}{4\pi}
\!\int\!d{\bf q}\,
\delta(\omega^t_{\bf k}+\omega^t_{{\bf k}_1}
-{\bf v}_{\beta}\cdot{\bf q}-{\bf v}_{\alpha}\cdot
({\bf k}+{\bf k}_1-{\bf q}))
\label{eq:7u}
\end{equation}
\[
\times
\biggl\{
|\,{\cal K}^{ii_1}({\bf v}_{\alpha},{\bf v}_{\beta}|\,k,k_1,q)
{\rm e}^i(\hat{\bf k},\zeta){\rm e}^{i_1}(\hat{\bf k}_1,\zeta_1)|^{\,2}
+|\,{\cal K}^{i_1i}({\bf v}_{\alpha},{\bf v}_{\beta}|\,k_1,k,q)
{\rm e}^{i_1}(\hat{\bf k}_1,\zeta_1){\rm e}^{i}(\hat{\bf k},\zeta)|^{\,2}
\]
\[
+{\rm Re}\Bigl[\Bigl(\,{\cal K}^{ii_1}
({\bf v}_{\alpha},{\bf v}_{\beta}|\,k,k_1,q)
{\rm e}^i(\hat{\bf k},\zeta){\rm e}^{i_1}(\hat{\bf k}_1,\zeta_1)
\Bigr)^{\!{\ast}}\!
\Bigl({\cal K}^{i_1i}({\bf v}_{\alpha},{\bf v}_{\beta}|\,k_1,k,q)
{\rm e}^{i_1}(\hat{\bf k}_1,\zeta_1){\rm e}^{i}(\hat{\bf k},\zeta)\Bigr)
\Bigr]\!\biggr\}_{\!q_0={\bf v}_{\beta}\cdot{\bf q}}.
\]
Here, the integrand can be made completely symmetric over permutation
${\bf k}\rightleftharpoons {\bf k}_1$ if we replace:
$\omega_{\bf k}^{t}N_{{\bf k}_1}^{t}\rightarrow
(\omega_{\bf k}^{t}N_{{\bf k}_1}^{t}+\omega_{{\bf k}_1}^{t}N_{\bf k}^{t})/2$.
Similar to (\ref{eq:3e}) one can define probability of two-gluon
bremsstrahlung
${\it w}_{{\bf p}_{\alpha},\,{\bf p}_{\beta}}({\bf k},{\bf k}_1;{\bf q})$
if radiation intensity is written as follows
\[
\left\langle\frac{dW_{2{\rm g}}({\bf b})}{dt}\right\rangle_{\!{\bf b}}\!=\!
\int\!\omega_{\bf k}^{t}N_{{\bf k}_1}^{t}
{\it w}_{{\bf p}_{\alpha},\,{\bf p}_{\beta}}({\bf k},{\bf k}_1;{\bf q})\,
\frac{f_{{\bf p}_{\beta}}d{\bf p}_{\beta}}{(2\pi)^3}\,
\frac{d{\bf k}d{{\bf k}_1}d{\bf q}}{(2\pi)^9}\,
\delta(\omega^t_{\bf k}+\omega^t_{{\bf k}_1}
-{\bf v}_{\beta}\cdot{\bf q}-{\bf v}_{\alpha}\cdot
({\bf k}+{\bf k}_1-{\bf q})).
\]
Comparing the last expression with (\ref{eq:7u}) it is not difficult to
obtain an explicit form of required probability.

In limit static case of target $({\bf v}_{\beta}=0)$ an expression
(\ref{eq:7u}) defines `net' energy loss of projectile $\alpha$.
Setting ${\bf v}_{\beta}=0$ in Eq.\,(A.1) we obtain considerable
simpler expression for partial coefficient function ${\cal K}^{ii_1}$
entering into integrand (\ref{eq:7u})
\[
{\cal K}^{ii_1}({\bf v}_{\alpha},0|\,k,k_1,q)=
\frac{1}{{\bf q}^2 + \mu_D^2}\,
\Biggl[
\,\frac{v_{\alpha}^i v_{\alpha}^{i_1}}
{(v_{\alpha}\cdot q)(v_{\alpha}\cdot (q-k_1))}
\]
\begin{equation}
+\,\frac{v_{\alpha}^i}{v_{\alpha}\cdot (q-k_1)}\,\,
v_{\alpha}^{\nu}\,^{\ast}{\cal D}_{C\nu\nu^{\prime}}\!(q-k_1)
\,^{\ast}\Gamma^{\nu^{\prime}i_10}(q-k_1,k_1,-q)
\label{eq:7i}
\end{equation}
\[
+\,^{\ast}\Gamma^{i\nu\lambda}(k,k_1-q,-k-k_1+q)
\!\,^{\ast}{\cal D}_{C\nu\nu^{\prime}}\!(q-k_1)
\,^{\ast}\Gamma^{\nu^{\prime}i_10}(q-k_1,k_1,-q)
\!\,^{\ast}{\cal D}_{C\lambda\lambda^{\prime}}\!(k+k_1-q)
v_{\alpha}^{\lambda^{\prime}}
\]
\[
+\,^{\ast}\Gamma^{ii_1\lambda}(k,k_1,-k-k_1)
\!\,^{\ast}{\cal D}_{C\lambda\lambda^{\prime}}\!(k+k_1)
\biggl\{\,\frac{v_{\alpha}^{\lambda^{\prime}}}{v_{\alpha}\cdot q}
+\,^{\ast}\Gamma^{\lambda^{\prime}0\nu}(k+k_1,-q,-k-k_1+q)
\!\,^{\ast}{\cal D}_{C\nu\nu^{\prime}}\!(k+k_1-q)v_{\alpha}^{\nu^{\prime}}
\biggr\}
\]
\[
-\,^{\ast}\Gamma^{i\nu0i_1}(k,-k-k_1+q,-q,k_1)
\!\,^{\ast}{\cal D}_{C\nu\nu^{\prime}}\!(k+k_1-q)v_{\alpha}^{\nu^{\prime}}
\Biggr]_{\!q_0=0}.
\]
Diagrammatic interpretation of different terms is presented in Fig.\,\ref{fig7}.
\begin{figure}[hbtp]
\begin{center}
\includegraphics[width=0.95\textwidth]{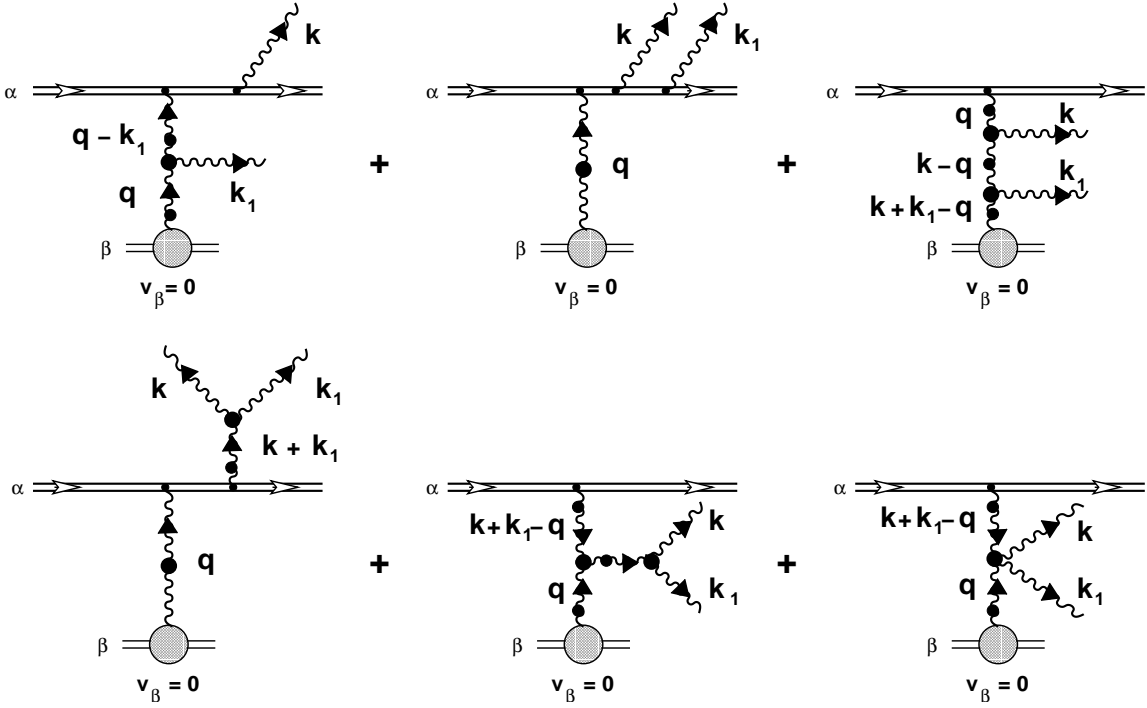}
\end{center}
\caption{\small The processes of bremsstrahlung of two soft gluons in
static potential model.}
\label{fig7}
\end{figure}
As in previous cases, here we have not presented diagrams of the
bremsstrahlung of two gluons a prior to one-gluon exchange. In
semiclassical approximation all these process are automatically taken
into account on the right-hand side of Eq.\,(\ref{eq:7i}).

The expression (\ref{eq:7u}) in region of soft momentum transfer
$(|{\bf q}|\leq gT)$ is distinct in value order from the expression 
(\ref{eq:3o}) associated with bremsstrahlung of one soft gluon by factor
\begin{equation}
\alpha_s\!\int\!d{\bf k}_1
\left(\frac{{\rm Z}_{t}({\bf k}_1)}{2\omega_{{\bf k}_1}^{t}}\right)
\frac{1}{(v_{\alpha}\cdot(k-q))^2}\,N^t_{{\bf k}_1}\sim
\alpha_s(gT)^3\frac{1}{gT}\,\frac{1}{(gT)^2}\,N^t_{{\bf k}_1}\sim
\alpha_s N^t_{{\bf k}_1}.
\label{eq:7o}
\end{equation}
Here, we take into account that for soft gluon $\omega_{\bf k}^t,\,
|{\bf k}|\sim gT$. For low density of soft gluon radiation in medium,
corresponding to the level of thermal fluctuations at soft scale $gT$
(Eq.\,(I.6.5)), when
\begin{equation}
|A_{\mu}(X)|\sim\sqrt{g}T,\quad N_{\bf k}^t\sim\frac{1}{g},
\label{eq:7p}
\end{equation}
from estimation (\ref{eq:7o}) it follows that bremsstrahlung of two
soft gluons is suppressed by more power $g$ in comparison with
bremsstrahlung of one soft gluon.

In opposite case for the field as strong as allowed (Eq.\,(I.6.4))
\begin{equation}
|A_{\mu}(X)|\sim T,\quad N_{\bf k}^t\sim\frac{1}{g^2},
\label{eq:7a}
\end{equation}
we see from (\ref{eq:7o}) that considered process (and also generally
speaking all other higher processes of bremsstrahlung of arbitrary
number of soft gluons) becomes of the same order in coupling
constant. Here, the problem of resummation of all relevant
contributions to soft gluon radiation intensity arises. As mentioned
above in the case of slowly inhomogeneous and nonstationarity QGP,
an explicit form of the number density $N_{\bf k}^t$ is determined by
solution of kinetic equation of Boltzmann type
\begin{equation}
\frac{\partial N_{\bf k}^t}{\partial t} +
{\bf v}_{\bf k}^t\cdot\frac{\partial N_{\bf k}^t}{\partial {\bf x}} =
- N_{\bf k}^t\Gamma_{\rm d}[N_{\bf k}^t ] + ( 1 + N_{\bf k}^t )
\Gamma_{\rm i}[N_{\bf k}^t],
\label{eq:7s}
\end{equation}
where ${\bf v}_{\bf k}^t = \partial\omega_{\bf k}^t/\partial{\bf k}$
is a group velocity of the transverse oscillations.
The functional dependence is denoted by argument of a
function in square brackets.

A generalized decay rate $\Gamma_{\rm d}$ and regenerating rate
$\Gamma_{\rm i}$ are written in the form of
functional expansion in powers of the number density $N_{\bf k}^t$
\begin{equation}
\Gamma_{\rm d} [N_{\bf k}^t] = \sum_{n = 0}^{\infty}
\Gamma_{\rm d}^{(n)} [N_{\bf k}^l] , \; \;
\Gamma_{\rm i} [N_{\bf k}^t] = \sum_{n = 0}^{\infty}
\Gamma_{\rm i}^{(n)} [N_{\bf k}^t].
\label{eq:7d}
\end{equation}
Here,
\[
\Gamma_{\rm d}^{(n)}[N_{\bf k}^t] = 
\sum\limits_{\beta=q\!,\,\bar{q}\!,\,{\rm g}}
\int\!
\frac{d{\bf p}_{\beta}}{(2\pi)^3}\,
f_{{\bf p}_{\beta}-\,{\bf q}}
\!\int\!{\it w}_{{\bf p}_{\alpha},\,{\bf p}_{\beta}}
({\bf k},{\bf k}_1,\ldots,{\bf k}_n;\,{\bf q})
N_{{\bf k}_1}^t \ldots N_{{\bf k}_n}^t
\]
\[
\times\,
\delta\Bigl(E_{{\bf p}_{\alpha}} + E_{{\bf p}_{\beta}}
- E_{{\bf p}_{\alpha}+{\bf q}-{\bf k}_{\rm tot}}
-E_{{\bf p}_{\beta}-{\bf q}} - {\cal E}_{\rm tot}\Bigr)
\,\frac{d{\bf q}}{(2\pi)^3}
\prod\limits_{i=1}^n\frac{d{\bf k}_i}{(2\pi)^3}\,,
\]
\[
\Gamma_{\rm d}^{(n)}[N_{\bf k}^t] = 
\sum\limits_{\beta=q\!,\,\bar{q}\!,\,{\rm g}}
\int\!
\frac{d{\bf p}_{\beta}}{(2\pi)^3}\,
f_{{\bf p}_{\beta}}
\!\int\!{\it w}_{{\bf p}_{\alpha},\,{\bf p}_{\beta}}
({\bf k},{\bf k}_1,\ldots,{\bf k}_n;\,{\bf q})
(1+N_{{\bf k}_1}^t) \ldots (1+N_{{\bf k}_n}^t)
\]
\[
\times\,
\delta\Bigl(E_{{\bf p}_{\alpha}} + E_{{\bf p}_{\beta}}
- E_{{\bf p}_{\alpha}+{\bf q}-{\bf k}_{\rm tot}}
-E_{{\bf p}_{\beta}-{\bf q}} - {\cal E}_{\rm tot}\Bigr)
\,\frac{d{\bf q}}{(2\pi)^3}
\prod\limits_{i=1}^n\frac{d{\bf k}_i}{(2\pi)^3}\,,
\]
where
\[
{\cal E}_{\rm tot}\equiv\omega_{\bf k}^t +\sum\limits_{i=1}^n\omega_{{\bf k}_i}^t,
\quad
{\bf k}_{\rm tot}\equiv{\bf k}+\sum\limits_{i=1}^n{\bf k}_i,
\]
and ${\it w}_{{\bf p}_{\alpha},\,{\bf p}_{\beta}} ({\bf k},{\bf
k}_1,\ldots,{\bf k}_n;\,{\bf q})$ is a probability of bremsstrahlung
of $(n+1)$ soft gluons. The particular expressions for the cases
$n=0$ and $n=1$ are given by Eqs.\,(\ref{eq:3s}) and (\ref{eq:7u}),
correspondingly. Note that, the expressions for generalized decay and
regenerating rates were simplified in the sense that a contributions
of more complicated `mixed' type being combination of the scattering
processes of Compton type and pure bremsstrahlung, are not taken into
account.

In conditions of (\ref{eq:7p}) each subsequent term in the functional
expansions (\ref{eq:7q}) is suppressed by more power $g$, and here we
can only restrict ourselves to the first leading term. By virtue of
the fact that $\Gamma_{\rm d,\,i}^{(0)}$ are not $N_{\bf k}^t$
dependent, the Boltzmann equation (\ref{eq:7s}) will be linear with
respect to the number density. For highly excited system,
Eq.\,(\ref{eq:7a}), all terms in the expansions (\ref{eq:7d}) become
of the same order in magnitude, and kinetic equation (\ref{eq:7s}) is
completely nonlinear one.

\section{Bremsstrahlung of two soft gluons (continuation)}
\setcounter{equation}{0}

Let us analyze the expression for radiation intensity
(\ref{eq:7u}) in more detail. For this purpose we introduce
new partial coefficient functions
\begin{equation}
{\cal K}^{(S)ii_1}({\bf v}_{\alpha},{\bf v}_{\beta}|\,k,k_1,q)
\equiv\frac{1}{2}\,\biggl(
{\cal K}^{ii_1}({\bf v}_{\alpha},{\bf v}_{\beta}|\,k,k_1,q) +
{\cal K}^{i_1i}({\bf v}_{\alpha},{\bf v}_{\beta}|\,k_1,k,q)\biggr),
\label{eq:8q}
\end{equation}
\[
{\cal K}^{(A)ii_1}({\bf v}_{\alpha},{\bf v}_{\beta}|\,k,k_1,q)
\equiv\frac{1}{2}\,\biggl(
{\cal K}^{ii_1}({\bf v}_{\alpha},{\bf v}_{\beta}|\,k,k_1,q) -
{\cal K}^{i_1i}({\bf v}_{\alpha},{\bf v}_{\beta}|\,k_1,k,q)\biggr),
\hspace{0.8cm}
\]
the former of which is symmetric function and the second one is
antisymmetric relative to permutation: $k\rightleftharpoons
k_1,\,i\rightleftharpoons i_1$. In terms of the functions
(\ref{eq:8q}) an expression within curly brackets in integrand of
equation (\ref{eq:7u}) is rewritten in the form (for brevity we
suppress arguments of functions)
\[
|\,{\cal K}^{ii_1}{\rm e}^i{\rm e}^{i_1}|^{\,2}
+|\,{\cal K}^{i_1i}{\rm e}^{i_1}{\rm e}^{i}|^{\,2}
+{\rm Re}\Bigl[\Bigl(\,{\cal K}^{ii_1}{\rm e}^i{\rm e}^{i_1}\Bigr)^{\!{\ast}}\!
\Bigl({\cal K}^{i_1i}{\rm e}^{i_1}{\rm e}^{i}\Bigr)\Bigr]
= 3|\,{\cal K}^{(S)ii_1}{\rm e}^i{\rm e}^{i_1}|^{\,2}
+|\,{\cal K}^{(A)i_1i}{\rm e}^{i_1}{\rm e}^{i}|^{\,2}.
\]
As we see from the right-hand side of this expression an interference term
is vanish when start using new functions, and
thus the probability of bremsstrahlung of two gluons decomposes
into two completely independent parts. This points to the
fact that the process can proceed through two physically
independent channels determined by parity
of final state of two gluon`s system with respect to permutation
of external soft-gluon legs. The functions ${\cal K}^{(S)ii_1}
{\rm e}^i{\rm e}^{i_1}_1$ and ${\cal K}^{(A)ii_1}{\rm e}^i{\rm e}^{i_1}_1$
define a matrix elements of bremsstrahlung
of two gluons being in even state and in odd state\footnote{The
classification of states of two photon`s system in their c.m.s. is
given in \cite{berestetski}. We assume that this classification is
true for system of two gluons also.}, correspondingly.

Note that going to functions (\ref{eq:8q}) having more clear
physical meaning can be already performed at the level of initial
effective current (\ref{eq:7q}) (with approximation (\ref{eq:7w})
for integrand) by rewriting coefficient function
$K_{\mu\mu_1}^{aa_1bc}$ in the form
\[
K_{\mu\mu_1}^{aa_1bc}
({\bf v}_{\alpha},\!{\bf v}_{\beta};\!{\bf b}\vert\,k,\!-k_1)\!=\!
\{T^a\!,T^{a_1}\}^{\!bc}
K^{(S)}_{\mu\mu_1}({\bf v}_{\alpha},\!{\bf v}_{\beta};\!{\bf b}\vert\,k,\!-k_1)+
[T^{a_1}\!,T^{a}]^{bc}
K^{(A)}_{\mu_1\mu}({\bf v}_{\alpha},\!{\bf v}_{\beta};{\bf b}\vert\,-k_1,k),
\]
where
\[
K^{(S\!,\,A)}_{\mu\mu_1}
({\bf v}_{\alpha},{\bf v}_{\beta};{\bf b}\vert\,k,\!-k_1)\equiv
\frac{1}{2}\,\biggl(
K_{\mu\mu_1}({\bf v}_{\alpha},{\bf v}_{\beta};{\bf b}\vert\,k,\!-k_1)\pm
K_{\mu_1\mu}({\bf v}_{\alpha},{\bf v}_{\beta};{\bf b}\vert\,\!-k_1,\!k)
\biggr)
\]
are symmetrized and anti-symmetrized partial coefficient functions
in `impact parameter representation'. A purely ``Abelian'' part of
the bremsstrahlung of two gluons is defined by a term
$\{T^a\!,T^{a_1}\}^{\!bc} K^{(S)}_{\mu\mu_1}
({\bf v}_{\alpha},\!{\bf v}_{\beta};\!{\bf b}\vert\,k,\!-k_1)$.

One would expect that after separation of `elastic' factor from
${\cal K}_{\mu\mu_1}({\bf v}_{\alpha},{\bf
v}_{\beta}|\,k,k_1,q)$ (i.e. the factor defining a scattering
process without radiation) at least symmetric function
${\cal K}^{(S)}_{\mu\mu_1}({\bf v}_{\alpha},{\bf v}_{\beta}|\,k,k_1,q)$
can be presented in factorized form -- in the form of product of
two functions separately depending on momenta $k$ and $k_1$.
Such factorization takes place in the case of QED interaction in
the semiclassical limit \cite{berestetski}. Further for simplicity
we restrict our consideration to static case ${\bf
v}_{\beta}=0\,(q_0=0)$. Making use explicit expression
(\ref{eq:7i}), we write out a matrix element
${\cal K}^{(S)ii_1}{\rm e}^i{\rm e}^{i_1}_1$ separating out `elastic' factor
\[
{\cal K}^{(S)ii_{\!1}}\!({\bf v}_{\alpha},0|\,k,k_1,q)
{\rm e}^i(\hat{\bf k},\zeta){\rm e}^{i_1}(\hat{\bf k}_1,\zeta_1)
=
\frac{1}{2({\bf q}^2+\mu_D^2)}\,
{\cal M}^{(S)ii_{\!1}}({\bf v}_{\alpha},0|\,k,k_1,q)
{\rm e}^i(\hat{\bf k},\zeta){\rm e}^{i_1}(\hat{\bf k}_1,\zeta_1),
\]
where
\begin{equation}
{\cal M}^{(S)ii_1}({\bf v}_{\alpha},0|\,k,k_1,q)=
\,\frac{v_{\alpha}^i v_{\alpha}^{i_1}}
{(v_{\alpha}\cdot k)(v_{\alpha}\cdot k_1)}
\label{eq:8w}
\end{equation}
\[
+\left\{\frac{v_{\alpha}^i}{(v_{\alpha}\cdot k)}\,\,
v_{\alpha}^{\nu}\,^{\ast}{\cal D}_{C\nu\nu^{\prime}}\!(q-k_1)
\,^{\ast}\Gamma^{\nu^{\prime}i_10}(q-k_1,k_1,-q)
+(k\rightleftharpoons k_1,\,i\rightleftharpoons i_1)\right\}
\]
\[
+\,\biggl[\,^{\ast}\Gamma^{i\nu\lambda}(k,k_1-q,-k-k_1+q)
\!\,^{\ast}{\cal D}_{C\nu\nu^{\prime}}\!(q-k_1)
\,^{\ast}\Gamma^{\nu^{\prime}i_10}(q-k_1,k_1,-q)
\!\,^{\ast}{\cal D}_{C\lambda\lambda^{\prime}}\!(k+k_1-q)
v_{\alpha}^{\lambda^{\prime}}
\]
\[
+\,(k\rightleftharpoons k_1,\,i\rightleftharpoons i_1)\biggr]
\]
\[
+\left(\,^{\ast}\Gamma^{ii_1\lambda}(k,k_1,-k-k_1)
+ \,^{\ast}\Gamma^{i_1i\lambda}(k_1,k,-k-k_1)\right)
\!\,^{\ast}{\cal D}_{C\lambda\lambda^{\prime}}\!(k+k_1)
\]
\[
\times\!
\left(\,\frac{v_{\alpha}^{\lambda^{\prime}}}{(v_{\alpha}\cdot q)}
+\,^{\ast}\Gamma^{\lambda^{\prime}0\nu}(k+k_1,-q,-k-k_1+q)
\!\,^{\ast}{\cal D}_{C\nu\nu^{\prime}}\!(k+k_1-q)v_{\alpha}^{\nu^{\prime}}
\right)
\]
\[
-\,\biggl\{\!\,^{\ast}\Gamma^{i\nu0i_1}(k,-k-k_1+q,-q,k_1)
+ \,^{\ast}\Gamma^{i_1\nu0i}(k_1,-k-k_1+q,-q,k)\biggr\}
\!\,^{\ast}{\cal D}_{C\nu\nu^{\prime}}\!(k+k_1-q)v_{\alpha}^{\nu^{\prime}}.
\]
On writing this expression we use energy-momentum conservation law
in the semiclassical limit: $v_{\alpha}\cdot(k+k_1-q)=0$. The
fourth term on the right-hand side of Eq.\,(\ref{eq:8w}) vanishes
owing to property
\[
\,^{\ast}\Gamma^{ii_1\lambda}(k,k_1,-k-k_1)=-
\,^{\ast}\Gamma^{i_1i\lambda}(k_1,k,-k-k_1).
\]
The last term on the right-hand side of Eq.\,(\ref{eq:8w}) contains
contributions associated with effective four-gluon vertex
$\,^{\ast}\Gamma_{4g}=\Gamma_{4g}+\delta\Gamma_{4g}$. If we drop four-gluon
HTL-correction $\delta\Gamma_{4g}$, then this term assumes the following form
in the static limit
\[
-2\delta^{ii_1}\,\frac{(k+k_1-q)^2}{({\bf k}+{\bf k}_1-{\bf q})^2}\,
\,^{\ast}{\!\Delta}^l(k+k_1-q).
\]
Thus a term containing bare four-gluon vertices $\Gamma_{4g}$ will be
proportional to longitudinal part of HTL-resummed gluon propagator.

From Eq.\,(\ref{eq:8w}) we see that even if we drop the contributions
containing longitudinal part of gluon propagators (in particular
contributions contained four-gluon vertex $\Gamma_{4g}$) and neglect by
HTL correction $\delta\Gamma_{4g}$, then we still can't present symmetrized
function ${\cal M}^{(S)ii_1}({\bf v}_{\alpha},0|\,k,k_1,q)$ in the
factorized form
\begin{equation}
{\cal M}^{(S)ii_1}({\bf v}_{\alpha},0|\,k,k_1,q)=
{\cal M}^{i}({\bf v}_{\alpha},0|\,k,q)
{\cal M}^{i_1}({\bf v}_{\alpha},0|\,k_1,q),
\label{eq:8e}
\end{equation}
where
\[
{\cal M}^{i}({\bf v}_{\alpha},0|\,k,q)=
\frac{v_{\alpha}^i}{(v_{\alpha}\cdot k)} +
v_{\alpha}^{\nu}\,^{\ast}{\cal D}_{C\nu\nu^{\prime}}\!(q-k)
\,^{\ast}\Gamma^{\nu^{\prime}i0}(q-k,k,-q)
\]
is function on which the probability of bremsstrahlung of soft transverse
gluon (Eqs.\,(\ref{eq:4q}), (\ref{eq:4e})) is determined. Let us suppose that
an equality (\ref{eq:8e}) is true for the time being and even moreover
a similar factorization is true for radiation process of arbitrary number
of soft gluons, i.e.
\begin{equation}
{\cal M}^{(S)ii_1\dots i_n}({\bf v}_{\alpha},0|\,k,k_1,\dots,q)\!=\!
{\cal M}^{i}({\bf v}_{\alpha},0|\,k,q)
{\cal M}^{i_1}({\bf v}_{\alpha},0|\,k_1,q)
\ldots
{\cal M}^{i_n}({\bf v}_{\alpha},0|\,k_n,q),
\label{eq:8r}
\end{equation}
\[
n=1,2,\dots.
\]
In this case it is easy to show that an expression for total energy loss
of high-energy parton induced by bremsstrahlung of arbitrary number of soft
gluons can be presented in highly compact form
\begin{equation}
\left\langle\frac{dW_{\rm tot}^{(S)}({\bf b})}{dt}\right\rangle_{\!\!{\bf b}}=
\sum\limits_{n=1}^{\infty}
\left\langle\frac{dW_{n{\rm g}}^{(S)}({\bf b})}{dt}\right\rangle_{\!\!{\bf b}}
\label{eq:8t}
\end{equation}
\[
\cong {\alpha}_s^2\!\sum_{\beta=q,\,\bar{q},\,{\rm g}}
\Biggl(\frac{C_2^{(\alpha)}C_2^{(\beta)}}{d_A}\Biggr)n_{\beta}
\!\int\!\frac{d{\bf q}}{({\bf q}^2+\mu_D^2)^2}
\int\!\frac{d\tau}{2\pi}\,{\rm e}^{i({\bf v}_{\alpha}\cdot {\bf q})\tau}
\]
\[
\times\Biggl\{\frac{\alpha_s}{\pi^2}\,C_A\!
\!\sum_{\zeta=1,2}\!\int\!d{\bf k}\,
{\rm e}^{i(v_{\alpha}\cdot k)\tau}\omega_{\bf k}^{t}\!
\left(\frac{{\rm Z}_{t}({\bf k})}{2\omega_{\bf k}^{t}}\right)\!
|\,{\cal M}^{i}({\bf v}_{\alpha},0|\,k,q)
{\rm e}^i(\hat{\bf k},\zeta)|^{\,2}\Biggr\}_{\!q_0=0}
\]
\[
\times\exp\Biggl[\frac{3\alpha_s}{2\pi^2}\,C_A\!
\!\sum_{\zeta_1=1,2}\!\int\!d{\bf k}_1\,
{\rm e}^{i(v_{\alpha}\cdot k_1)\tau}N_{{\bf k}_1}^{t}\!
\left(\frac{{\rm Z}_{t}({\bf k}_1)}{2\omega_{{\bf k}_1}^{t}}\right)\!
|\,{\cal M}^{i_1}({\bf v}_{\alpha},0|\,k_1,q)
{\rm e}^{i_1}(\hat{\bf k}_1,\zeta_1)|^{\,2}_{\,q_0=0}
\Biggr].
\]
In the second line of this equation we separate out an `elastic' part. 
In the deriving of
Eq.\,(\ref{eq:8t}) we making use the fact that the momentum-energy constraint
can be factorized through the Fourier representation
\[
\delta(v_{\alpha}\cdot(k+k_1+\ldots+k_n-q))|_{q_0=0}=
\int\limits^{+\infty}_{-\infty}\frac{d\tau}{2\pi}\,
{\rm e}^{i({\bf v}_{\alpha}\cdot {\bf q})\tau}
{\rm e}^{i(v_{\alpha}\cdot k)\tau}
\prod\limits_{l=1}^{n}\!{\rm e}^{i(v_{\alpha}\cdot k_l)\tau}.
\]
At the soft momentum scale for low-excited QGP (an estimation
(\ref{eq:7p})) exponential factor on the right-hand side of
Eq.\,(\ref{eq:8t}) is of order one and thus the energy loss
in the leading order is defined by bremsstrahlung of one gluon. In opposite
case high-excited QGP (an estimation (\ref{eq:7a})) argument of exponential
function is not perturbatively small, and therefore in this case we can expect
essential increase of radiation energy loss.

Unfortunately for QCD interaction even within semiclassical
approximation and for ``Abelian'' part of soft gluon emission
factorizations (\ref{eq:8e}), (\ref{eq:8r}) don`t take place, and
therefore a simple and descriptive formula (\ref{eq:8t}) for
energy loss here is not correct. This points to the fact that
multiple soft gluon radiation is not in general {\it Poisson
process} and in this way the Poisson approximation for multiple
soft gluon radiation used in a number of works (e.g.
\cite{matinyan, baier5, gyulassy6}), is not correct.

Let us derive an expression for symmetrized part of matrix element of
two soft-gluon bremsstrahlung ${\cal K}^{(S)ii_1}({\bf
v}_{\alpha},0|\,k,k_1,q) {\rm e}^i(\hat{\bf k},\zeta){\rm
e}^{i_1}(\hat{\bf k}_1,\zeta_1)$ in high-frequency and small-angle
approximations, similar to expression ${\cal M}^{i}({\bf
v}_{\alpha},0|\,k,q){\rm e}^i(\hat{\bf k},\zeta)$ defined in Section
4, i.e. in conditions when we can neglect by HTL-corrections to
vertices of interaction and longitudinal part of gluon propagators on
the right-hand side of Eq.\,(\ref{eq:8w}).

The Coulomb factor in this approximation reads
\[
\frac{1}{{\bf q}^2+\mu_D^2}\simeq
\frac{1}{{\bf q}_{\perp}^{\,2}+\mu_D^2+\displaystyle\frac{1}{v_{\alpha}^2}
\left(\frac{1}{\tau_f}+\frac{1}{\tau_{\!1f}}\right)^2}\,,
\]
where
\begin{equation}
{\tau}_f\equiv
\frac{2\omega}{({\bf k}_{\perp}^2 + m_g^2 + x^2M^2)},\quad
{\tau}_{\!1 f}\equiv
\frac{2\omega_1}{({\bf k}_{1\perp}^2 + m_g^2 + x_1^2M^2)},
\quad x=\frac{\omega}{E},\,\,x_1=\frac{\omega_1}{E},
\label{eq:8y}
\end{equation}
are formation times. Furthermore the functions $({\bf e}\cdot{\bf
v}_{\alpha})/(v_{\alpha}\cdot k)$ and $({\bf e}_1\cdot{\bf
v}_{\alpha})/(v_{\alpha}\cdot k_1)$ entering into the first term
on the right-hand side of Eq.\,(\ref{eq:8w}) is approximated by an
expression of (\ref{eq:4u}) type. The second term\footnote{By
terms hereafter we meant all expression standing within curly,
square or round brackets.} is approximated by a similar way as it
was made for term (\ref{eq:4o}), where except usual requirements
in the form
\[
\omega\tau_f\gg 1,\quad \omega_1\tau_{1f}\gg 1
\]
it is necessary to consider that inequalities
\[
\omega_1\tau_f\gg 1,\quad \omega\tau_{1f}\gg 1
\]
are also fulfilled. Uniquely, the approximation of the transverse
scalar propagator $\,^{\ast}{\!\Delta}^t(q-k)$ requires more
accuracy. Instead of approximation (\ref{eq:4oo}) in this case we
will have
\begin{equation}
({\bf k}-{\bf q})^2\simeq ({\bf k}-{\bf q})^2_{\perp} +
\frac{1}{v_{\alpha}^2}\left(\omega + \frac{1}{\tau_{1f}}\right)^2
\simeq
{\bf k}^2 + ({\bf k}-{\bf q})^2_{\perp} +
m_g^2 + x^2M^2 + \frac{2}{v_{\alpha}^2}\,\frac{\omega}{\tau_{1f}}\,,
\label{eq:8u}
\end{equation}
and hence expression (\ref{eq:4ooo}) must be replaced by
\begin{equation}
^{\ast}\!\Delta^{t}(q-k)\simeq
-\frac{1}{({\bf k}-{\bf q})^2_{\perp} + m_g^2 + x^2M^2
+\biggl(\displaystyle\frac{x}{x_1}\biggr)
({\bf k}_{1\perp}^2 + m_g^2 + x_1^2M^2)}\,.
\label{eq:8i}
\end{equation}

The third term on the right-hand side of Eq.\,(\ref{eq:8w}) is more
complicated for analysis. Neglecting by HTL-correction to bare
three-gluon vertices and longitudinal part of gluon propagators,
contracting with ${\rm e}^i(\hat{\bf k},\zeta){\rm e}^{i_1}(\hat{\bf
k}_1,\zeta_1)$ and taking into account the equalities ${\bf
e}\cdot{\bf k}={\bf e}_1\cdot{\bf k}_1=0$, we have an exact expression
initial for analysis
\[
2\omega_1\,^{\ast}\!\Delta^{t}(q-k_1)\,^{\ast}\!\Delta^{t}(k+k_1-q)
\Biggl\{
-2({\bf e}\cdot ({\bf k}_1-{\bf q}))({\bf e}_1\cdot{\bf v}_{\alpha})+
({\bf e}_1\cdot(2{\bf k}-{\bf q}))({\bf e}\cdot {\bf v}_{\alpha})
\]
\[
-\,({\bf v}_{\alpha}\cdot({\bf k}-{\bf k}_1+{\bf q}))({\bf e}\cdot{\bf e}_1)
+\frac{\omega+\omega_1}{({\bf k}+{\bf k}_1-{\bf q})^2}\,
\!\biggl[
\Bigl(\,{\bf k}^2-({\bf k}_1-{\bf q})^2\Bigr)({\bf e}\cdot{\bf e}_1)
-({\bf e}\cdot ({\bf k}_1-{\bf q}))({\bf e}_1\cdot{\bf q})\biggr]
\]
\[
+\,v_{\alpha}^2\frac{({\bf e}_1\cdot{\bf q})}{({\bf k}_1-{\bf q})^2}
\biggl[
({\bf e}\cdot {\bf v}_{\alpha})((2{\bf k}+{\bf k}_1-{\bf q})
\cdot({\bf k}_1-{\bf q}))
-(\omega + \omega_1)({\bf e}\cdot ({\bf k}_1-{\bf q}))
\biggl(1-
\frac{{\bf k}^2}{({\bf k}+{\bf k}_1-{\bf q})^2}\biggr)\biggr]\!\Biggr\}.
\]
In the small-angle approximation next expressions hold
\[
{\bf e}\cdot{\bf k}\simeq {\bf e}_{\perp}\cdot
\Bigl({\bf k}_1-\frac{x_1}{x}{\bf k}_{\perp}\Bigr),\quad
{\bf e}_1\cdot{\bf k}\simeq {\bf e}_{1\perp}\cdot
\Bigl({\bf k}-\frac{x}{x_1}{\bf k}_{1\perp}\Bigr),
\]
\[
{\bf e}\cdot{\bf q}\simeq{\bf e}_{\perp}\cdot{\bf q}_{\perp},\quad
{\bf e}_1\cdot{\bf q}\simeq{\bf e}_{1\perp}\cdot{\bf q}_{\perp},\quad
{\bf e}\cdot{\bf e}_1\simeq{\bf e}_{\perp}\cdot{\bf e}_{1\perp},
\]
\[
(2{\bf k}+{\bf k}_1-{\bf q})\cdot({\bf k}_1-{\bf q})\simeq
\omega_1(2\omega+\omega_1),
\]
etc. Let us consider in more detail an approximation of coefficient before
scalar product $({\bf e}\cdot{\bf e}_1)$, that is equals to
\begin{equation}
-\,{\bf v}_{\alpha}\cdot({\bf k}-{\bf k}_1+{\bf q})
+\frac{\omega+\omega_1}{({\bf k}+{\bf k}_1-{\bf q})^2}\,
\Bigl(\,{\bf k}^2-({\bf k}_1-{\bf q})^2\Bigr).
\label{eq:8o}
\end{equation}
By the use of approximation (\ref{eq:8u}) it is easy to see that the expression
within round brackets in the second term of Eq.\,({\ref{eq:8o}) can be
approximated as
\[
\,{\bf k}^2-({\bf k}_1-{\bf q})^2\simeq
-\,\Bigl[{({\bf k}_1-{\bf q})^2_{\perp} + m_g^2 + x_1^2M^2
+\biggl(\displaystyle\frac{x_1}{x}\biggr)
({\bf k}_{\perp}^2 + m_g^2 + x^2M^2)}\Bigr]+
(\omega^2-\omega_1^2)
\]
\[
\equiv
\,^{\ast}\!\Delta^{-1t}(q-k_1)+(\omega^2-\omega_1^2).
\]

Futhermore for denominator in the second term of Eq.\,(\ref{eq:8o}) we have
\[
\frac{1}{({\bf k}+{\bf k}_1-{\bf q})^2}=
\frac{1}{(\omega+\omega_1)^2/v_{\alpha}^2 +
({\bf k}+{\bf k}_1-{\bf q})^2_{\perp}}\simeq
\frac{v_{\alpha}^2}{(\omega+\omega_1)^2} -
v_{\alpha}^4\,\frac{({\bf k}+{\bf k}_1-{\bf q})^2_{\perp}}
{(\omega+\omega_1)^4}
\]
and such now we can write instead of Eq.\,(\ref{eq:8o})
\begin{equation}
-v_{\alpha}({\bf k}-{\bf k}_1)_{\parallel} - v_{\alpha}q_{\parallel}
+v_{\alpha}^2(\omega-\omega_1)-
v_{\alpha}^4\,
\frac{({\bf k}+{\bf k}_1-{\bf q})^2_{\perp}}
{(\omega+\omega_1)^4}\,(\omega-\omega_1)+
\frac{v_{\alpha}^2}{\omega+\omega_1}\,\,^{\ast}\!\Delta^{\!-1t}(q-k_1).
\label{eq:8p}
\end{equation}
Here, in two last terms within accepted accuracy we can set
$v_{\alpha}\simeq 1$. In high-frequency approximation the third term in
Eq.\,(\ref{eq:8p}) is approximated as
\[
v_{\alpha}^2(\omega-\omega_1)\simeq
v^2_{\alpha}({\bf k}-{\bf k}_1)_{\parallel}+
\frac{{\bf k}_{\perp}^2+m_g^2}{2\omega}-
\frac{{\bf k}_{1\perp}^2+m_g^2}{2\omega_1}.
\]
Let us subtitute this expression into Eq.\,(\ref{eq:8p}). Taking into account
estimations
\[
1-v_{\alpha}\simeq\frac{M^2}{2E^2},\quad
v_{\alpha}\,q_{\parallel}\simeq
-\Biggl(\frac{1}{\tau_f}+\frac{1}{\tau_{\!1f}}\Biggr)
\]
and definition of inverse propagator $\!\,^{\ast}\!\Delta^{\!-1t}(q-k_1)$,
it is easy to see that all terms containing $m_g^2$ and $M^2$ in
Eq.\,(\ref{eq:8p}) exactly cancel, and such we obtain the following simple
approximation of coefficient (\ref{eq:8o})
\[
\frac{1}{\omega+\omega_1}\,
\Biggl[\,{\bf k}_{\perp}^2-({\bf k}_1-{\bf q})^2_{\perp}-
\Biggl(\frac{\omega-\omega_1}{\omega+\omega_1}\Biggr)
({\bf k}+{\bf k}_1-{\bf q})^2_{\perp}\Biggr]
\]
\[
\equiv\frac{2}{(\omega+\omega_1)^2}\,
\Bigl[\,\omega_1{\bf k}_{\perp}^2-\omega({\bf k}_1-{\bf q})^2_{\perp}-
(\omega-\omega_1)({\bf k}_{\perp}\cdot({\bf k}_1-{\bf q})_{\perp})\Bigr].
\]

Finally we write out an approximation for transverse scalar propagator
$^{\ast}\!\Delta^{t}(k+k_1-q)$:
\begin{equation}
^{\ast}\!\Delta^{t}(k+k_1-q)\simeq
-\frac{1}{({\bf k}+{\bf k}_1-{\bf q})^2_{\perp} + m_g^2 + (x+x_1)^2M^2}\,.
\label{eq:8a}
\end{equation}
Taking into account above mentioned we derive complete expression for
matrix element ${\cal K}^{(S)ii_1}{\rm e}^i{\rm e}^{i_1}$ in a small-angle
approximation
\begin{equation}
{\cal K}^{(S)ii_1}({\bf v}_{\alpha},0|\,k,k_1,q)
{\rm e}^i(\hat{\bf k},\zeta){\rm e}^{i_1}(\hat{\bf k}_1,\zeta_1)|_{\,q_0=0}
\label{eq:8s}
\end{equation}
\[
\simeq\frac{2}{{\bf q}^2_{\perp}+\mu_D^2+\displaystyle\frac{1}{v_{\alpha}^2}
\left(\frac{1}{\tau_f}+\frac{1}{\tau_{\!1f}}\right)^{\!2}}\,
\Biggl[\,
\frac{{\bf e}_{\perp}\cdot{\bf k}_{\perp}}
{{\bf k}_{\perp}^2 + m_g^2 + x^2M^2}\,\,
\frac{{\bf e}_{1\perp}\cdot{\bf k}_{1\perp}}
{{\bf k}_{1\perp}^2 + m_g^2 + x_1^2M^2}
\]
\[
+\,\Biggl\{\!
\,^{\ast}\!\Delta^{t}(q-k)\,
\frac{({\bf e}_{\perp}\cdot({\bf k}-{\bf q})_{\perp})
({\bf e}_{1\perp}\cdot{\bf k}_{1\perp})}
{{\bf k}_{1\perp}^2 + m_g^2 + x_1^2M^2}\,+
\,^{\ast}\!\Delta^{t}(q-k_1)\,
\frac{({\bf e}_{1\perp}\cdot({\bf k}_1-{\bf q})_{\perp})
({\bf e}_{\perp}\cdot{\bf k}_{\perp})}
{{\bf k}_{\perp}^2 + m_g^2 + x^2M^2}
\Biggr\}
\]
\[
+\bigg\{\!\!\,^{\ast}\!\Delta^{t}(q-k_1)
\biggl[\Bigl(
{\bf e}_{\perp}\cdot\Bigl({\bf k}_1-\frac{x_1}{x}{\bf k}
-{\bf q}\Bigr)_{\!\perp}\Bigr)
({\bf e}_{1\perp}\cdot({\bf k}_1-{\bf q})_{\perp})
+\Bigl({\bf e}_{1\perp}\cdot\Bigl({\bf k}_1-\frac{x_1}{x}{\bf k}
-{\bf q}\Bigr)_{\!\perp}\Bigr)
({\bf e}_{\perp}\cdot{\bf k}_{\perp})
\]
\[
+\,\frac{x_1}{(x+x_1)^2}\,({\bf e}_{\perp}\cdot{\bf e}_{1\perp})
\Bigl(x_1{\bf k}_{\perp}^2-x({\bf k}_1-{\bf q})^2_{\perp}
-(x-x_1)({\bf k}_{\perp}\cdot({\bf k}_1-{\bf q})_{\perp})\Bigr)\biggr]
\]
\[
+\,({\bf k}_{\perp}\rightleftharpoons {\bf k}_{1\perp},\,
{\bf e}_{\perp}\rightleftharpoons{\bf e}_{1\perp},\,
x\rightleftharpoons x_1)\biggr\}\,^{\ast}\!\Delta^{t}(k+k_1-q)\Biggr].
\]
Here, the propagators $\,^{\ast}\!\Delta^{t}(q-k)$,
$\,^{\ast}\!\Delta^{t}(q-k_1)$ and $^{\ast}\!\Delta^{t}(k+k_1-q)$ are defined
by Eqs.\,(\ref{eq:8i}) and (\ref{eq:8a}) correspondingly. By similar way
matrix element ${\cal K}^{(A)ii_1}{\rm e}^i{\rm e}^{i_1}$ is calculated.
Here, a new contribution defined by fourth term on the right-hand
side of Eq.\,(\ref{eq:8w}) is appeared, where now a difference
\[
\,^{\ast}\Gamma^{ii_1\lambda}(k,k_1,-k-k_1)-
\,^{\ast}\Gamma^{i_1i\lambda}(k_1,k,-k-k_1)=
2\,^{\ast}\Gamma^{ii_1\lambda}(k,k_1,-k-k_1).
\]
will be stay instead of a sum of two three-gluon vertices.
The explicit expression for this matrix element in high-energy and small-angle
approximations is written out in Appendix A.

Matrix elements (\ref{eq:8s}) and (A.2) can be used in particular in
simulation of on-shell parton cascade at the build of the elliptic
flow at RHIC. They enables the inelastic $gg\leftrightarrow gggg$
pQCD processes to be included in the consideration. Up to now only
elastic $gg\leftrightarrow gg$ \cite{molnar1} and the inelastic
$gg\leftrightarrow ggg$ \cite{molnar2,xu} gluon interactions are
included in simulation.

\section{Soft gluon bremsstrahlung in the case of two-scatte\-ring thermal
partons}
\setcounter{equation}{0}

In this Section we extend consideration of the bremsstrahlung process
to the case of scattering of a high-energy incident parton $\alpha$
off two hard thermal partons $\beta_1$ and $\beta_2$ moving with
velocities ${\bf v}_{\beta_1}$ and ${\bf v}_{\beta_2}$,
correspondingly. As initial currents we choose the expression in the
momentum space
\[
\tilde{J}^{(0)a\mu}_{Q\alpha}(k)=
\frac{g}{(2\pi)^3}\,\,Q_{0\alpha}^av_{\alpha}^{\mu}
\delta(v_{\alpha}\cdot k),\quad
\tilde{J}^{(0)a\mu}_{Q{\beta}_{1,\,2}}(k)=
\frac{g}{(2\pi)^3}\,\,Q_{0{\beta}_{1,\,2}}^av^{\mu}_{{\beta}_{1,\,2}}
\delta(v_{{\beta}_{1,\,2}}\cdot k)
{\rm e}^{i{\bf k}\cdot{\bf x}_{10,20}}.
\]
Here, ${\bf x}_{10},\;{\bf x}_{20}$ are coordinates of partons $\beta_1$ and
$\beta_2$ at time $t=t_0$. In this case the energy of radiation field is
defined (instead of (\ref{eq:3q})) by equation
\begin{equation}
W({\bf x}_{10},{\bf x}_{20})\!=\!
-(2\pi)^4\!\!\int\!\!d{\bf k}d\omega
\!\!\int\!\!dQ_{0\alpha}dQ_{0\beta_1}dQ_{0\beta_2}
\,\omega
{\rm Im}\,\langle
\tilde{J}^{({\rm br})\ast a}_{Q\mu}(k,{\bf x}_{10},{\bf x}_{20})
\!\,^{\ast}{\cal D}^{\mu\nu}_C(k)
\tilde{J}^{({\rm br})a}_{Q\nu}(k,{\bf x}_{10},{\bf x}_{20})
\rangle\,,
\label{eq:9q}
\end{equation}
and gluon radiation intensity -- by equation
\begin{equation}
{\cal I}=
\int\!\omega({\bf k})
{\it w}_{{\bf p}_{\alpha},\,{\bf p}_{\beta_1},\,{\bf p}_{\beta_2}}
({\bf k};{\bf q}_1,{\bf q}_2)\,
\frac{f_{{\bf p}_{\beta_1}}d{\bf p}_{\beta_1}}{(2\pi)^3}\,
\frac{f_{{\bf p}_{\beta_2}}d{\bf p}_{\beta_2}}{(2\pi)^3}\,
\frac{d{\bf k}d{\bf q}_1d{\bf q}_2}{(2\pi)^9}\,
\label{eq:9w}
\end{equation}
\[
\times\,
\delta(\omega({\bf k})-{\bf v}_{\beta_1}\cdot{{\bf q}_1}
-{\bf v}_{\beta_2}\cdot{{\bf q}_2}
-({\bf v}_{\alpha}\cdot({\bf k}-{\bf q}_1-{\bf q}_2)).
\]
Here, ${\it w}_{{\bf p}_{\alpha},\,{\bf p}_{\beta_1},\,{\bf p}_{\beta_2}}$ is
a probability of bremsstrahlung in the case of two-scattering hard thermal
partons. The standard calculations result in following expression for an
effective current generating this process
\[
\tilde{J}^{a}_{Q\mu}(k,{\bf x}_{10},{\bf x}_{20})=
\frac{g^5}{(2\pi)^9}\,\Bigl[
(T^{d_1}T^{d_2})^{ab}\,Q_{0\alpha}^bQ_{0\beta_1}^{d_1}Q_{0\beta_2}^{d_2}
K_{\mu}({\bf v}_{\alpha},{\bf v}_{\beta_1},{\bf v}_{\beta_2};
{\bf x}_{10},{\bf x}_{20}\vert\,k)
\]
\begin{equation}
+\,
(T^{d_2}T^{d_1})^{ab}\,Q_{0\alpha}^bQ_{0\beta_2}^{d_2}Q_{0\beta_1}^{d_1}
K_{\mu}({\bf v}_{\alpha},{\bf v}_{\beta_2},{\bf v}_{\beta_1};
{\bf x}_{20},{\bf x}_{10}\vert\,k)\,\Bigr],
\label{eq:9e}
\end{equation}
where partial coefficient function
$K_{\mu}({\bf v}_{\alpha},{\bf v}_{\beta_1},{\bf v}_{\beta_2};
{\bf x}_{10},{\bf x}_{20}\vert\,k)$ is
$$
K_{\mu}({\bf v}_{\alpha},{\bf v}_{\beta_1},{\bf v}_{\beta_2};
{\bf x}_{10},{\bf x}_{20}|\,k)=\int\biggl\{
-\frac{v_{\beta_1\mu}v_{\beta_1\nu}}{v_{\beta_1}\cdot(q_1-k)}\,
\,^{\ast}{\cal D}_C^{\nu\nu^{\prime}}\!(q_1-k)
{\cal K}_{\nu^{\prime}}({\bf v}_{\alpha},{\bf v}_{\beta_2}|\,q_1-k,-q_2)
$$
$$
+\,\frac{v_{\beta_1\mu}}{(v_{\beta_1}\cdot q_2)
(v_{\beta_1}\cdot(k-q_1))}
\,(v_{\beta_1\nu}\!\,^{\ast}{\cal D}_C^{\nu\nu^{\prime}}\!(q_2)
v_{\beta_2\nu^{\prime}})
\,(v_{\beta_1\lambda}\!
\,^{\ast}{\cal D}_C^{\lambda\lambda^{\prime}}\!(k-q_1-q_2)
v_{\alpha\lambda^{\prime}})
$$
$$
-\,2\,\frac{v_{\beta_2\mu}}{(v_{\beta_2}\cdot q_1)
(v_{\beta_2}\cdot(k-q_2))}\,
\,(v_{\beta_1\nu}\!\,^{\ast}{\cal D}_C^{\nu\nu^{\prime}}\!(q_1)
v_{\beta_2\nu^{\prime}})
\,(v_{\beta_2\lambda}\!
\,^{\ast}{\cal D}_C^{\lambda\lambda^{\prime}}\!(k-q_1-q_2)
v_{\alpha\lambda^{\prime}})
$$
$$
-\,\frac{v_{\alpha\mu}v_{\alpha\nu}}{v_{\alpha}\cdot(q_1+q_2)}\,
\,^{\ast}{\cal D}_C^{\nu\nu^{\prime}}\!(q_1+q_2)
{\cal K}_{\nu^{\prime}}({\bf v}_{\beta_1},{\bf v}_{\beta_2}|\,q_1+q_2,q_2)
$$
$$
+\,\frac{v_{\alpha\mu}}{(v_{\alpha}\cdot(q_1+q_2))(v_{\alpha}\cdot q_2)}\,
(v_{\alpha\nu}\,^{\ast}{\cal D}_C^{\nu\nu^{\prime}}\!(q_1)
v_{\beta_1\nu^{\prime}})
(v_{\alpha\lambda}\,^{\ast}{\cal D}_C^{\lambda\lambda^{\prime}}\!(q_2)
v_{\beta_2\lambda^{\prime}})
$$
$$
-\,^{\ast}\Gamma_{\mu\nu\lambda}(k,-q_1-q_2,-k+q_1+q_2)
\Bigl[\!
\,^{\ast}{\cal D}_C^{\nu\nu^{\prime}}\!(q_1+q_2)
{\cal K}_{\nu^{\prime}}({\bf v}_{\beta_1},{\bf v}_{\beta_2}|\,q_1+q_2,q_2)
\,^{\ast}{\cal D}_C^{\lambda\lambda^{\prime}}\!(k-q_1-q_2)
v_{\alpha\lambda^{\prime}}
$$
$$
-\,^{\ast}{\cal D}_C^{\nu\nu^{\prime}}\!(q_1+q_2)
v_{\beta_1\nu^{\prime}}
\,^{\ast}{\cal D}_C^{\lambda\lambda^{\prime}}\!(k-q_1-q_2)
{\cal K}_{\lambda^{\prime}}
({\bf v}_{\beta_1},{\bf v}_{\beta_2}|\,k-q_1-q_2,q_2)\Bigr]
$$
$$
+
\,^{\ast}\Gamma_{\mu\nu\lambda\sigma}(k,-k+q_1+q_2,-q_2,-q_1)
\,^{\ast}{\cal D}_C^{\nu\nu^{\prime}}\!(k-q_1-q_2)v_{\alpha\nu^{\prime}}
\,^{\ast}{\cal D}_C^{\lambda\lambda^{\prime}}\!(q_1)
v_{\beta_1\lambda^{\prime}}
\,^{\ast}{\cal D}_C^{\sigma\sigma^{\prime}}\!(q_2)v_{\beta_2\sigma^{\prime}}
\!\biggr\}
$$
$$
\times\,{\rm e}^{i{\bf q}_1\cdot\,{\bf x}_{10}}
{\rm e}^{i{\bf q}_2\cdot\,{\bf x}_{20}}
\delta(v_{\alpha}\cdot(k-q_1-q_2))\delta(v_{\beta_1}\cdot q_1)
\delta(v_{\beta_2}\cdot q_2)\,dq_1dq_2.
$$
For the reason of large number of terms on the right-hand side their
complete diagrammatic interpretation is omitted. Here, we
discuss at qualitative level a contribution defining transition
bremsstrahlung induced by four-gluon HTL amplitude
(Fig.\,\ref{fig8}):
$^{\ast}\Gamma_{4g}=\Gamma_{4g}+\delta\Gamma_{4g}$, where
$\Gamma_{4g}$ is bare four-gluon vertex, and $\delta\Gamma_{4g}$ is
HTL correction.
\begin{figure}[hbtp]
\begin{center}
\includegraphics*[height=8cm]{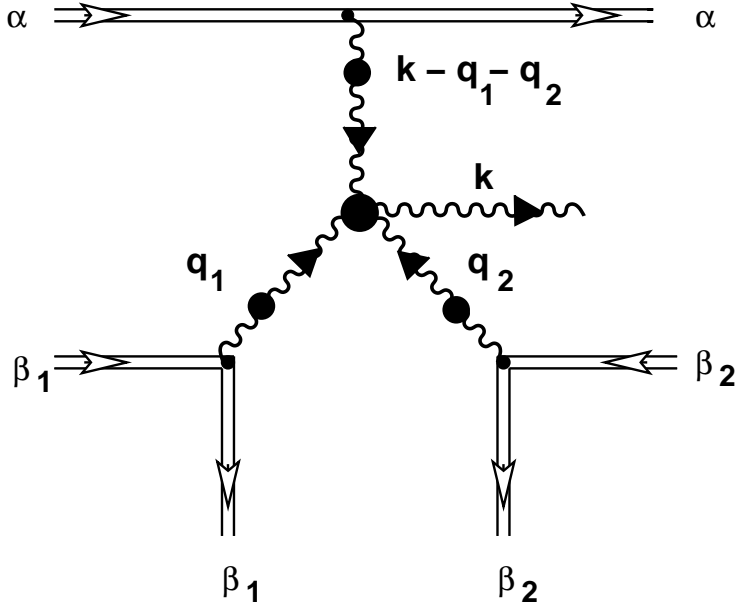}
\end{center}
\caption{\small The transition bremsstrahlung induced by
$\delta\Gamma_{4g}$ vertex HTL correction.}
\label{fig8}
\end{figure}
In the case of bare four-gluon vertex $\Gamma_{4g}$ this contribution is 
usually dropped for kinematic reason associated with absence of momentum 
dependence of $\Gamma_{4g}$ \cite{gyulassy1,wang1}. The momentum dependence of
$\delta\Gamma_{4g}$ on incoming lines is very nontrivial, and here we can
expect an appreciable contribution of new mechanism of bremsstrahlung to gluon
radiation intensity. If we restrict our consideration to this case, then under
partial coefficient function in Eq.\,(\ref{eq:9e}) we mean expression
\[
K_{\mu}({\bf v}_{\alpha},{\bf v}_{\beta_1},{\bf v}_{\beta_2};
{\bf x}_{10},{\bf x}_{20}|\,k)=
\int\!{\cal K}_{\mu}({\bf v}_{\alpha},{\bf v}_{\beta_1},{\bf v}_{\beta_2}
\vert\,k,q_1,q_2)\delta(v_{\alpha}\cdot(k-q_1-q_2))
\delta(v_{{\beta}_1}\cdot q_1)\delta(v_{{\beta}_2}\cdot q_2)
\]
\begin{equation}
\times\,{\rm e}^{i{\bf x}_{10}\cdot\,{\bf q}_1}
{\rm e}^{i{\bf x}_{20}\cdot\,{\bf q}_2}
dq_{10}dq_{20}d{\bf q}_1d{\bf q}_2,
\label{eq:9r}
\end{equation}
where
\[
{\cal K}_{\mu}({\bf v}_{\alpha},{\bf v}_{\beta_1},{\bf v}_{\beta_2}
\vert\,k,q_1,q_2)\equiv
\delta\Gamma_{\mu\nu\lambda\sigma}(k,-q_1,-q_2,-k+q_1+q_2)
\,^{\ast}{\cal D}_{C}^{\nu\nu^{\prime}}(q_1)v_{\beta_1\nu^{\prime}}
\!\,^{\ast}{\cal D}_{C}^{\lambda\lambda^{\prime}}(q_2)
v_{\beta_2\lambda^{\prime}}
\]
\begin{equation}
\times\,^{\ast}{\cal D}_{C}^{\sigma\sigma^{\prime}}
(k-q_1-q_2)v_{\alpha\sigma^{\prime}}.
\label{eq:9t}
\end{equation}

Now we substitute an effective current (\ref{eq:9e}) into Eq.\,(\ref{eq:9q})
and average over initial values of color charges, taking into account a relation
for color algebra (\ref{eq:7y}). Keeping in propagator
$\!\,^{\ast}{\cal D}_{C}^{\mu\nu}(k)$ a transverse part only with allowance for
(\ref{eq:3t}) and going over to mass-shell of radiated transverse gluon
(Eq.\,(\ref{eq:3p})), after simple algebraic transformations we lead
to the expression for radiation field energy
\begin{equation}
W({\bf x}_{10},{\bf x}_{20})=
\frac{1}{(2\pi)^4}\,
\biggl(\frac{{\alpha}_s}{\pi}\biggr)^{\!5}\,C_A^{\,2}
\Biggl(\frac{C_2^{(\alpha)}C_2^{(\beta_1)} C_2^{(\beta_2)}}{d_A^2}\Biggr)
\!\sum_{\zeta=1,\,2}\!\int\!d{\bf k}\,\omega_{\bf k}^{t}
\left(\frac{{\rm Z}_{t}({\bf k})}{2\omega_{\bf k}^{t}}\right)
\label{eq:9y}
\end{equation}
\[
\times
\biggl\{
|\,K^{i}({\bf v}_{\alpha},{\bf v}_{\beta_1},{\bf v}_{\beta_2};
{\bf x}_{10},{\bf x}_{20}|\,k){\rm e}^i(\hat{\bf k},\zeta)|^{\,2}
+ |\,K^{i}({\bf v}_{\alpha},{\bf v}_{\beta_2},{\bf v}_{\beta_1};
{\bf x}_{20},{\bf x}_{10}|\,k){\rm e}^i(\hat{\bf k},\zeta)|^{\,2}
\]
\[
+\,{\rm Re}
\Bigl[\Bigl(\,K^{i}({\bf v}_{\alpha},{\bf v}_{\beta_1},{\bf v}_{\beta_2};
{\bf x}_{10},{\bf x}_{20}|\,k){\rm e}^i(\hat{\bf k},\zeta)\Bigr)^{\!\ast}
\Bigl(K^{j}({\bf v}_{\alpha},{\bf v}_{\beta_2},{\bf v}_{\beta_1};
{\bf x}_{20},{\bf x}_{10}|\,k){\rm e}^j(\hat{\bf k},\zeta)\Bigr)\Bigr]
\!\biggr\}.
\]
Let us consider module squared $|K^i{\rm e}^i|^2$ in integrand. Substituting
an expression (\ref{eq:9r}) instead of $K^i$ and performing integration over
$dq_{10}dq_{20}dq_{10}^{\prime}dq_{20}^{\prime}$, we obtain
\begin{equation}
|\,K^{i}({\bf v}_{\alpha},{\bf v}_{\beta_1},{\bf v}_{\beta_2};
{\bf x}_{10},{\bf x}_{20}|\,k){\rm e}^i(\hat{\bf k},\zeta)|^{\,2}=
\!\int\!
\Bigl[\,{\cal K}^{i}({\bf v}_{\alpha},{\bf v}_{\beta_1},{\bf v}_{\beta_2}
|\,k,q_1,q_2){\rm e}^i(\hat{\bf k},\zeta)\Bigr]_{q^0_{1,2}
={{\bf v}_{\beta_{1,2}}}\cdot\,{{\bf q}_{1,2}}}
\label{eq:9uu}
\end{equation}
\[
\times
\Bigl[\,{\cal K}^{j}({\bf v}_{\alpha},{\bf v}_{\beta_1},{\bf v}_{\beta_2}
|\,k,q_1^{\prime},q_2^{\prime}){\rm e}^j(\hat{\bf k},\zeta)
\Bigr]^{\ast}_{q^{\prime0}_{1,2}={{\bf v}_{\beta_{1,2}}}
\cdot\,{{\bf q}_{1,2}^{\prime}}}
\delta(\omega_{\bf k}^t-{\bf v}_{\beta_1}\cdot{{\bf q}_1}
-{\bf v}_{\beta_2}\cdot{{\bf q}_2}
-{\bf v}_{\alpha}\cdot({\bf k}-{\bf q}_1-{\bf q}_2))
\]
\[
\times\,
\delta(({\bf v}_{\alpha}-{\bf v}_{\beta_1})\!\cdot\!
({\bf q}_1-{\bf q}_1^{\prime})+
({\bf v}_{\alpha}-{\bf v}_{\beta_2})\!\cdot\!
({\bf q}_2-{\bf q}_2^{\prime}))
{\rm e}^{i({\bf q}_1-{\bf q}_1^{\prime})\cdot{\bf x}_{10}}
{\rm e}^{i({\bf q}_2-{\bf q}_2^{\prime})\cdot{\bf x}_{20}}
d{\bf q}_1d{\bf q}_2d{\bf q}_1^{\prime}d{\bf q}_2^{\prime}.
\]
Furthermore we analyze the expression in the last line of Eq.\,(\ref{eq:9uu}). We
present vector ${\bf x}_{10}$ in the form of an expansion in terms of vectors
system: the vector along relative velocity ${\bf v}_{\alpha}-{\bf v}_{\beta_1}$
and the vector orthogonal to ${\bf v}_{\alpha}-{\bf v}_{\beta_1}$, i.e.
${\bf x}_{10}=({\bf b}_1\equiv({\bf x}_{10})_{\perp}, ({\bf x}_{10})_\|).$
In this case a scalar product in the argument of the first exponential can be
presented as
\[
({\bf q}_1-{\bf q}_1^{\prime})\cdot{\bf x}_{10}=
({\bf q}_1-{\bf q}_1^{\prime})_{\|}({\bf x}_{10})_{\|}+
({\bf q}_1-{\bf q}_1^{\prime})_{\perp}\cdot {\bf b}_{1}.
\]
The vector ${\bf b}_1$ is impact parameter of particle $\alpha$ relative to
particle $\beta_1$. In a similar way we present vector ${\bf x}_{20}$ in the
form of an expansion in terms of basis associated with vector of relative
velocity ${\bf v}_{\alpha}-{\bf v}_{\beta_2}$:
${\bf x}_{20}=({\bf b}_2\equiv({\bf x}_{20})_{\perp}, ({\bf x}_{20})_\|).$
Then a scalar product in the argument of second exponential can be written in
the form
\[
({\bf q}_2-{\bf q}_2^{\prime})\cdot{\bf x}_{20}=
({\bf q}_2-{\bf q}_2^{\prime})_{\|}({\bf x}_{20})_{\|}+
({\bf q}_2-{\bf q}_2^{\prime})_{\perp}\cdot {\bf b}_{2}.
\]
In this case vector ${\bf b}_2$ is impact parameter of particle $\alpha$ relative
to particle $\beta_2$. By virtue of the fact that vectors ${\bf v}_{\alpha},\;
{\bf v}_{{\beta}_1}$ and ${\bf v}_{{\beta}_2}$ are fixed, such representation
of scalar products is uniquely determined.

With allowance for above-mentioned the following equality will be hold
\[
\delta(({\bf v}_{\alpha}-{\bf v}_{\beta_1})\!\cdot\!
({\bf q}_1-{\bf q}_1^{\prime})+
({\bf v}_{\alpha}-{\bf v}_{\beta_2})\!\cdot\!
({\bf q}_2-{\bf q}_2^{\prime}))
{\rm e}^{i({\bf q}_1-{\bf q}_1^{\prime})\cdot{\bf x}_{10}}
{\rm e}^{i({\bf q}_2-{\bf q}_2^{\prime})\cdot{\bf x}_{20}}
\]
\begin{equation}
=\delta(|\,{\bf v}_{\alpha}-{\bf v}_{\beta_1}\!|
({\bf q}_1-{\bf q}_1^{\prime})_{\|}+
|\,{\bf v}_{\alpha}-{\bf v}_{\beta_2}\!|
({\bf q}_2-{\bf q}_2^{\prime})_{\|})
{\rm e}^{i({\bf q}_1-{\bf q}_1^{\prime})_{\perp}\!\cdot{\bf b}_{1}}
{\rm e}^{i({\bf q}_2-{\bf q}_2^{\prime})_{\perp}\!\cdot{\bf b}_{2}}
\label{eq:9uuu}
\end{equation}
\[
\times\frac{1}{2}\left\{
\exp\left[i\,l\,\frac{{\cal V}}{|\,{\bf v}_{\alpha}-{\bf v}_{\beta_1}\!|}
\,({\bf q}_2-{\bf q}_2^{\prime})_{\|}\right] +
\exp\left[-i\,l\,\frac{{\cal V}}{|\,{\bf v}_{\alpha}-{\bf v}_{\beta_2}\!|}
\,({\bf q}_1-{\bf q}_1^{\prime})_{\|}\right]
\right\},
\]
where in the last line we have
\[
{\cal V}
\equiv 2\left(\frac{1}{|\,{\bf v}_{\alpha}-{\bf v}_{\beta_1}\!|}
+ \frac{1}{|\,{\bf v}_{\alpha}-{\bf v}_{\beta_2}\!|}\right)^{\!\!-1}\!,\quad
l\equiv\frac{|\,{\bf v}_{\alpha}-{\bf v}_{\beta_1}\!|({\bf x}_{20})_{\|}-
|\,{\bf v}_{\alpha}-{\bf v}_{\beta_2}\!|({\bf x}_{10})_{\|}}{\cal V}\,.
\]
After integrating Eq.\,(\ref{eq:9uuu}) over $d{\bf b}_1$, $d{\bf
b}_2$ and $dl$, we derive
\[
\frac{(2\pi)^5}{\cal V}\,
\delta^{(2)}(({\bf q}_1-{\bf q}_1^{\prime})_{\perp})
\delta(({\bf q}_1-{\bf q}_1^{\prime})_{\|})
\delta^{(2)}(({\bf q}_2-{\bf q}_2^{\prime})_{\perp})
\delta(({\bf q}_2-{\bf q}_2^{\prime})_{\|}),
\]
that enables us to perform integration in Eq.\,(\ref{eq:9uu}) over
$d{\bf q}_1^{\prime}d{\bf q}_2^{\prime}$.
Now we write an expression for the gluon radiation
intensity for the case of two nonrelativisic scatterers
\[
\left\langle\frac{dW({\bf b}_1,{\bf b}_2,l)}{dt}\right
\rangle_{\!{\bf b}_1\!,\,{\bf b}_2\!,\,l}=
\int\!d{\bf b}_1d{\bf b}_2\!\int\!\!dl\!
\int\!f_{{\bf p}_{{\beta}_1}}\,\frac{d{\bf p}_{{\beta}_1}}{(2\pi)^3}\,
\int\!f_{{\bf p}_{{\beta}_2}}\,\frac{d{\bf p}_{{\beta}_2}}{(2\pi)^3}\,
W({\bf b}_1,{\bf b}_2,l)\,{\cal V}.
\]
Substituting (\ref{eq:9y}) in this expression\footnote{One can slightly
extend an expression for radiation intensity, if replace
\[
\int\!f_{{\bf p}_{{\beta}_1}}\,\frac{d{\bf p}_{{\beta}_1}}{(2\pi)^3}\,
\int\!f_{{\bf p}_{{\beta}_2}}\,\frac{d{\bf p}_{{\beta}_2}}{(2\pi)^3}
\rightarrow
\int\!f_{{\bf p}_{{\beta}_1},{\bf p}_{{\beta}_2}}
\,\frac{d{\bf p}_{{\beta}_1}}{(2\pi)^3}\,
\frac{d{\bf p}_{{\beta}_2}}{(2\pi)^3}\,,
\]
where $f_{{\bf p}_{{\beta}_1},{\bf p}_{{\beta}_2}}$ is
two-particle distribution function. This enables us to take into
account more subtle effects connected with a possible correlation
in motion of two thermal partons $\beta_1$ and $\beta_2$.}, taking
into account (\ref{eq:9uu}) and (\ref{eq:9uuu}), we obtain a final
expression for gluon radiation intensity
\[
\left\langle\frac{dW({\bf b}_1,{\bf b}_2,l)}{dt}\right
\rangle_{\!{\bf b}_1\!,\,{\bf b}_2\!,\,l}=
\frac{{\alpha}_s^5}{\pi^4}\,C_A^2\!\sum_{{\beta}_1,{\beta}_2}\!
\Biggl(\frac{C_2^{(\alpha)}C_2^{(\beta_1)}C_2^{(\beta_2)}}{d_A}\Biggr)\!
\Biggl(\int\!{\bf p}_{\beta_1}^2f_{|{\bf p}_{\beta_1}|}
\,\frac{d|{\bf p}_{\beta_1}|}{2\pi^2}\Biggr)\!
\Biggl(\int\!{\bf p}_{\beta_2}^2f_{|{\bf p}_{\beta_2}|}
\,\frac{d|{\bf p}_{\beta_2}|}{2\pi^2}\Biggr)
\]
\[
\times
\int\!\frac{d\Omega_{{\bf v}_{\beta_1}}}{4\pi}
\!\int\!\frac{d\Omega_{{\bf v}_{\beta_2}}}{4\pi}
\!\sum_{\zeta=1,\,2}
\!\int\!d{\bf k}\,\omega_{\bf k}^{t}
\left(\frac{{\rm Z}_{t}({\bf k})}{2\omega_{\bf k}^{t}}\right)
\!\!\int\!d{\bf q}_1d{\bf q}_2\,
\delta(\omega_{\bf k}^t-{\bf v}_{\beta_1}\cdot{{\bf q}_1}
-{\bf v}_{\beta_2}\cdot{{\bf q}_2}
-{\bf v}_{\alpha}\cdot({\bf k}-{\bf q}_1-{\bf q}_2))
\]
\begin{equation}
\times
\biggl\{
|\,{\cal K}^{i}({\bf v}_{\alpha},{\bf v}_{\beta_1},{\bf v}_{\beta_2}
|\,k,q_1,q_2){\rm e}^i(\hat{\bf k},\zeta)|^{\,2}
+|\,{\cal K}^{i}({\bf v}_{\alpha},{\bf v}_{\beta_2},{\bf v}_{\beta_1}
|\,k,q_2,q_1){\rm e}^i(\hat{\bf k},\zeta)|^{\,2}
\label{eq:9u}
\end{equation}
\[
+{\rm Re}\Bigl[\Bigl(
{\cal K}^{i}({\bf v}_{\alpha},{\bf v}_{\beta_1},{\bf v}_{\beta_2}
|\,k,q_1,q_2){\rm e}^i(\hat{\bf k},\zeta)\Bigr)^{\ast}\!
\Bigl(
{\cal K}^{i}({\bf v}_{\alpha},{\bf v}_{\beta_2},{\bf v}_{\beta_1}
|\,k,q_2,q_1){\rm e}^i(\hat{\bf k},\zeta)\Bigr)\Bigr]\!
\biggr\}_{\omega=\omega_{\bf k}^t\!,\,q^0_{1,2}
={{\bf v}_{\beta_{1,2}}}\cdot\,{{\bf q}_{1,2}}}.
\]

Furthermore we restrict our consideration to limit of static color
centers, i.e. we set ${\bf v}_{\beta_1}={\bf v}_{\beta_2}=0$. (In this
case variable $l$ in Eq.\,(\ref{eq:9uuu}) coincides with the longitudinal
separation between two color static center $\beta_1$ and $\beta_2$.)
Taking into account Eq.\,(\ref{eq:4w}) and keeping in the effective
propagator $\,^{\ast}{\cal D}_{C}^{\sigma\sigma^{\prime}}
(k-q_1-q_2)v_{\alpha\sigma^{\prime}}$ transverse part only, we derive
for special case (\ref{eq:9t})
\[
{\cal K}^{i}({\bf v}_{\alpha},0,0|\,k,q_1,q_2)
{\rm e}^i(\hat{\bf k},\zeta)|_{q_1^0=q_2^0=0}=
\frac{1}{({\bf q}_1^2+\mu_D^2)}\,
\frac{1}{({\bf q}_2^2+\mu_D^2)}
\,^{\ast}{\!\Delta}^t(\omega,{\bf k}-{\bf q}_1-{\bf q}_2)
\]
\[
\times
\Bigl[\,{\rm e}^i(\hat{\bf k},\zeta)
\delta\Gamma^{i00j}(k,-q_1,-q_2,-k+q_1+q_2)v_{\alpha}^j
\Bigr]_{q_1^0=q_2^0=0},
\]
where four-gluon HTL vertex correction is defined by \cite{frenkel,pisarski}
\begin{equation}
\delta\Gamma^{i00j}(k,-q_1,-q_2-k+q_1+q_2)|_{\,q_1^0=q_1^0=0}
\label{eq:9i}
\end{equation}
\[
=-
\mu_D^2\,{\omega}\!\int\!\frac{d{\Omega}_{\bf v}}{4\pi}\,
\frac{v^iv^j}{v\cdot k}\,
\,\frac{1}{v\cdot (k-q_1) + i\epsilon}
\,\frac{1}{v\cdot (k-q_1-q_2) + i\epsilon}.
\]
Now our problem  is to show  that function (\ref{eq:9i}) contains singularity
similar to that of vertex function (\ref{eq:5e}) considered above,
associated with
zeros of function $\delta$, Eq.\,(\ref{eq:5y}). The existence of such
singularity enables us at least in principle in spirit of Section 5 to
simplify analysis of expression for energy loss\footnote{Remind that for
potential model of medium the expression (\ref{eq:9u}) defines radiation energy
loss of energetic parton $\alpha$.} (\ref{eq:9u}) and perform it to a great
extent by analytic.

Let us define an explicit form of integral on the right-hand side of
Eq.\,(\ref{eq:9i}) replacing at the moment for convenience
\begin{equation}
k\rightarrow k_1,\quad k-q_1\rightarrow k_2,\quad
k-q_1-q_2\rightarrow k_3.
\label{eq:9o}
\end{equation}
In this case the analog of expansion (\ref{eq:5e}) is
\begin{equation}
\int\!\frac{d{\Omega}_{\bf v}}{4\pi}\,
\frac{v^iv^j}{(v\cdot k_1)(v\cdot k_2)(v\cdot k_3)}=
Y_1n_1^in_1^j + Y_2n_2^in_2^j + Y_3n_3^in_3^j
\label{eq:9p}
\end{equation}
\[
+\,Z_{12}(n_1^in_2^j + n_1^jn_2^i) + Z_{13}(n_1^in_3^j + n_1^jn_3^i)
+ Z_{23}(n_2^in_3^j + n_2^jn_3^i).
\]
Here, ${\bf n}_1=[{\bf k}_2,{\bf k}_3],\;{\bf n}_2=[{\bf k}_3,{\bf
k}_1]$ and ${\bf n}_3=[{\bf k}_1,{\bf k}_2]$. Chosen
basis\footnote{If we enter vector ${\bf N}\equiv [{\bf k}_1,{\bf
n}_1] + [{\bf k}_2,{\bf n}_2] + [{\bf k}_3,{\bf n}_3]$ orthogonal to
vectors ${\bf k}_1,\,{\bf k}_2,$ and ${\bf k}_3$, then this expansion
basis should be added to tensor $N^iN^j$. However by Jacobi identity
we have ${\bf N}\equiv0$. We also can enter identity tensor
$X\delta^{ij}$in expansion (\ref{eq:9p}). As shown in Appendix B a
coefficient of expansion $X$ without loss of generality can be set
equal to zero. In this sense chosen expansion basis is complete.} is
very convenient by virtue of properties
\[
{\bf k}_s\cdot {\bf n}_l = 0,\;\, {\rm if}\;s\neq l;\quad
{\bf k}_1\cdot {\bf n}_1= {\bf k}_2\cdot {\bf n}_2={\bf k}_3\cdot {\bf n}_3.
\]
The explicit form of coefficient functions $Y_s,Z_{sl}$,
$s\!<\!l,\,s,l=1,2,3$ is defined in Appendix B. They have following
general structure
\begin{equation}
Y_s=\frac{1}{\delta^{\prime}}\,\tilde{Y}_s,\quad
Z_{sl}=\frac{1}{\delta^{\prime}}\,\tilde{Z}_{sl},
\label{eq:9a}
\end{equation}
where functions $\tilde{Y}_s$ and $\tilde{Z}_{sl}$ are given by Eqs.\,(B.1)
and (B.2), and determinant
\begin{equation}
\delta^{\prime}=
\left|
\begin{array}{lcr}
{\bf k}^2_1 & {\bf k}_1\cdot{\bf k}_2 & {\bf k}_1\cdot{\bf k}_3 \\
{\bf k}_1\cdot{\bf k}_2 & {\bf k}_2^2 & {\bf k}_2\cdot{\bf k}_3 \\
{\bf k}_1\cdot{\bf k}_3 & {\bf k}_2\cdot{\bf k}_3 & {\bf k}^2_3
\end{array}
\right|
\label{eq:9s}
\end{equation}
represents Gramian for three-dimensional vectors ${\bf k}_1,\,{\bf
k}_2,$ and ${\bf k}_3$. As known vanishing of this determinant gives
necessary and sufficient condition of complanarity of these vectors.
Note that condition (\ref{eq:5y}) also can be written in the form of
Gramian, and such in this case condition $\delta=0$ will yields
condition of collinearity of vectors ${\bf k}$ and ${\bf
q}^{\prime}$.

We return now to variables (\ref{eq:9o}) and consider equation
\begin{equation}
\delta^{\prime}\equiv
\left|
\begin{array}{lcr}
{\bf k}^2 & {\bf k}\cdot{\bf q}_1^{\prime} & {\bf k}\cdot{\bf q}_2^{\prime} \\
{\bf k}_1\cdot{\bf q}_1^{\prime} & {\bf q}_1^{\prime2}
& {\bf q}_1^{\prime}\cdot{\bf q}_2^{\prime} \\
{\bf k}_1\cdot{\bf q}_2^{\prime}
& {\bf q}_1^{\prime}\cdot{\bf q}_2^{\prime} & {\bf q}_2^{\prime 2}
\end{array}
\right|=0\,,
\label{eq:9d}
\end{equation}
where ${\bf q}_1^{\prime}\equiv {\bf k}-{\bf q}_1$,
${\bf q}_{\,2}^{\prime}\equiv {\bf k}-{\bf q}_1-{\bf q}_2$. Delta-function in the
integrand (\ref{eq:9u}) enables us to fix only
$q_{2\|}^{\prime}=\omega_{\bf k}^t/v_{\alpha}$. We choose coordinate system
as depicted in Fig.\,\ref{fig2}, axis $0Z$ is aligned with the velocity
${\bf v}_{\alpha}$. In chosen coordinate system we have
${\bf k}=(|{\bf k}|,\theta,\varphi),\;
{\bf q}_1^{\prime}=(|{\bf q}_1^{\prime}|,\vartheta,\beta_1),\;
{\bf q}_{\,2\perp}^{\prime}=(|{\bf q}_{2\perp}^{\prime}|,\pi/2,\beta_2)$, and
correspondingly integration measures are
\[
d{\bf k}={\bf k}^2d|{\bf k}|\sin\theta d\theta d\varphi,\quad
d{\bf q}_1^{\prime}={\bf q}_1^{\prime\, 2}d|{\bf q}_1^{\prime}|
\sin\vartheta d\vartheta d\beta_2,\quad
d{\bf q}_{2\perp}^{\prime}= |{\bf q}_{2\perp}^{\prime}|
d|{\bf q}_{2\perp}^{\prime}|d\beta_2.
\]
On the conditions ${\bf k}^2\neq 0$, ${\bf q}_1^{\prime\, 2}\neq 0$
equation (\ref{eq:9d}) can be presented as
\begin{equation}
(1+\chi^2)\Bigl[1-(\sin\theta\sin\vartheta\cos(\varphi-\beta_1)
+\cos\theta\cos\vartheta)^2\,\bigr]
\label{eq:9f}
\end{equation}
\[
+\,\Bigl[\,(1+\chi^2)-(\chi\sin\theta\cos(\varphi-\beta_2)+\cos\theta)^2
\,\Bigr]
+\,\Bigl[\,(1+\chi^2)-(\chi\sin\vartheta\cos(\beta_1-\beta_2)+\cos\vartheta)^2
\,\Bigr]=0.
\]
Here, $\chi\equiv v_{\alpha}|{\bf q}_{2\perp}^{\prime}|/\omega$. This
equation represents appropriate extension of Eq.\,(\ref{eq:5u}). Note
that angle $\beta_2$ will be enter everywhere as difference with
azimuth angles $\varphi$ and $\beta_1$, and therefore it can
be set equal to zero and integral is $\int d\beta_2 =2\pi$.

We can obtain simplest nontrivial solution of this equation if we set
expressions in two last square brackets on the left-hand side of
Eq.\,(\ref{eq:9f}) equal to zero. The solution have a form
\begin{equation}
\cos\varphi_{\pm}=-\,\frac{1}{\chi\sin\theta}\,
\Bigl(\cos\theta\mp\sqrt{1+{\chi}^2}\,\Bigr),\quad
\cos\beta_{1\pm}=-\,\frac{1}{\chi\sin\vartheta}\,
\Bigl(\cos\vartheta\mp\sqrt{1+{\chi}^2}\,\Bigr).
\label{eq:9g}
\end{equation}
It is not difficult to see that in deciding of (\ref{eq:9g}) the expression
in first square brackets vanishes. This solution is correspondent to the case
already considered in Section 5, and therefore performed analysis of
transition bremsstrahlung can be extended to this problem. However solution
(\ref{eq:9g}) represents very hard condition for vanishing of function
$\delta^{\prime}$. It corresponds to condition of collinearity of three
vectors ${\bf k},\;{\bf q}_1^{\prime}$ and ${\bf q}_2^{\prime}$. In general
case it is necessary to solve algebraic equation (\ref{eq:9f}),
that in notation $x\equiv\cos\varphi,\;y\equiv\cos\beta_1$, can be written in
more clear form
\[
\Bigl[\,a\Bigl(xy-\sqrt{(1-x^2)(1-y^2)}\,\Bigr)+b\,\Bigr]^{\,2}
+(c_{\theta}x + d_{\theta})^2 + (c_{\vartheta}y + d_{\vartheta})^2-3=0,
\]
where
\[
a=\sin\theta\sin\vartheta,\;b=\cos\theta\cos\vartheta,\;
c_{\theta}=\tanh\xi\sin\theta,\;c_{\vartheta}=\tanh\xi\sin\vartheta,
\]
\[
d_{\theta}=\frac{\cos\theta}{\cosh\xi},\;\quad
d_{\vartheta}=\frac{\cos\vartheta}{\cosh\xi},\quad
\xi\equiv\ln\Bigl(\chi + \sqrt{1+\chi^2}\,\Bigr),\quad\chi=\sinh\xi.
\]

At the end of this Section we briefly analyze a case of transition
bremsstrahlung in scattering of energetic parton $\alpha$ off three static
color centers, the process proceeding through five soft gluon HTL induced
vertex, as depicted on Fig.\,\ref{fig9}. In the case of radiation of transverse
\begin{figure}[hbtp]
\begin{center}
\includegraphics*[height=7cm]{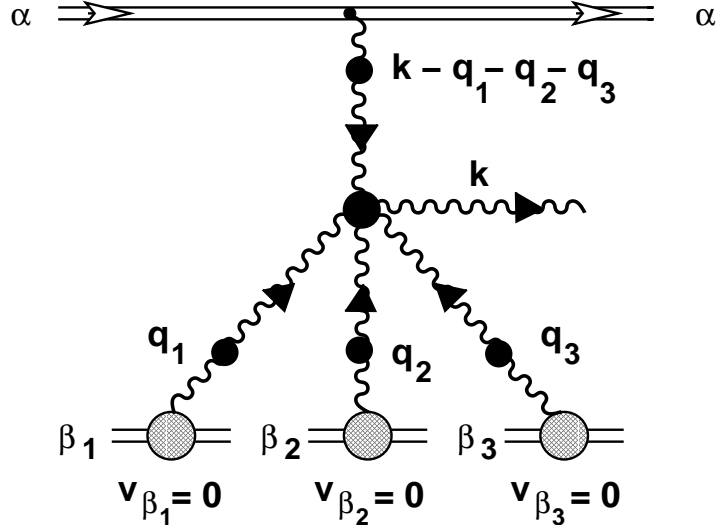}
\end{center}
\caption{\small Transition bremsstrahlung induced by
$\delta\Gamma_{5g}$ vertex HTL correction.}
\label{fig9}
\end{figure}
soft gluon, calculation of matrix element of this process is reduced to 
calculation of integral in the form
\[
\int\!\frac{d{\Omega}_{\bf v}}{4\pi}\,
\frac{v^iv^j}{(v\cdot k_1)(v\cdot k_2)(v\cdot k_3)(v\cdot k_4)},
\]
where $k_1=k,\;k_2=k-q_1,\;k_3=k-q_1-q_2$ and $k_4=k-q_1-q_2-q_3$. According
reasoning of Frenkel and Taylor \cite{frenkel}, this integral represents pure
spatial part of Lorentz-covariant tensor of 2-rank
\[
U^{\mu\nu}=
\int\!\frac{d{\Omega}_{\bf v}}{4\pi}\,
\frac{v^{\mu}v^{\nu}}{(v\cdot k_1)(v\cdot k_2)(v\cdot k_3)(v\cdot k_4)}.
\]
This integral can be presented in the form of expansion in basis
\begin{equation}
U^{\mu\nu}=\sum\limits_{s=1}^4\!
A_sn_s^{\mu}n_s^{\nu} +\sum\limits_{s<l}\!B_{s\,l}\,(n_s^{\mu}n_l^{\nu} +
n_l^{\mu}n_s^{\nu}),\quad s,\,l=1,2,3,4\,,
\label{eq:9h}
\end{equation}
where we enter 4-vectors
\[
n_1^{\mu}=\varepsilon^{\mu\nu\lambda\sigma}
k_2^{\nu}k_3^{\lambda}k_4^{\sigma},\quad
n_2^{\mu}=\varepsilon^{\mu\nu\lambda\sigma}
k_3^{\nu}k_4^{\lambda}k_1^{\sigma},\quad
n_3^{\mu}=\varepsilon^{\mu\nu\lambda\sigma}
k_4^{\nu}k_1^{\lambda}k_2^{\sigma},\quad
n_4^{\mu}=\varepsilon^{\mu\nu\lambda\sigma}
k_1^{\nu}k_2^{\lambda}k_3^{\sigma},
\]
and $\varepsilon^{\mu\nu\lambda\sigma}$ is absolute antisymmetric tensor of
4-rank. In Appendix C calculation of coefficients $A_s$ and $B_{s\,l},\;s<l$
are given. They have following general structure analogous to (\ref{eq:9a})
\begin{equation}
A_s=\frac{1}{\Delta^{\prime\prime}}\,\tilde{A}_s,\quad
Z_{sl}=\frac{1}{\Delta^{\prime\prime}}\,\tilde{B}_{s\,l},
\label{eq:9j}
\end{equation}
where the functions $\tilde{A}_s$ and $\tilde{B}_{sl}$ are given by
expressions (C.1) and (C.2), correspondingly, and
\begin{equation}
\Delta^{\prime\prime}=(\varepsilon^{\mu\nu\lambda\sigma}
k_{1\mu}k_{2\nu}k_{3\lambda}k_{4\sigma})^2=-
\left|
\begin{array}{lccr}
k_1^2 & k_1\cdot k_2 & k_1\cdot k_3 & k_1\cdot k_4 \\
k_1\cdot k_2 & k_2^2 & k_2\cdot k_3 & k_2\cdot k_4 \\
k_1\cdot k_3 & k_2\cdot k_3 & k_3^2 & k_3\cdot k_4 \\
k_1\cdot k_4 & k_2\cdot k_4 & k_3\cdot k_4 & k_4^2
\end{array}
\right|
\label{eq:9k}
\end{equation}
represents Gramian for 4-vectors $k_1,\,k_2,\,k_3$ and $k_4$. Zeros
of determinant (\ref{eq:9k}) define singular surface in a space of
kinematic variables, on which it should be defined all functions in
matrix element of this transition bremsstrahlung.

If we consider  medium induced bremsstrahlung of higher-order than in
Fig.\,\ref{fig9}, i.e. one in scattering off four, five etc. static
color centers, then we can expect an absence of general singular
factor as it occurs in coefficient functions (\ref{eq:5r}), (\ref{eq:9a})
and (\ref{eq:9j}) by virtue of the fact that it is impossible to
define nontrivial Gramian for five and more 4-vectors. Therefore we
can suppose that these higher processes of transition bremsstrahlung
will be strongly suppressed in comparison with lower-order processes (i.e.
containing gluon vertices with smaller number of soft external legs)
and thus they are not of our interest.

\section{\bf Summary and outlook}
\setcounter{equation}{0}

In this work in the framework of semiclassical approximation we have
proposed an approach to calculation of radiation intensity of
high-energy parton travelling through hot QCD medium. The deriving
of some effective color currents generating bremsstrahlung
processes of arbitrary number of soft gluons for scattering
energetic parton off arbitrary number of thermal particles of medium
is a great moment for this consideration.

A suggested scheme for calculation of these effective currents
tacitly implies by virtue of initial dynamical equations that hot QCD
medium represents a weakly coupled system $(g\ll 1)$, allowing to use
perturbative approach. However theoretical analysis of
experimental data, especially concerning with collective flow,
unambiguously shows that properties of `new matter' produced at RHIC
experiments for moderate temperatures $(1\div3)T_c$ substantially
different from the expectation of a weakly coupled colored plasma
(see the last reviews on this subject \cite{gyulassy7, shuryak2,
heinz,thoma1}). A major revision of proposed scheme to calculation of
radiative energy loss can be one of the consequences of such
conclusion. For example within standard Gyulassy-Wang model multiple
scattering in medium is considered as totality of subsequent acts
of pair interaction of particles, and in particular an influence of
simultaneous collision of several (more then two) particles on
formation of bremsstrahlung spectrum is neglected. It is clear
that effects of
such type can play an important role in the processes of gluon
multiplication in rather dense scattering medium (e.g., liquid type),
when plasma parameter $\Gamma$ becomes more or of the order of one. The
attempt of taking into account of such effects in dense matter was
made in work \cite{koshelkin} for the case of bremsstrahlung of fast
electric charged particles. Here, diagrammatic formalism for
two-particle Green functions was used. It was shown that in the case of
radiation in sufficiently dense medium in a distant long wavelength region of
spectrum coherent multiple scattering of particles results in
(supplementary to LPM effect) suppression of the yield of
bremsstrahlung photons. We can expect that similar phenomenon has to
take place and for the case of
dense\footnote{There is some ambiguity
in definition of $s$QGP, obeying the classical statistics, in terms of
plasma parameter $\Gamma$. The ambiguity of such a kind was first
noted in work \cite{litim}. The initial definition of $\Gamma$ is
\begin{equation}
\Gamma = \frac{1}{n r_D^3} ,
\label{eq:10q}
\end{equation}
where $n = \sum_{\beta} n_{\beta}$ is a mean density of system and
$r_D$ is Debye length, that in the case of the classical non-Abelian
plasma reads
\begin{equation}
r_D^2 = \frac{k_B T}{4 \pi g^2\!\!
\sum\limits_{\beta=q,\,\bar{q},\,g}\!\! C_2^{(\beta)}n_{\beta}},
\label{eq:10w}
\end{equation}
$k_B$ is a Boltzmann constant. In the theory of usual plasma strong
coupling regime is defined by condition $\Gamma \geq 1$. In the case
of QED interaction the coupling is weak, and thus the condition
$\Gamma \geq 1$ can be realized only for large value of plasma
density. For QCD interaction a situation is more complicated. As it
is seen from Eqs.\,(\ref{eq:10q}), (\ref{eq:10w}) the last condition
can fulfilled in general and for a rarefied plasma, but with the
strong coupling between thermal particles and for very dense plasma
but with weak coupling. For the quantum non-Abelian plasma, when
\[
r_D^2 \sim \frac{1}{g^2}\,\biggl(\frac{\hbar c}{k_B T}\biggr)^{\!2},
\quad n \sim\! \biggl(\frac{k_B T} {\hbar c}\biggr)^{\!3}
\]
such ambiguity does not exist. Here, the plasma parameter is more or
equals one if and only if $ g\geq 1$ \cite{litim}.

To avoid the ambiguity in definition of $s$QGP for classical statistics we
will assume the medium is dense and the coupling is strong.} strongly
coupled quark-gluon plasma ($s$QGP) for description of that a number of
approaches is already proposed \cite{letessier,shuryak3,brown,peshier}.

The approach to calculation of effective currents suggested in this
work (and in Papers I and II also) can be extended to the case of
strongly coupled QGP. The use of collisionless Blaizot-Iancu equation
(I.3.3) correctly describing behavior of QGP in the weakly coupling
regime and at soft momentum scale of plasma density fluctuations is
one of the most key moments for the derivation of effective color currents. 
Therefore as a first step of construction
of the radiation energy loss theory in $s$QGP this kinetic equation
is necessary to be extended in a way. As one of the leading ideas
representations developed in attempt to describe thermodynamical
properties, the transport coefficients and the dynamical correlations
of strongly coupled usual plasma, can be used.

The extensive research  of properties of usual plasma, when plasma
parameter $\Gamma$ changes within huge region: from slowly
coupling regime $\Gamma \ll 1$ and up to $\Gamma \sim 1000$ were
undertaken more than 20 years ago. In particular, for a strongly
coupled one-component plasma Vieillefosse and Hansen \cite{hansen}
and Wallenborn and Baus \cite{baus} have advanced theoretical
estimations of the shear $\eta$ and the bulk  viscosities on the
basis of hydrodynamic and kinetic-theoretical arguments,
correspondingly. The two calculations are in agreement shown that
the transverse viscosity $\eta$ (exactly its dimensionless
combination) exhibits a minimum as a function of $\Gamma$ (for
values $\Gamma \sim 10\div20$) and $\zeta$ is negligible compared
to $\eta$. The value $\eta$ in a point of minimality is more
little then a value for the weakly coupled plasma, when $\Gamma
\ll 1$. This very important and striking observation was confirmed
through molecular-dynamics computer simulations \cite{bernu}. The
position of the minimum is, however, rather sensitive to the
various approximations.

Furthermore Tanaka and Ichimaru \cite{tanaka,ichimaru} have developed
a new kinetic theory of dynamic correlations for a strongly coupled,
classical one-component plasma within the generalized viscoelastic
formalism \cite{frenkel1}. The strong Coulomb-coupling effects beyond
random phase approximation are described in the present theory
through the dynamic local-field correction $G({\bf k},\omega)$
introduced via linear-response relation between the external
potential $\varphi_{\rm ext}({\bf k}, \omega)$ and the induced
density fluctuation $\delta n({\bf k},\omega)$. Making use 
a fully convergent kinetic equation originally proposed by them, 
where higher-order scattering
processes are included through the factor $1 - G({\bf k}, \omega)$ in
integrand of collision term, Tanaka, and Ichimaru have established a
self-consistent scheme to calculate $\eta$ and $G({\bf k}, \omega)$
simultaneously. Their predictions showed a good agreement with early
obtained results of different approaches \cite{hansen, baus, bernu}.

Performed researches for strongly coupled usual plasma suggest that
in a case of classical $s$QGP we can also expect an existence of
minimal value of the shear viscosity for some finite value of QCD
plasma parameter  (\ref{eq:10q}) and this value can be appreciably smaller
than value following from perturbative estimations \cite{baym, arnold1}
for weakly coupled QGP. It is represented extremely important by
virtue of very small value of $\eta$ for a new QCD matter discovered
at RHIC experiments \cite{teaney}. The requirement of minimality of
shear viscosity has to lead to definite fixing of generalized
collision term for kinetic equation, that in weakly coupled limit
must turn to kinetic equation (I.3.3). Having in hand such an
equation we can further develop an extended scheme for calculations of
color effective currents generating bremsstrahlung processes (and
radiative energy losses produced by them) of
high-energy partons in the $s$QGP. The exact consideration of this
approach will be the subject of a further research.

\section*{\bf Acknowledgments}
Yu.M. is grateful for inspiring and helpful discussion to Edward
Shuryak, Larry McLerran, Dmitri Kharzeev, Eugene Levin, Alfred
Mueller, Ismail Zahed, Kirill Tuchin, Peter Petreczky, Derek Teaney
and The Nuclear Physics Groups at Brookhaven National Laboratory and
Stone Brook University for warm hospitality. This work was supported
by the Russian Foundation for Basic Research (project no 03-02-16797).

\newpage
\section*{Appendix A}
\setcounter{equation}{0}

Here, we give complete expression for partial coefficient function
$K_{\mu\mu_1}({\bf v}_{\alpha},{\bf v}_{\beta};{\bf b}\vert\,k,k_1)$
defining bremsstrahlung of two soft gluons in scattering of energetic
color particle $\alpha$ off hard thermal parton $\beta$
$$
K_{\mu\mu_1}({\bf v}_{\alpha},{\bf v}_{\beta};{\bf b}\vert\,k,k_1)=
\int\!\biggl\{
\frac{v_{\alpha\mu}v_{\alpha\nu}}{v_{\beta}\cdot(q-k_1)}
\,^{\ast}{\cal D}_C^{\nu\nu^{\prime}}\!(q-k_1)
K_{\nu^{\prime}\mu_1}({\bf v}_{\beta}|\,q-k_1,k_2)
$$
$$
+\,\frac{v_{\alpha\mu}v_{\alpha\mu_1}}
{(v_{\alpha}\cdot(q-k_1))(v_{\alpha}\cdot q)}
\,(v_{\alpha\nu}\!\,^{\ast}{\cal D}_C^{\nu\nu^{\prime}}\!(q)
v_{\beta\nu^{\prime}})
\eqno{({\rm A}.1)}
$$
$$
+\,\frac{v_{\beta\mu}v_{\beta\mu_1}}
{(v_{\beta}\cdot k_1))(v_{\beta}\cdot(k+k_1-q))}
\,(v_{\beta\nu}\!\,^{\ast}{\cal D}_C^{\nu\nu^{\prime}}\!(k+k_1-q)
v_{\alpha\nu^{\prime}})
$$
$$
+\,^{\ast}\Gamma_{\mu\nu\lambda}(k,k_1-q,-k-k_1+q)
\,^{\ast}{\cal D}_C^{\nu\nu^{\prime}}\!(q-k_1)
K_{\nu^{\prime}\mu_1}({\bf v}_{\beta}|\,q-k_1,k_1)
\,^{\ast}{\cal D}_C^{\lambda\lambda^{\prime}}\!(k+k_1-q)
v_{\alpha\lambda}
$$
$$
+\,^{\ast}\Gamma_{\mu\mu_1\lambda}(k,k_1,-k-k_1)
\,^{\ast}{\cal D}_C^{\lambda\lambda^{\prime}}\!(k+k_1)
{\cal K}_{\lambda^{\prime}}({\bf v}_{\alpha},{\bf v}_{\beta}|\,k+k_1,q)
$$
$$
-\,^{\ast}\Gamma_{\mu\nu\lambda\mu_1}(k,-k-k_1+q,-q,k_1)
\,^{\ast}{\cal D}_C^{\nu\nu^{\prime}}\!(k+k_1-q)v_{\beta\nu^{\prime}}
\,^{\ast}{\cal D}_C^{\lambda\lambda^{\prime}}\!(q)
v_{\beta\lambda^{\prime}}
\biggr\}
$$
$$
\times
\,{\rm e}^{-i{\bf b}\cdot\,{\bf q}}\,
\delta(v_{\alpha}\cdot (k+k_1-q))\delta(v_{\beta}\cdot q)dq.
$$
Here, the functions $K_{\mu\mu_1}({\bf v}|\,k,k_1)$  and 
${\cal K}_{\mu}({\bf v}_{\alpha},{\bf v}_{\beta}|\,k,q)$ are defined by
Eq.\,(\ref{eq:2p}) and Eq.\,(\ref{eq:3u}), correspondingly. The
graphic interpretation of various terms in this expression is
presented on Fig.\,\ref{fig10}. Remind that graphs, where radiation
\begin{figure}[hbtp]
\begin{center}
\includegraphics[width=0.95\textwidth]{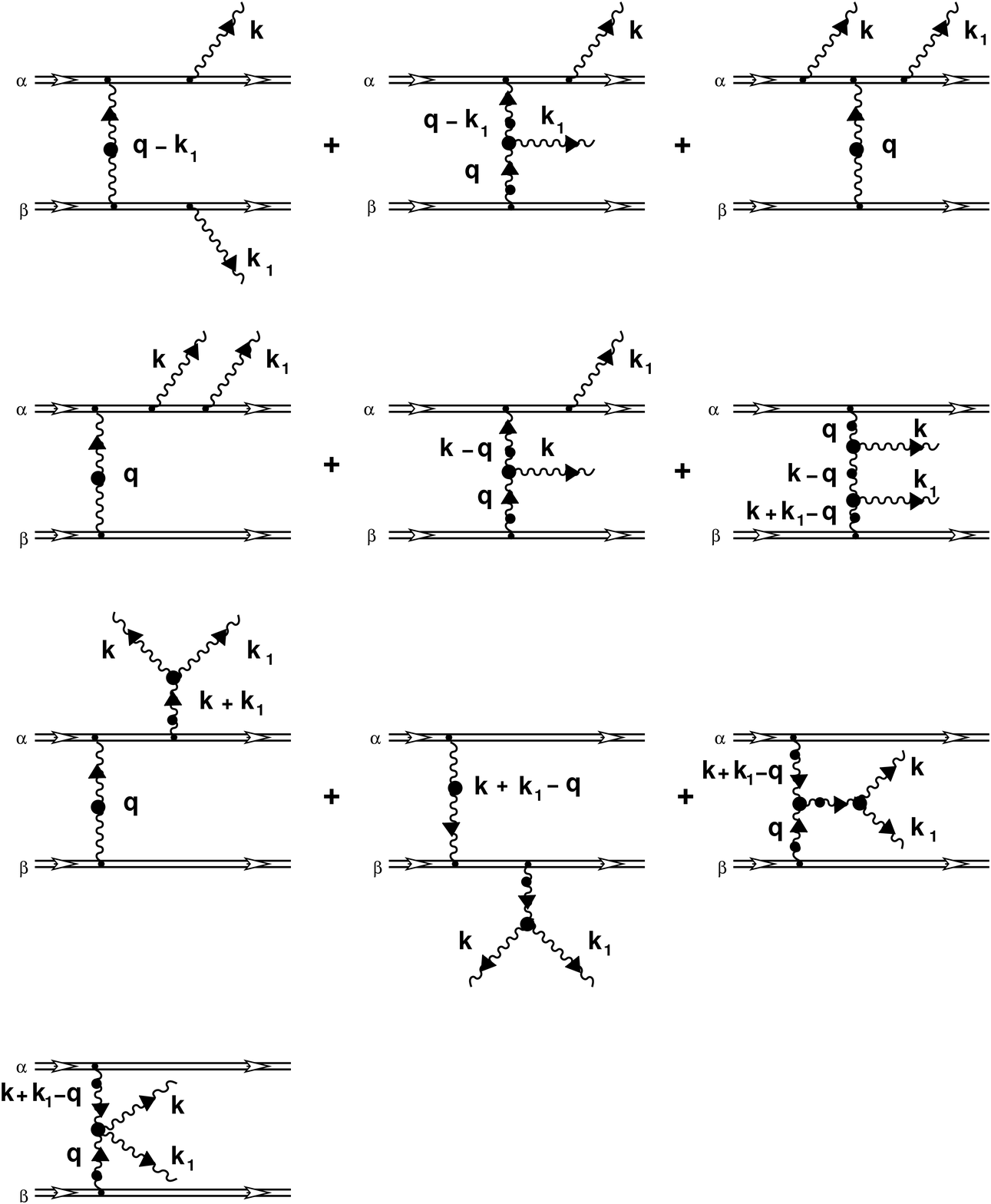}
\end{center}
\caption{\small Two soft gluon bremsstrahlung.}
\label{fig10}
\end{figure}
lines from hard lines are located a prior one-gluon exchange, are not given
here.

\newpage

Below we adduce a complete expression for matrix element 
${\cal K}^{(A)ii_1}{\rm e}^i{\rm e}^{i_1}$ in the high-frequency and 
small-angle approximations. This matrix element defines bremsstrahlung of two
gluons forming odd state system (in the c.m.s. of gluons) with respect to space 
inversion 
$$
{\cal K}^{(A)ii_1}({\bf v}_{\alpha},0|\,k,k_1,q)
{\rm e}^i(\hat{\bf k},\zeta){\rm e}^{i_1}(\hat{\bf k}_1,\zeta_1)|_{\,q_0=0}
$$
$$
\simeq\frac{2}{{\bf q}_{\perp}+\mu_D^2+\displaystyle\frac{1}{v_{\alpha}^2}
\left(\frac{1}{\tau_f}+\frac{1}{\tau_{\!1f}}\right)^{\!2}}\,
\Biggl[\,\Biggl(\frac{\tau_f-\tau_{1f}}{\tau_f+\tau_{1f}}\Biggr)
\frac{{\bf e}_{\perp}\cdot{\bf k}_{\perp}}
{{\bf k}_{\perp}^2 + m_g^2 + x^2M^2}\,\,
\frac{{\bf e}_{1\perp}\cdot{\bf k}_{1\perp}}
{{\bf k}_{1\perp}^2 + m_g^2 + x_1^2M^2}
$$
$$
+\,\Biggl\{
\,^{\ast}\!\Delta^{t}(q-k)\,
\frac{({\bf e}_{\perp}\cdot({\bf k}-{\bf q})_{\perp})
({\bf e}_{1\perp}\cdot{\bf k}_{1\perp})}
{{\bf k}_{1\perp}^2 + m_g^2 + x_1^2M^2}\,-
\,^{\ast}\!\Delta^{t}(q-k_1)\,
\frac{({\bf e}_{1\perp}\cdot({\bf k}_1-{\bf q})_{\perp})
({\bf e}_{\perp}\cdot{\bf k}_{\perp})}
{{\bf k}_{\perp}^2 + m_g^2 + x^2M^2} 
\Biggr\}
$$
$$
+\bigg\{\!\!\,^{\ast}\!\Delta^{t}(q-k_1)
\biggl[\Bigl(
{\bf e}_{\perp}\cdot\Bigl({\bf k}_1-\frac{x_1}{x}{\bf k}
-{\bf q}\Bigr)_{\!\perp}\Bigr)
({\bf e}_{1\perp}\cdot({\bf k}_1-{\bf q})_{\perp})
+\Bigl({\bf e}_{1\perp}\cdot\Bigl({\bf k}_1-\frac{x_1}{x}{\bf k}
-{\bf q}\Bigr)_{\!\perp}\Bigr)
({\bf e}_{\perp}\cdot{\bf k}_{\perp})
$$
$$
+\,x_1\,\frac{({\bf e}_{\perp}\cdot{\bf e}_{1\perp})}{(x+x_1)^2}\,
\Bigl[\,x_1{\bf k}_{\perp}^2-x({\bf k}_1-{\bf q})^2_{\perp}
-(x-x_1)({\bf k}_{\perp}\cdot({\bf k}_1-{\bf q})_{\perp})\Bigr]\biggr]
$$
$$
+\,({\bf k}_{\perp}\rightleftharpoons {\bf k}_{1\perp},\,
{\bf e}_{\perp}\rightleftharpoons{\bf e}_{1\perp},\,
x\rightleftharpoons x_1)\biggr\}\,^{\ast}\!\Delta^{t}(k+k_1-q)
\eqno{({\rm A}.2)}
$$
$$
+\,\,^{\ast}\!\Delta^{t}(k+k_1)\Biggl(
\frac{\omega+\omega_1}{\tau_f}+\frac{\omega+\omega_1}{\tau_{\!1f}}
\Biggr)^{\!\!-1}\Biggr\{\Biggl(1+\frac{x}{x_1}\Biggr)
\Bigl({\bf e}_{\perp}
\cdot\Bigl({\bf k}_{1}-\frac{x_1}{x}{\bf k}\Bigr)_{\!\perp}\Bigr)
({\bf e}_{1\perp}\cdot{\bf k}_{1\perp})
$$
$$
-\,\Biggl(1+\frac{x_1}{x}\Biggr)
\Bigl({\bf e}_{1\perp}
\cdot\Bigl({\bf k}-\frac{x}{x_1}{\bf k}_{1}\Bigr)_{\!\perp}\Bigr)
({\bf e}_{\perp}\cdot{\bf k}_{\perp})+
\frac{({\bf e}_{\perp}\cdot{\bf e}_{1\perp})}{x+x_1}
\Bigl[\,x_1{\bf k}_{\perp}^2 - x{\bf k}_{1\perp}^2 -
(x-x_1)({\bf k}_{\perp}\cdot{\bf k}_{1\perp})\Bigr]
\!\Biggr\}
$$
$$
+\,2\,^{\ast}\!\Delta^{t}(k+k_1)\,^{\ast}\!\Delta^{t}(k+k_1-q)
\biggr\{\Bigl({\bf e}_{\perp}
\cdot\Bigl({\bf k}_{1}-\frac{x_1}{x}\,{\bf k}\Bigr)_{\!\perp}\Bigr)
\Bigl({\bf e}_{1\perp}\cdot\Bigl(\Bigl(1+\frac{x}{x_1}\Bigr){\bf k}_{1}
-{\bf q}\Bigr)_{\!\perp}\Bigr)
$$
$$
-\,\Bigl({\bf e}_{1\perp}
\cdot\Bigl({\bf k}-\frac{x}{x_1}\,{\bf k}_{1}\Bigr)_{\!\perp}\Bigr)
\Bigl({\bf e}_{\perp}\cdot\Bigl(\Bigl(1+\frac{x_1}{x}\Bigr){\bf k}
-{\bf q}\Bigr)_{\!\perp}\Bigr)
$$
$$
+\,\frac{({\bf e}_{\perp}\cdot{\bf e}_{1\perp})}{x+x_1}\,
\Bigl[\,x_1{\bf k}_{\perp}^2 - x{\bf k}_{1\perp}^2 -
(x-x_1)({\bf k}_{\perp}\cdot{\bf k}_{1\perp}) -
((x_1{\bf k} - x{\bf k}_{1})_{\perp}\cdot{\bf q}_{\perp})\Bigr]
\biggr\}\Biggr],
$$
where
$$
^{\ast}\!\Delta^{t}(k+k_1)\simeq
-\frac{1}{({\bf k}+{\bf k}_1)^2_{\perp} + m_g^2 
-\displaystyle\Biggl[\Biggl(1+\frac{x_1}{x}\Biggr)
({\bf k}_{\perp}^2+m_g^2)
+\Biggl(1+\frac{x}{x_1}\Biggr)
({\bf k}_{1\perp}^2+m_g^2)\Biggr]}\,.
$$
The functions $\tau_f$, $\tau_{1f}$ are defined by Eq.\,(\ref{eq:8i}) and
the propagators $\,^{\ast}\!\Delta^{t}(q-k)$, 
$\,^{\ast}\!\Delta^{t}(q-k_1)$ and $^{\ast}\!\Delta^{t}(k+k_1-q)$ are defined
by Eqs.\,(\ref{eq:8i}) and (\ref{eq:8a}), correspondingly.



\newpage
\section*{Appendix B}
\setcounter{equation}{0}

The coefficients $Y_s,\;s=1,2,3$ in the expansion of (\ref{eq:9p}) are defined
by contracting with $k_s^ik_s^j$. Such, for example a contracting with
$k_1^ik_1^j$ with use of identity
$$
({\bf v}\cdot{\bf k}_1)^2=\omega_1^2-2\omega_1(v\cdot k_1) + (v\cdot k_1)
(\omega_1-{\bf v}\cdot{\bf k}_1)
$$
and formulas (3.24), (3.30) and (4.15) from \cite{frenkel} yields,
$$
\delta^{\prime}\,Y_1=\tilde{Y}_1\equiv\omega_1^2[\,a\omega_1 + b\,\omega_2
+ c\,\omega_3] - \omega_1M(k_2,k_3)
\eqno{({\rm B}.1)}
$$
$$
+\,
\frac{1}{{\bf n}_1^2}\,\Bigl\{\Bigl(\omega_2({\bf n}_1\cdot {\bf n}_2)+
\omega_3({\bf n}_1\cdot {\bf n}_3)\Bigr)M(k_2,k_3)
-({\bf n}_1\cdot {\bf n}_2)L(k_3)
-({\bf n}_1\cdot {\bf n}_3)L(k_2)
\Bigr\},
$$
where $\delta^{\prime}$ is defined by Eq.\,(\ref{eq:9s}) and
$$
a\equiv a(k_1,k_2,k_3)=\frac{\Delta_1}{\Delta^{\prime}},\quad b\equiv
b(k_1,k_2,k_3)=\frac{\Delta_2}{\Delta^{\prime}},\quad c\equiv
c(k_1,k_2,k_3)=\frac{\Delta_3}{\Delta^{\prime}},
\eqno{({\rm B}.2)}
$$
$$
\Delta_1=\left|
\begin{array}{lcr}
M(k_2,k_3) & k_1\cdot k_2 & k_1\cdot k_3 \\
M(k_1,k_3) & k_2^2 & k_2\cdot k_3 \\
M(k_1,k_2) & k_2\cdot k_3 & k_3^2
\end{array}
\right|\,,\quad
\Delta_2=\left|
\begin{array}{lcr}
k_1^2 & M(k_2,k_3) & k_1\cdot k_3 \\
k_1\cdot k_2 & M(k_1,k_3) & k_2\cdot k_3 \\
k_1\cdot k_3 & M(k_1,k_2) & k_3^2
\end{array}
\right|\,,
$$
\vspace{0.5cm}
$$
\Delta_3=\left|
\begin{array}{lcr}
k_1^2 & k_1\cdot k_2 & M(k_2,k_3) \\
k_1\cdot k_2 &k_2^2 & M(k_1,k_3)  \\
k_1\cdot k_3 & k_2\cdot k_3 & M(k_1,k_2)
\end{array}
\right|\,,\quad
\Delta^{\prime}=\left|
\begin{array}{lcr}
k_1^2 & k_1\cdot k_2 & k_1\cdot k_3 \\
k_1\cdot k_2 &k_2^2 & k_2\cdot k_3  \\
k_1\cdot k_3 & k_2\cdot k_3 & k_3^2
\end{array}
\right|\,,
\hspace{0.9cm}
$$
and functions $M(k,k_1),\,L(k)$ are defined by Eq.\,(\ref{eq:5t}). The
coefficients $Y_2$ and $Y_3$ are obtained in a similar way. They can be
obtained from ({\rm B}.1) by cyclic permutation of 1,2,3.

The coefficients $Z_{sl},\,s<l$ are defined by contracting (\ref{eq:9p})
with $k_s^ik_l^j$. For example, a contracting with $k_1^ik_2^j$ yields
an expression for $Z_{12}$
$$
\delta^{\prime}Z_{12}=\tilde{Z}_{12}\equiv
\omega_1\omega_2[\,a\omega_1 + b\,\omega_2 + c\,\omega_3] -
\omega_1M(k_1,k_3)-\omega_2M(k_2,k_3)+L(k_3).
\eqno{({\rm B}.3)}
$$
The remaining coefficients $Z_{13}$ and $Z_{23}$ follow from ({\rm
B}.3) by cyclic permutation of 1,\,2,\,3.

Now we add a term $X\delta^{ij}$ to the right-hand side of (\ref{eq:9p}).
In this case instead of relations (B.1) and (B.3) we will have
$$
\delta^{\prime}Y_s + {\bf k}_s^2X=\tilde{Y}_s,\quad
\delta^{\prime}Z_{ls}+({\bf k}_s\cdot{\bf k}_l)X=\tilde{Z}_{ls},\quad
s<l;\;s,l=1,2,3.
\eqno{({\rm B}.4)}
$$
The equation for $X$ is derived by taking of trace (\ref{eq:9p})
$$
\tilde{X}\equiv
\int\!\frac{d{\Omega}_{\bf v}}{4\pi}\,
\frac{1}{(v\cdot k_1)(v\cdot k_2)(v\cdot k_3)}=
\,a\omega_1 + b\,\omega_2 + c\,\omega_3
\eqno{({\rm B}.5)}
$$
$$
=3X+\sum\limits_{s=1}^{3}{\bf n}_s^2Y_s + 2\sum\limits_{s<l}
({\bf n}_s\cdot {\bf n}_l)Z_{sl}.
$$
If we substitute (B.4) into the rightmost expression in (B.5) and collect
similar terms, then we will have an equation for $X$
$$
\tilde{X}=\frac{1}{\delta^{\prime}}\,X
\biggl[\,3\delta^{\,\prime} - \Bigl(
\sum\limits_{s=1}^{3}{\bf k}_s^2{\bf n}_s^2
+ 2\sum\limits_{s<l}({\bf k}_s\cdot {\bf k}_l)
({\bf n}_s\cdot {\bf n}_l)\Bigl)\biggr]
+\frac{1}{\delta^{\prime}}
\biggl[\,
\sum\limits_{s=1}^{3}{\bf n}_s^2\tilde{Y}_s + 2\sum\limits_{s<l}
({\bf n}_s\cdot {\bf n}_l)\tilde{Z}_{sl}\biggr].
\eqno{({\rm B}.6)}
$$
Let us consider an expression in square brackets in the first term on
the right-hand side of Eq.\,(B.6). By virtue of the fact that scalar
production represents algebraical adjunct to element $(ij)$ in Gramian
(\ref{eq:9s}), then
$$
{\bf k}_s^2{\bf n}_s^2 + 2\sum\limits_{l\neq s}({\bf k}_s\cdot {\bf k}_l)
({\bf n}_s\cdot {\bf n}_l) = \delta^{\prime} \quad
(\mbox{no summation over $s$!})
\eqno{({\rm B}.7)}
$$
for all $s=1,2,3$. Therefore a factor after $X$ exactly vanishes. Let us show
that the following identity is true
$$
\tilde{X}=
\frac{1}{\delta^{\prime}}
\biggl[\,
\sum\limits_{s=1}^{3}{\bf n}_s^2\tilde{Y}_s + 2\sum\limits_{s<l}
({\bf n}_s\cdot {\bf n}_l)\tilde{Z}_{sl}\biggr].
\eqno{({\rm B}.8)}
$$
If one substitute instead of $\tilde{Y}_s$ and $\tilde{Z}_{sl}$ their
integral representation, then the right-hand side of Eq.\,(\ref{eq:6i})
can be written as
$$
\frac{1}{\delta^{\prime}}
\int\!\frac{d{\Omega}_{\bf v}}{4\pi}\,
\frac{\biggl(\,\sum\limits_{s=1}^{3}{\bf n}_s({\bf v}\cdot {\bf k}_s)\biggr)^2}
{(v\cdot k_1)(v\cdot k_2)(v\cdot k_3)}\,.
\eqno{({\rm B}.9)}
$$
Let us expand unit vector ${\bf v}$ in basis $\{{\bf n}_s\}$:
$$
{\bf v}=\sum\limits_{s=1}^3\kappa_s{\bf n}_s,
$$
where $\kappa_s$ are coefficients of expansion. Then by the fact of
independence of scalar product ${\bf k}_s\cdot {\bf n}_s$ from $s$ and
condition ${\bf v}^2=1$, we will have
$$
\biggl(\,\sum\limits_{s=1}^{3}{\bf n}_s({\bf v}\cdot {\bf k}_s)\biggr)^2=
\biggl(\,\sum\limits_{s=1}^{3}\kappa_s{\bf n}_s({\bf k}_s\cdot {\bf n}_s)
\biggr)^2
=\delta^{\prime}\biggl(\,\sum\limits_{s=1}^3\kappa_s{\bf n}_s\biggr)^2=
\delta^{\prime}.
$$
Comparing (B.9) with (B.5) we see that equality (B.8) is really true.
Incidentally its correctness can be shown by direct substitution of the
functions (B.1), (B.3) and (B.5), and by use of equality\footnote{
This equality is generalization of 
$
\Delta=\delta-(\omega_1{\bf k}_2-\omega_2{\bf k}_1)^2
$
defined for functions $\Delta=k_1^2k_2^2-(k_1\cdot k_2)^2,\;
\delta = {\bf k}_1^2{\bf k}_2^2-({\bf k}_1\cdot {\bf k}_2)^2$
(Eq.\,(\ref{eq:5t})).}
$$
\Delta^{\prime}=
({\bf n}_1\omega_1+{\bf n}_2\omega_2 + {\bf n}_3\omega_3)^2
-\delta^{\prime}.
$$
Such we see that without loss of generality one can set coefficient of expansion
$X$ equal to zero, although in particular calculations it may be more
convenient to use its explicit value.


\newpage
\section*{Appendix C}
\setcounter{equation}{0}

The expression for coefficients $A_s$ in the expansion (\ref{eq:9h}) is
obtained by contracting with $k_s^{\mu}k_s^{\nu},$ where now $s=1,2,3,4$ and
using properties
$$
k_s\cdot n_l=0,\;{\rm for}\;s\neq l,\quad
(k_s\cdot n_s)^2=\Delta^{\prime\prime},
$$
where $\Delta^{\prime\prime}$ is defined by Eq.\,(\ref{eq:9k}). Thus for $A_1$
we derive equation
$$
\Delta^{\prime\prime}A_1=\tilde{A}_1\equiv
\int\!\frac{d{\Omega}_{\bf v}}{4\pi}\,
\frac{(v\cdot k_1)}{(v\cdot k_2)(v\cdot k_3)(v\cdot k_4)}
=a_1(k_1\cdot k_2) + b_1(k_1\cdot k_3) +c_1(k_1\cdot k_4).
\eqno{({\rm C}.1)}
$$
Here, $a_1\equiv a_1(k_2,k_3,k_4),\;b_1\equiv b_1(k_2,k_3,k_4),\;
c_1\equiv c_1(k_2,k_3,k_4)$ and functions $a_1,\,b_1$ and $c_1$ are
defined by Eq.\,(B.2). The other coefficients $A_2,\,A_3$ and $A_4$
are calculated in similar way.

The coefficients $B_{sl}$ are derived by contracting (\ref{eq:9h}) with
$k_s^{\mu}k_l^{\nu}$. Here, we will have, for example, for $B_{12}$
$$
\Delta^{\prime\prime}B_{12}=\tilde{B}_{12}\equiv
\int\!\frac{d{\Omega}_{\bf v}}{4\pi}\,
\frac{1}{(v\cdot k_3)(v\cdot k_4)}
=M(k_3,k_4).
\eqno{({\rm C}.2)}
$$

Now we add term $Cg^{\mu\nu}$ to the right-hand side of (\ref{eq:9h}). In this
case instead of (C.1) and (C.2), we will have equations
$$
\Delta^{\prime\prime}A_s+Ck_s^2=\tilde{A}_s,\quad
\Delta^{\prime\prime}B_{sl} + C(k_s\cdot k_l)
=\tilde{B}_{12},\;s<l;\;s,l=1,2,3,4.
\eqno{({\rm C}.3)}
$$
The equation for $C$ is defined by taking the trace (\ref{eq:9h}) and allowing
for $v^2=0$. Substituting the coefficients $A_s$ and $B_{sl}$ in obtained
equation and collecting similar terms, we derive desired equation for
coefficient $C$
$$
-\frac{1}{\Delta^{\prime\prime}}
\biggl[\,
\sum\limits_{s=1}^{4}n_s^2\tilde{A}_s + 2\sum\limits_{s<\,l}
(n_s\cdot n_l)\tilde{B}_{sl}\biggr]
=\frac{1}{\Delta^{\prime\prime}}
\biggl[\,4\Delta^{\,\prime\prime} - \Bigl(
\sum\limits_{s=1}^{4} k_s^2 n_s^2
+ 2\sum\limits_{s<\,l}(k_s\cdot k_l)
(n_s\cdot n_l)\Bigl)\biggr]C.
\eqno{({\rm C}.4)}
$$

Let us show that expression in square brackets on the left-hand side of
Eq.\,(C.4) vanishes. Substituting instead of $\tilde{A}_s$ and
$\tilde{B}_{sl}$ their integral representations of (C.1) and (C.2) type,
we obtain
$$
\sum\limits_{s=1}^{4}n_s^2\tilde{A}_s + 2\sum\limits_{s<\,l}
(n_s\cdot n_l)\tilde{B}_{sl} =
\int\!\frac{d{\Omega}_{\bf v}}{4\pi}\,
\frac{\biggl(\,\sum\limits_{s=1}^{4} n_s(v\cdot k_s)\biggr)^2}
{(v\cdot k_1)(v\cdot k_2)(v\cdot k_3)(v\cdot k_4)}\,.
\eqno{({\rm C}.5)}
$$
Furthermore we use the same trick as supposed in Appendix B. We present
4-vector $v$ in the form of expansion over basis $\{n_s^{\mu}\}$:
$$
v^{\mu}=\sum_{s=1}^{4}\eta_sn_s^{\mu},
$$
where $\eta_s$ are coefficients of expansion. Then we can read
$$
\biggl(\,\sum\limits_{s=1}^{4} n_s(v\cdot k_s)\biggr)^2=
\biggl(\,\sum\limits_{s=1}^{4}\eta_s n_s(k_s\cdot n_s)
\biggr)^2
=\Delta^{\prime\prime}
\biggl(\,\sum\limits_{s=1}^4\eta_s n_s\biggr)^2=
\Delta^{\prime\prime}v^2=0,
$$
by virtue of condition $v^2=0$ and independence of $k_s\cdot n_s$ on $s$. Such 
we show that expression on the left-hand side of Eq.\,(C.4) vanishes. It can be also
shown by direct computing with use of explicit expression for $\tilde{A}_s$
and $\tilde{B}_{sl}$ of (C.1) and (C.2) type.

The expression in square brackets on the right-hand side of Eq.\,(C.4)
vanishes by equality
$$
k_s^2 k_s^2 + 2\sum\limits_{l\neq s}(k_s\cdot k_l)
(n_s\cdot n_l) = \Delta^{\prime\prime} \quad
(\mbox{no summation over $s$!}),
$$
being analog of identity (B.7). Such we show that coefficient of expansion
$C$ by virtue of its arbitrariness can be assumed equal to zero.\\

\end{document}